\documentclass[12pt,titlepage]{article}

\usepackage[default]{jasa_harvard}    % for formatting citations in text
\usepackage{JASA_manu}

\newcommand {\ctn}{\citeasnoun} % change to \citet if using natbib
\usepackage{graphicx,subfigure,amsmath,latexsym,amssymb}
\usepackage{float,epsfig,multirow,rotating,times}
\usepackage{upgreek,wrapfig}
\usepackage{comment}
\usepackage{slashbox}

\newcommand{\blambda}{\boldsymbol{\lambda}}

\newcommand{\bLambda}{\boldsymbol{\Lambda}}

\newcommand{\bSigma}{\boldsymbol{\Sigma}}

\newcommand{\bpi}{\boldsymbol{\pi}}

\newcommand{\bmu}{\boldsymbol{\mu}}

\newcommand{\bC}{\boldsymbol{C}}

\newcommand{\bG}{\boldsymbol{G}}

\newcommand{\bA}{\boldsymbol{A}}

\newcommand{\bp}{\boldsymbol{p}}
\newcommand{\bP}{\boldsymbol{P}}

\newcommand{\bv}{\boldsymbol{v}}

\newcommand{\bx}{\bm{x}}
\newcommand{\bX}{\boldsymbol{X}}
\newcommand{\by}{\boldsymbol{y}}
\newcommand{\bY}{\boldsymbol{Y}}

\newcommand{\bzero}{\boldsymbol{0}}

\newcommand{\bm}{\mathbf}

\numberwithin{equation}{section}
\numberwithin{algo}{section}
\numberwithin{table}{section}
\numberwithin{figure}{section}

\begin{document}

\normalsize

\title{\vspace{-0.8in}
A Bayesian Semiparametric Approach to Learning About Gene-Gene Interactions
in Case-Control Studies}
\author{Durba Bhattacharya and Sourabh Bhattacharya\thanks{
Durba Bhattacharya is an Assistant Professor in St. Xavier's College, Kolkata, pursuing PhD in 
Interdisciplinary Statistical Research Unit, Indian Statistical
Institute, 203, B. T. Road, Kolkata 700108. Sourabh Bhattacharya is an Assistant Professor in
Interdisciplinary Statistical Research Unit, Indian Statistical Institute, 203, B. T. Road, Kolkata 700108.
Corresponding e-mail: sourabh@isical.ac.in.
}}
\date{\vspace{-0.5in}}
\maketitle%

\begin{abstract}
Gene-gene interactions are often regarded as playing significant roles in influencing variabilities of complex traits. Although much research has been devoted to this area, 
to date a comprehensive statistical model that addresses the various sources of uncertainties, 
%that adequately addresses the highly dependent structures associated with the interactions between the genes, multiple loci of every gene, 
%various and unknown number of sub-populations that the subjects arise from, 
seem to be lacking. In this paper, we propose and develop a novel Bayesian semiparametric approach 
composed of finite mixtures based on Dirichlet processes and a hierarchical matrix-normal distribution that can comprehensively account for the unknown number of sub-populations 
and gene-gene interactions. Then, by formulating novel and suitable Bayesian tests of hypotheses 
%using a suitable metric for comparing clusterings in conjunction with the 
%interaction parameters of the matrix-normal, 
we attempt to single out the roles of the genes, individually, and in interaction with other genes, in case-control studies. 
We also attempt to identify the significant loci associated with the disease. Our model facilitates a highly efficient parallel computing methodology, combining Gibbs sampling 
and Transformation based MCMC (TMCMC). Application of our ideas to biologically realistic data sets revealed quite encouraging performance. We also applied our 
ideas to a real, myocardial infarction dataset, and obtained interesting results that partly agree with, and also complement, the existing works in this area, 
to reveal the importance of sophisticated and realistic modeling of gene-gene interactions. 
\\[2mm]

\end{abstract}

\section{{\bf Introduction}}
\label{sec:intro}

%\subsection{{\bf Importance of studying gene-gene interactions}}
%\label{subsec:importance}
Isolated evaluation of individual single nucleotide polymorphisms (SNPs) for their association with
complex diseases in genome wide association studies (GWAS) have succeeded in explaining
only small proportions of genetic inheritances; see \ctn{Larson13} and the references therein. 
It is now well-known that some genes interact with one another in complex networks, and that
such interactions may influence genetic variations of complex traits (see \ctn{Bonetta10},
\ctn{Moore03}, \ctn{Howard02}, \ctn{Li12}, \ctn{Moore02}). It is hence
anticipated that the want of significant success of GWAS may perhaps be due to the lack of a sophisticated 
statistical model incorporating the scientific understanding about gene-gene interactions (see \ctn{Cordell09}) 
into genomic profiling, that may help in understanding the biological and biochemical pathways behind complex diseases; 
see \ctn{Yi2011}.

%\subsection{{\bf Defining gene-gene interactions}}
%\label{subsec:statistical_definition}

One of the main challenges faced at the very onset of investigating genetic interactions is that of non-uniqueness in the definition of epistasis or gene-gene interactions. According to \ctn{Phillips08}, biologically, epistasis can be categorised as functional epistasis and compositional epistasis, both of which differ widely from the statistical definition of epistasis.
While functional epistasis indicates protein-protein interactions at molecular level, any disruption of which is explained by a genetic consequence, compositional epistasis refers to blocking of one allelic effect by an allele at another locus.
\ctn{Fisher18} (see also \ctn{Kempthorne54}) defined statistical interaction among the genes as deviations from additive marginal effects of individual genes.  
Although \ctn{VanderWeele09} derived some strong empirical conditions under which statistical interactions correspond to compositional epistasis, a prevailing opinion reflected in both genetic and epidemiological literature suggests limitations of the tests based on Fisher's statistical definition of epistasis in explaining the gene-gene interactions in the biological sense of the term.
According to \ctn{Cordell02} and \ctn{Wang10}, although statistical
and biological interpretations of interaction need not be compatible with each other, quantification of biological interaction should be based on statistical concepts of interactions.

%\subsection{{\bf Gene-level versus SNP-level interaction}} 
Several case-control studies, defining gene-gene interactions via SNP-SNP interactions have been performed (\ctn{Yi11}), 
assuming that interaction between any two genes are in fact caused by
interactions between their respective SNPs. Apart from not considering the genes as functional units, these linear 
model-based statistical analyses suffer from very high testing dimensionality and become computationally burdensome 
because of the very large number of marginal and 
pairwise interaction effects to be incorporated in the linear model when dealing with a large number of SNPs. 
The relevant statistical interaction models existing in the case-control literature
point towards the following trade-off: dimension reduction by working at the gene-level makes
computation feasible, but at the cost of useful SNP-level information (see \ctn{Larson13}), while working at SNP-level
promises all the necessary information but at the cost of enormous computational burden (see also \ctn{Musameh15}).

%\subsection{{\bf Association of the trade-offs to the additive modeling strategy 
%for both gene and SNP levels}}
%\label{subsec:additive_problems}

The aforementioned difficulties in the forms of trade-offs can be traced back to additive modeling strategies. 
Indeed, even with a small number of SNPs, the traditional linear model is constituted of a very large
number of terms consisting of the marginal and interaction effects at the SNP level. Attempts to incorporate the genes as functional units in the model necessarily calls for sacrifice of useful SNP-level information,
while principal components analysis for dimension reduction make genetic interpretation difficult.
The linear modeling strategy based on Fisher's statistical definition of epistasis can also be questioned on 
the ground of oversimplicity, since functionally, gene-gene interactions may involve very complex physical 
interplay among proteins as gene products (\ctn{Wang10}). 
Moreover, in linear models the main effects and the interaction effects are estimated
from the genotype data and then onwards assumed to be non-random covariates. 
More holistic approaches should be concerned with postulating highly structured joint distributions 
of the complex genotype data.

%\subsection{{\bf Population stratification effect on gene-gene interaction}}
%\label{subsec:stratification}

A further drawback of the existing interaction models is that they often ignore multiple sub-populations
that the genotype data usually arise from. Indeed, for different sub-populations, the genes (or SNPs)
may interact differently, which adds further complexity to the complicated functional form of epistasis. 
\ctn{Bhattacharjee10} empirically demonstrate that methods ignoring population
sub-structures can incur severe bias leading to large-scale false positives. 
%because of the presence of 
%population stratification that creates long-range linkage disequilibrium in the genome.
%
The fact that the number of sub-populations is not 
usually known is a further challenging issue that needs to be considered, as one must coherently and carefully account for the uncertainty associated with the unknown number of sub-populations.

The criticisms of the interaction models existing in the case-control literature motivated us to 
propose a new and general Bayesian model composed of
mixture distributions based on Dirichlet processes, incorporating the effects due to complex genetic interactions
through hierarchical matrix-normal-inverse-Wishart distribution.
Furthermore, we develop novel Bayesian 
hypotheses testing procedures and associated methodologies to investigate the effects of genes
on complex diseases in the context of case-control dataset arising from a possibly stratified population of genotypes. 
In what follows we investigate only the genetic effects on complex diseases, 
without taking into account the environmental effects (but see \ctn{Bhattacharya17a} and \ctn{Bhattacharya17b}).

The rest of our paper is structured as follows. In Section \ref{sec:proposal} we introduce our
proposed Bayesian semiparametric model,
and in Section \ref{sec:detection} we propose and develop a novel Bayesian hypothesis testing procedure
for detecting the roles of genes in case-control studies.
%We validate our model and methodologies with biologically realistic simulated data sets in 
%Section \ref{sec:simulation_study}. 
In Section \ref{sec:simstudy_briefing} we present a brief discussion on validation of our model
and methodologies with biologically realistic simulated data sets, the details of which are
provided in the supplement, described below.
In Section \ref{sec:realdata} we conduct a detailed analysis
of case-control data on early onset of Myocardial Infarction obtained from dbGap, comparing and 
contrasting our findings with the existing results. 
We summarize our work and make concluding remarks in Section \ref{sec:conclusion}.

Additional details are provided in the supplement, whose sections and figures have the
prefix ``S-" when referred to in this paper. 
%Specifically, in Section S-1,
%we propose a fast and efficient parallel MCMC methodology for fitting our model.
%The methodology combines efficient Gibbs sampling strategies with an efficient TMCMC scheme, while
%exploiting the conditional independence structure of our model to build a highly efficient parallel MCMC
%algorithm. In Section S-2, we include a detailed study on 
%validation of our model and methodologies with biologically realistic simulated data sets.

\section{{\bf A new Bayesian semiparametric model for gene-gene interactions}}
\label{sec:proposal}
Before we introduce our proposed model, we first detail the type of genotype and phenotype data
that we are interested in.

\subsection{{\bf Genotype data}}
\label{subsec:data}

For $s=1,2$ denoting the two chromosomes, let $x^s_{ijkr}=1$ and $x^s_{ijkr}=0$ indicate 
the presence and absence of the minor allele of 
the $i$-th individual, $j$-th gene, the $k$-th group, and $r$-th locus; 
$i=1,\ldots,N_k$; $j=1,\ldots,J$; $k=0,1$, with $k=1$ denoting case, and $r=1,\ldots,L_j$.

In this paper, we shall concern ourselves with data sets of the aforementioned type. However,
for our model, which we introduce below, it is obvious that data sets consisting of only minor allele counts at each locus
contains exactly the same information as the above described data type.

\subsection{{\bf Mixture models driven by Dirichlet processes}}
\label{subsec:mixtures}
Given any $(j,k)$, let $\bx_{ijkr}=(x^1_{ijkr},x^2_{ijkr})$, and
$\bX_{ijk}=(\bx_{ijk1},\bx_{ijk2},\ldots,\bx_{ijkL_j})$.
We assume that for every triplet $(i,j,k)$,  $\bX_{ijk}$ are independently distributed with
mixture probability mass function with a {\it maximum} of $M$ components, given by 
\begin{equation}
[\bX_{ijk}]=\sum_{m=1}^M\pi_{m jk}\prod_{r=1}^{L_j}f\left(\bx_{ijkr}\vert p_{m jkr}\right),
\label{eq:mixture1}
\end{equation}
where $f\left(\cdot\vert p_{m jkr}\right)$ is the probability mass function of independent Bernoulli
distributions, given by
\begin{equation}
f\left(\bx_{ijkr}\vert p_{m jkr}\right)=
\left\{p_{m jkr}\right\}^{x^1_{ijkr}+x^2_{ijkr}}
\left\{1-p_{m jkr}\right\}^{2-(x^1_{ijkr}+x^2_{ijkr})}.
\label{eq:pmf1}
\end{equation}
In (\ref{eq:mixture1}) and throughout the paper we use the notation $[\cdot]$ to denote the probability distribution as well as the 
probability mass or density functon.
Using allocation variables $z_{ijk}$, with probability distribution
\begin{equation}
[z_{ijk}=m]=\pi_{mjk},
\label{eq:alloc_z}
\end{equation}
for $i=1,\ldots,N_k$ and $m=1,\ldots,M$, (\ref{eq:mixture1}) can be represented as
\begin{equation}
[\bX_{ijk}|z_{ijk}]=\prod_{r=1}^{L_j}f\left(\bx_{ijkr}\vert p_{z_{ijk} jkr}\right).
\end{equation}
We may assume appropriate Dirichlet distribution priors on $\left(\pi_{1jk},\ldots,\pi_{Mjk}\right)$
for $j=1,\ldots,J$; $k=0,1$. 
However, as investigated in \ctn{Majumdar13}, the Dirichlet distribution often yields very small values
of the probabilities $\bpi_{mjk}$, thereby tending to underestimate the true number of mixture components.
On the other hand, setting $\bpi_{mjk}=1/M$ exhibited much better performance.
Therefore, in this work, we set $\pi_{mjk}=1/M$, for $m=1,\ldots,M$,
and for all $(j,k)$. 

Letting $\bp_{m jk}=\left(p_{m jk1},p_{m jk2},\ldots,p_{m jkL_j}\right)$, we further assume that 
\begin{align}
\bp_{1jk},\bp_{2jk},\ldots,\bp_{Mjk}&\stackrel{iid}{\sim} \bG_{jk};\label{eq:dp1}\\
\bG_{jk}&\sim \mbox{DP}\left(\alpha_{jk}\bG_{0,jk}\right),\label{eq:dp2}
\end{align}
where $\mbox{DP}\left(\alpha_{jk}\bG_{0,jk}\right)$ stands for Dirichlet process
with expected probability measure $\bG_{0,jk}$ having precision parameter $\alpha_{jk}$.
We assume that under $\bG_{0,jk}$, for $m=1,\ldots,M$ and $r=1,\ldots,L_j$, 
\begin{equation}
%\bG_{0,jk}=\prod_{r=1}^{L_j}\mbox{Beta}\left(\nu_{1jkr},\nu_{2jkr}\right).
p_{mjkr}\stackrel{iid}{\sim} \mbox{Beta}\left(\nu_{1jkr},\nu_{2jkr}\right).
\label{eq:dp3}
\end{equation}

Thus, given a particular pair $(j,k)$, our mixture model has the same structure as adopted by
\ctn{Majumdar13} for inference on population structure. Discreteness of Dirichlet processes
cause coincidences among the parameter vectors of $\bP_{Mjk}=\left\{ \bp_{1jk},\bp_{2jk},\ldots,\bp_{Mjk}\right\}$ 
with positive probability, so that, with positive probability, the actual number of mixture
components in (\ref{eq:mixture1}) falls below $M$, the maximum number of components, the
mixing probabilities taking the form $M^*/M$, where $1\leq M^*\leq M$. See \ctn{Majumdar13},
\ctn{Sabya11}, \ctn{Sabya12}, \ctn{Bhattacharya08}, for the details. In fact, we marginalize
over $\bG_{jk}$ to arrive at the well-known Polya urn distribution of $\bP_{Mjk}$:
\begin{equation}
\left[\bp_{mjk}|\bP_{Mjk}\backslash \{\bp_{mjk}\}\right]
\sim\frac{\alpha_{jk}}{\alpha_{jk}+M-1}\bG_{0,jk}\left(\bp_{mjk}\right)
+\frac{1}{\alpha_{jk}+M-1}\sum_{m'\neq m=1}^M\delta_{\bp_{m'jk}}\left(\bp_{mjk}\right),
\label{eq:polya}
\end{equation}
where $\delta_{\bp_{m'jk}}(\cdot)$ denotes point mass at $\bp_{m'jk}$.
The property of coincidences among the parameter vectors is clearly preserved by the Polya urn scheme.

%It is important to remark that the mixtures associated with different pairs $(j,k)$, are independent.
Observe that, after coincidences among the mixture components, the pairs $(j,k)$ come to be
associated with different mixtures, with different numbers of components. This is reasonable, because, the distributions of the genotypes for the gene $j$ of any two individuals belonging to the same subpopulation but with different case-control status, that is, $(j,k=0)$ and $(j,k=1)$ are expected to correspond to different mixtures under significant genetic effect on the disease (see \ctn{Antonyuk09}). 
%For any two genes indexed by $j$ and $j'$,
%$(j,k)$ and $(j',k)$ may also correspond to different mixtures being the
%genotype distributions of two different genes.

%Following \ctn{Majumdar13}, \ctn{Sabya12}, \ctn{Bhattacharya08} and \ctn{Antoniak74}, we set $M=30$ and $\alpha_{jk}=10$ in our applications. 
We have discussed a rule of thumb for the choice of $M$ as well as $\alpha$ in Section S-1 of
the supplement, following whch we set $M=30$ and $\alpha_{jk}=10$ in our applications.

%Following \ctn{Majumdar13}, \ctn{Sabya12}, \ctn{Sabya11}, \ctn{Bhattacharya08}, we set $M=30$ in our applications.
%It follows from \ctn{Antoniak74} that the mean and variance of the distinct parameter vectors in the set
%$\bp_{1jk},\bp_{2jk},\ldots,\bp_{Mjk}$ are both given by approximately $\alpha_{jk}\log\left(1+\frac{M}{\alpha_{jk}}\right)$.
%When prior information regarding the true number of mixture components is lacking, 
%it may be reasonable to specify the expected number of distinct components
%to be close to half of the maximum number of components possible, namely, close to $M/2$. 
%With $M=30$, we fix $\alpha_{jk}=10$, so that about $14$ distinct mixture components
%in (\ref{eq:mixture1}) are to be expected {\it a priori}. Apart from this choice, we also considered
%the possibilities $\alpha_{jk}=1$, $\alpha_{jk}\sim\mbox{Gamma}\left(0.1,0.1\right)$, that is, the gamma
%distribution with mean $1$ and variance $10$, and $\alpha_{jk}\sim\mbox{Gamma}\left(1,0.1\right)$
%(so that the mean and variance are $10$ and $100$, respectively); however, the choice $\alpha_{jk}=10$
%for all $(j,k)$ outperformed the other choices with regard to capturing the true number of mixture components.
%Hence, in this paper, we report all our results associated with $M=30$ and $\alpha_{jk}=10$.
%According to this specification, the prior mean and variance of the number of distinct components are approximately
%$14$. Thus, compared to smaller values of $\alpha_{jk}$, this choice ensures greater variability
%so that data-driven inference on the number of components receives greater weight. 

\subsection{{\bf Incorporating the gene-gene and SNP-SNP interactions through appropriate modeling of the parameters }}

\subsubsection{{\bf Modeling the parameters of $\bG_{0,jk}$}}
\label{subsubsec:model_G_0}
Taking into consideration the SNP-SNP dependence, which may exist within each gene and also among the genes, 
we model the Beta parameters $\nu_{1jkr}$ and $\nu_{2jkr}$ of (\ref{eq:dp3}) as follows:

For $r=1,\ldots,L$, where $L=\max\{L_j;~j=1,\ldots,J\}$, and for every $(j,k)$, 
\begin{align}
\nu_{1jkr}&=\exp\left(u_{r}+\lambda_{jk}\right);\label{eq:nu_1}\\
\nu_{2jkr}&=\exp\left(v_{r}+\lambda_{jk}\right).\label{eq:nu_2}
\end{align}
Allowing $u_r$ and $v_r$ to be different ensures that the mean of $p_{mjkr}$ under $\bG_{0,jk}$ depends upon the $r$-th SNP.
We further assume that for $r=1,\ldots,L$,
\begin{align}
u_{r} &\stackrel{iid}{\sim} N(0,1);\label{eq:u_r}\\
v_{r} &\stackrel{iid}{\sim} N(0,1).\label{eq:v_r}
\end{align}
We found that the Gaussian priors on $u_{r}$ and $v_{r}$ with other means and variances did not yield significantly 
different results, thus pointing towards in-built prior robustness in our modeling strategy.

Subsequently, using matrix-normal distribution as a prior on $\blambda=\left\{\lambda_{jk};~j=1,\ldots,J,~k=0,1\right\}$  
we incorporate the SNP-wise dependence in a gene. Moreover, the SNPs associated with
different genes are also dependent through the dependence structure among the genes imposed by the matrix-normal prior.

Note that allowing $u_{r}$ and $v_{r}$ to be shared by all the genes creates the impression that the labels of the loci
are not exchangeable. However, our matrix-normal-inverse-Wishart prior ensures that this is not the case, and that
for any two genes $j_1$ and $j_2$, and $k_1,k_2\in\{0,1\}$, $u_r+\lambda_{j_1k_1}$ 
(or $v_{r}+\lambda_{j_1k_1}$)
and $u_{r}+\lambda_{j_2k_2}$ (or $v_{r}+\lambda_{j_2k_2}$), and hence their exponentiated versions, 
are independent with positive probability, given the data.
This is elucidated in Section S-3 of the supplement in light of the matrix-normal-inverse-Wishart prior.
We vindicated this mathematical argument with simulation studies; see Section \ref{sec:simstudy_briefing} 
(details provided in Section S-7 of the supplement).
Specifically, we conducted extra simulation studies after randomly permuting the labels of the loci of each gene and  
re-analyzed each such data set. The results, provided in Section S-7 of the supplement, 
are consistent with those obtained without permuting the labels.

\subsubsection{{\bf Matrix normal prior for $\blambda$}}
\label{subsubsec:matrix_normal}

We consider the following model for $\blambda$:
\begin{equation}
\blambda\sim %N\left(\bzero,\bC\right), 
N\left(\bmu,\bA\otimes\bSigma\right).
\label{eq:mn1}
\end{equation}
Re-writing the $2J$-dimensional vector $\blambda$ as a $J\times 2$ matrix $\bLambda$, 
(\ref{eq:mn1}) can be represented as a matrix normal distribution with mean matrix $\bmu^{J\times 2}$, %$\bzero^{J\times 2}$,
left covariance matrix $\bA$ and right covariance matrix $\bSigma$, having probability density function
\begin{equation}
\pi(\bLambda)=\frac{\exp\left[-tr\left\{\bSigma^{-1}\left(\bLambda-\bmu\right)^T\bA^{-1}\left(\bLambda-\bmu\right)\right\}\right]}
{\left(2\pi\right)^J\left|\bA\right|^2\left|\bLambda\right|^J}.
\label{eq:pi_Lambda}
\end{equation}
We note that the $k$-th column of $\bLambda$, which we denote by $\bLambda^{col,k}$, follows the multivariate 
normal distribution:
\begin{equation}
\bLambda^{col,k}\sim N_J\left(\bmu^{col,k},\sigma_{kk}\bA\right),
\label{eq:mvn_col}
\end{equation}
where $\bmu^{col,k}$ is the $k$-th column of $\bmu$.
The covariance matrix between $\bLambda^{col,k_1}$ and $\bLambda^{col,k_2}$ is given by
\begin{equation}
cov\left(\bLambda^{col,k_1},\bLambda^{col,k_2}\right)=\sigma_{k_1k_2}\bA.
\label{eq:mvn_cov}
\end{equation}
Similarly, the $j$-th row of $\bLambda$, which we denote by $\bLambda^{row,j}_{s}$, has the following
multivariate normal distribution:
\begin{equation}
\bLambda^{row,j}\sim N_2\left(\bmu^{row,j},a_{jj}\bSigma\right),
\label{eq:mvn_row}
\end{equation}
$\bmu^{row,j}$ being the $j$-th row of $\bmu$.
Also,
\begin{equation}
cov\left(\bLambda^{row,j_1},\bLambda^{row,j_2}\right)=a_{j_1j_2}\bSigma.
\label{eq:mvn_cov2}
\end{equation}
In our applications we chose $\bmu=\bzero$.

The essence of the matrix normal distribution is to offer a dependence structure among the genes. Given case-control status $k$, the dependence structure associated with the genes
is provided by $\bA$, while the matrix $\bSigma$ represents the dependence between the genotype distribution of the cases and controls, given
any particular gene.

\subsubsection{{\bf Priors on $\bA$ and $\bSigma$}}
\label{subsubsec:other_priors}
We assume that
\begin{equation}
\bA\sim\mathcal{I}\mathcal{W}\left(\xi,\bA_0\right),
\label{eq:iw1}
\end{equation}
where $\mathcal{I}\mathcal{W}\left(\xi,\bA\right)$ stands for Inverse-Wishart distribution with degrees of freedom
$\xi~(\geq J)$ and positive definite scale matrix $\bA$. 
The density function is given by
\begin{equation}
\pi(\bA) %=\frac{\left|\bA_0\right|^{\frac{\xi}{2}}}{2^{\frac{\xi J}{2}}\Gamma_J\left(\frac{\xi}{2}\right)}
\propto \left|\bA\right|^{-\left(\frac{\xi+J+1}{2}\right)}\times\exp\left\{-\frac{1}{2}tr\left(\bA_0\bA^{-1}\right)\right\}.
\label{eq:pi_A}
\end{equation}
We further assume that
\begin{equation}
\bSigma\sim\mathcal{I}\mathcal{W}\left(\zeta,\bSigma_0\right),
\label{eq:iw2}
\end{equation}
where the degrees of freedom $\zeta$ satisfies $\zeta~(\geq 2)$ and $\bSigma_0$ is a $2\times 2$ 
positive definite matrix; the density function is given by
\begin{equation}
\pi(\bSigma) %=\frac{\left|\bSigma_0\right|^{\frac{\zeta}{2}}}{2^{\frac{2\zeta }{2}}\Gamma_2\left(\frac{\zeta}{2}\right)}
\propto \left|\bSigma\right|^{-\left(\frac{\zeta+3}{2}\right)}
\times\exp\left\{-\frac{1}{2}tr\left(\bSigma_0\bSigma^{-1}\right)\right\}.
\label{eq:pi_Sigma}
\end{equation}
For our applications, we set $\xi=J+2$ and $\zeta=4$.
These choices are the minimum values such that the prior expectations of $\bA$ and $\bSigma$ are well-defined.
Choices of $\bA_0$ and $\bSigma_0$ are detailed in Section S-2 of the supplement.

A schematic representation of our model and the parallel processing algorithm is provided in Figure \ref{fig:schematic}.
Details of our parallel processing algorithm are provided in Section S-4 of the supplement.
\begin{figure}%[htp]
\centering
\includegraphics[width=10cm,height=10cm]{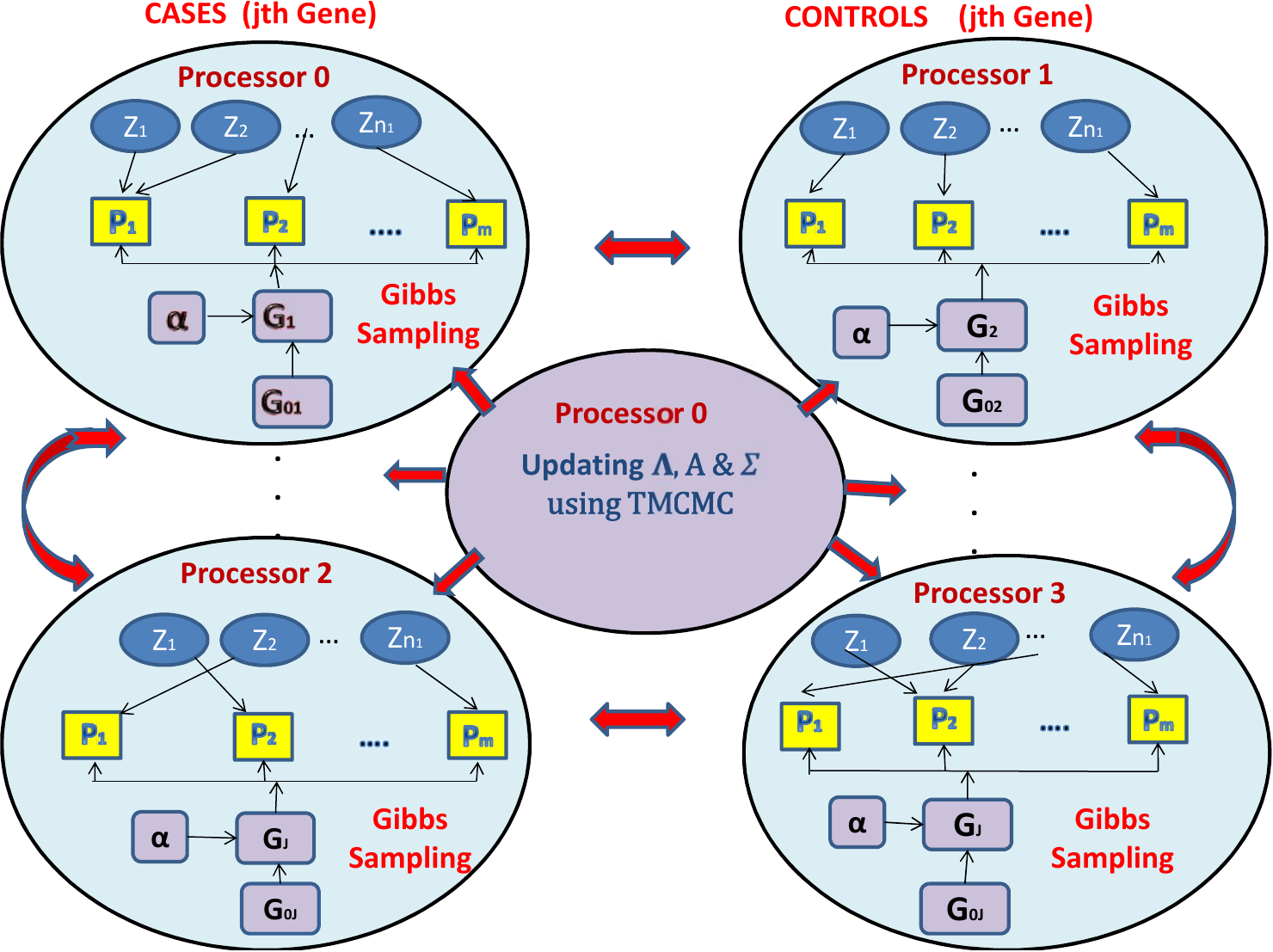}
\caption{{\bf Schematic diagram for our model and parallel processing idea:} The arrows in the diagram
represent dependence between the variables. The ranks of the processors updating the sets of parameters in parallel 
using Gibbs sampling are also
shown. Once the other parameters are updated in parallel, the interaction parameters are updated using TMCMC 
by the processor with rank zero.}
\label{fig:schematic}
\end{figure}

\section{{\bf Detection of the roles of genes in case-control studies}}
\label{sec:detection}

\subsection{{\bf Formulation of a Bayesian hypothesis testing procedure}}
\label{subsec:test_formulation}

%Recall the definition of $\bar{w}_{ijk}$ from (\ref{eq:w_bar1}). 
%%Let $\bar{W}_{ijk}$ denote the random variable associated with the observation $\bar{w}_{ijk}$.
%Let $h_{0j}$ and $h_{1j}$ denote the distributions of $\bar{w}_{ijk=0}$ and $\bar{w}_{ijk=1}$, respectively.
%Clearly, $h_{0j}$ and $h_{1j}$ are also $M$-component mixtures, where the $m$-th mixture component of the respective
%distributions is characterized by $\bp_{mjk=0}$ and $\bp_{mjk=1}$, haing the same mixing probability $1/M$.
%%If, for case ($k=1$) and control ($k=0$), the distributions of the form (\ref{eq:mixture1}) are

%If $h_{0j}$ and $h_{1j}$ are not significantly different, then it is plausible to conclude that the role of genes is not
%significant in the case-control study. As such,
In order to investigate if genes have any significant effect on case-control, it is pertinent to test
\begin{equation}
H_0: h_{0j}=h_{1j}; ~j=1,\ldots,J,
%H_0:\sum_{m=1}^M\pi_{m jk=0}\prod_{r=1}^{L_j}f\left(\cdot\vert p^r_{m jk=0}\right)
%=\sum_{m=1}^M\pi_{m jk=1}\prod_{r=1}^{L_j}f\left(\cdot\vert p^r_{m jk=1}\right);~j=1,\ldots,J,
\label{eq:H_0}
\end{equation}
versus
\begin{equation}
H_1:\mbox{not}~H_0,
\label{eq:H_1}
\end{equation}
where
\begin{align}
h_{0j}(\cdot)&=\sum_{m=1}^M\pi_{m jk=0}\prod_{r=1}^{L_j}f\left(\cdot\vert p^r_{m jk=0}\right)\label{eq:h_0}\\
h_{1j}(\cdot)&=\sum_{m=1}^M\pi_{m jk=1}\prod_{r=1}^{L_j}f\left(\cdot\vert p^r_{m jk=1}\right).\label{eq:h_1}
\end{align}
If $h_{0j}$ and $h_{1j}$ are not significantly different, then it is plausible to conclude that the role of genes is not
significant in the case-control study. 

In a nutshell, testing the hypothesis in (\ref{eq:H_0})-(\ref{eq:h_1}) requires some appropriate 
divergence measure between $h_{0j}$ and $h_{1j}$ and if $d(h_{0j},h_{1j})$
denotes the divergence, then $H_0$ is to be accepted for appropriately large posterior probability of 
the event that $\underset{1\leq j\leq J}{\max}~d(h_{0j},h_{1j})$ is small. 

In our situation, simultaneous consideration of a large number of genes, involving thousands of SNPs 
renders the existing measures of divergence practically infeasible to compute. 
For details, see Section S-5 of the supplement. 
This compels us to seek alternative measures of divergence that are also amenable to efficient computation. 
In the mixture context, a natural measure is the discrepancy between the clusterings associated with 
the two mixture distributions 
(\ctn{Sabya11}, \ctn{Ghosh08}). 
However, since clusterings do not account for the magnitudes of the parameters, insignificant difference between 
the clusterings does not necessarily imply 
insignificant difference between the associated mixture densities. 
It is worth noting that even the Euclidean metric alone is not appropriate -- since mixture densities are
invariant with respect to permutations of the parameter components, large Euclidean distances between
parameter vectors need not imply large difference between the densities.
Thus, we propose to use the clustering
based ideas in conjunction with ideas based on the Euclidean metric, appropriately modified for mixture densities.
Indeed, if the clusterings associated with the mixture
densities are known to be of insignificant difference, then insignificant Euclidean divergence between the
parameter vectors does imply insignificant difference between the mixture densities. 
Details of all these issues are provided in Section S-6 of the supplement.

%Thus, if based on the clustering metric $H_0$ is not implausible, we perform a significance test 
%based on the Euclidean distance;
%we reject $H_0$ for significantly large Euclidean distance between the two logit-transformed parameter vectors. 
 
%Next, we provide some details regarding an appropriate clustering metric, which we shall consider in our applications.

\subsection{{\bf Formal Bayesian hypothesis testing procedure integrating the above developments}}
\label{subsec:testing}

With the clustering metric $\hat d$ provided in Section S-6 of the supplement, %(\ref{eq:approx_final}), 
let us define
\begin{equation*}
d^*=\max_{1\leq j\leq J}d_j,
\label{eq:d_star}
\end{equation*}
where
\begin{equation*}
d_j=\hat d\left(\bP_{Mjk=0},\bP_{Mjk=1}\right)
\label{eq:d_j}
\end{equation*}
is the distance between the clusterings
$\bP_{Mjk=0}=\left\{ \bp_{1jk=0},\bp_{2jk=0},\ldots,\bp_{Mjk=0}\right\}$
and $\bP_{Mjk=1}=\left\{ \bp_{1jk=1},\bp_{2jk=1},\ldots,\bp_{Mjk=1}\right\}$, for $j=1,\ldots,J$.
With this, we first test
\begin{equation}
H_{0d^*}:~d^*< \varepsilon\hspace{2mm}\mbox{versus}\hspace{2mm}H_{1d^*}:~d^*\geq\varepsilon,
\label{eq:hypothesis2}
\end{equation}
for reasonably small choice of $\varepsilon$ ($>0$).
Acceptance of $H_{0d^*}$ indicates that clusterings associated with $h_{0j}$ and $h_{1j}$ have
insignificant difference, for every $j=1,\ldots,J$. If $H_{0d^*}$ is rejected, then it follows that
for at least one $j\in\{1,\ldots,J\}$, the clustering difference is significant.

If $H_{0d^*}$ is rejected, then it entails that the clusterings associated with the mixture densities
for case and control are significantly different. This implies that the mixture densities themselves are
significantly different, so that rejection of $H_{0d^*}$ leads to rejection of $H_0$ given by (\ref{eq:H_0}).

But whenever $H_{0d^*}$ is accepted based on the ``$0-1$" loss and the clustering metric, 
as already argued, %in Section \ref{subsubsec:clustering_shortcoming}, 
this does not necessarily 
imply that the mixture densities have insignificant difference. 
All one can infer in this case is that differences between the associated clusterings are insignificant.  
Hence, when $H_{0d^*}$ is accepted, we consider a second test of the form
\begin{equation}
H_{0d^*_E}:~d^*_E< \varepsilon\hspace{2mm}\mbox{versus}\hspace{2mm}H_{1d^*_E}:~d^*_E\geq\varepsilon,
\label{eq:hypothesis_euclidean}
\end{equation}
where $d^*_E=\underset{1\leq j\leq J}{\max}~d_{E,j}$; here $d_{E,j}$ is the Euclidean distance between
$$\mbox{logit}\left(\bar{\bP}_{Mjk=0}\right)=\left\{\mbox{logit}\left(\bar{p}_{1jk=0}\right),
\mbox{logit}\left(\bar{p}_{2jk=0}\right),
\ldots, \mbox{logit}\left(\bar{p}_{Mjk=0}\right)\right\}$$
and 
$$\mbox{logit}\left(\bar{\bP}_{Mjk=1}\right)=\left\{\mbox{logit}\left(\bar{p}_{1jk=1}\right),
\mbox{logit}\left(\bar{p}_{2jk=1}\right),
\ldots, \mbox{logit}\left(\bar{p}_{Mjk=1}\right)\right\},$$
with
$\bar{p}_{mjk}=\sum_{r=1}^{L_j}p_{m,jkr}/L_j$, and
$\mbox{logit}\left(\bar{p}_{mjk}\right)=\log\left\{\bar{p}_{mjk}/(1-\bar{p}_{mjk})\right\}$; see Section S-6
of the supplement. 

If $H_{0d^*_E}$ is also accepted, then one can safely accept $H_0$. 
If $H_{0d^*_E}$ is rejected, %as argued in Section \ref{subsubsec:avoid_permutations},
we then consider a third test of the form
\begin{equation}
H_{0d^*_{E,\min}}:~d^*_{E,\min}< \varepsilon\hspace{2mm}\mbox{versus}\hspace{2mm}H_{1d^*_{E,\min}}:~d^*_{E,\min}\geq\varepsilon,
\label{eq:hypothesis_euclidean2}
\end{equation}
where $d^*_{E,\min}=\underset{1\leq j\leq J}{\max}~d_{E,\min,j}$, with
$d_{E,\min,j}=d_{E,\min}\left(\mbox{logit}\left(\bar{\bP}_{Mjk=0}\right),
\mbox{logit}\left(\bar{\bP}_{Mjk=1}\right)\right)$. 
Here $d_{E,\min}$ is a pseudo-metric based on the Euclidean distance;
see Section S-6 for details.

\begin{figure}%[htp]
\centering
\includegraphics[width=10cm,height=10cm]{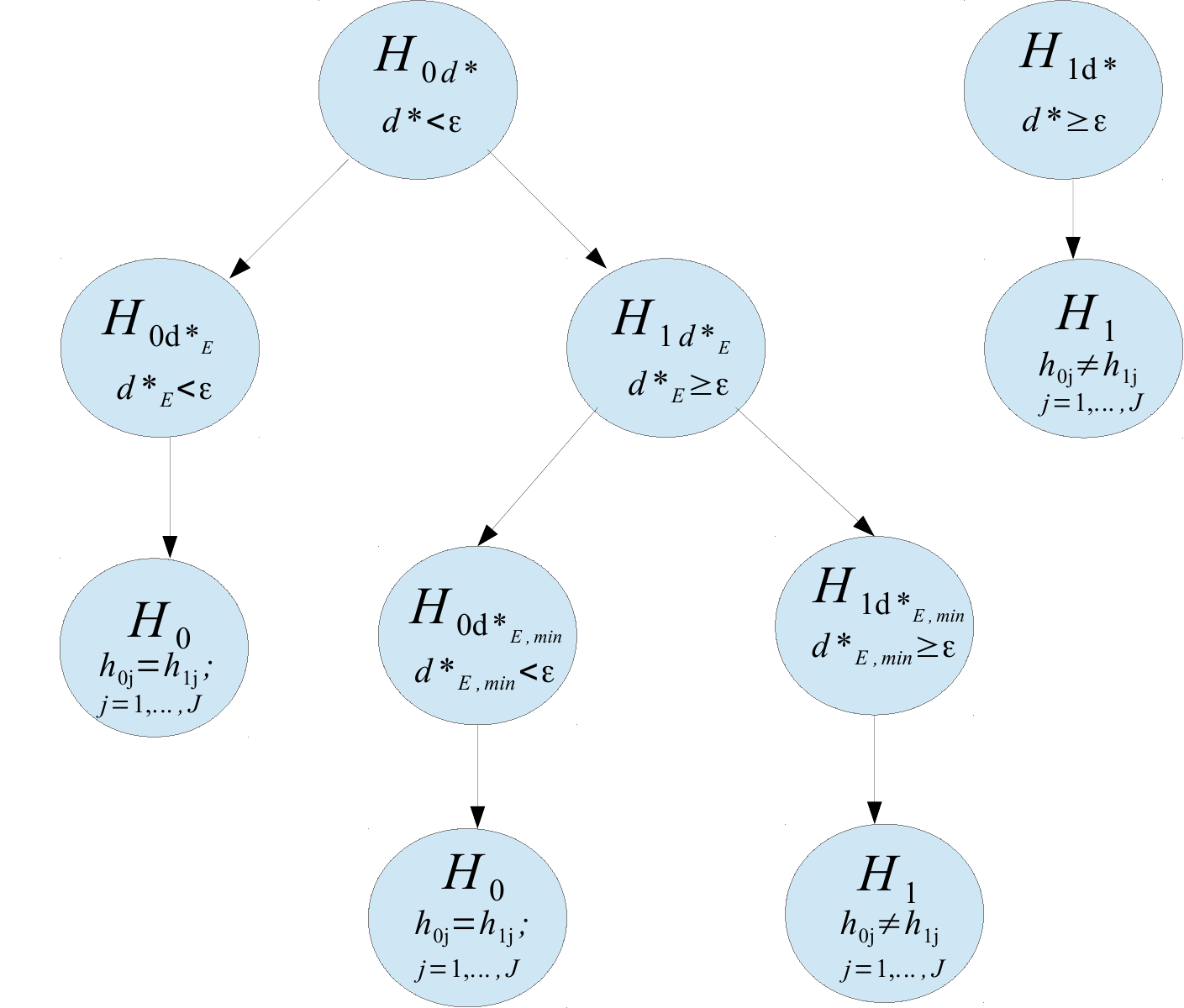}
\caption{{\bf Schematic diagram for the hierarchy of Bayesian tests.} 
}
\label{fig:hypothesis_schematics}
\end{figure}

If $H_{0d^*_{E,\min}}$ is accepted, then this implies acceptance of $H_0$ given by (\ref{eq:H_0}).
Else, $H_0$ must be rejected.
For clarity, we present a schematic diagram of the hierarchy of the hypotheses tests in Figure \ref{fig:hypothesis_schematics}. 

\subsubsection{{\bf Choice of loss function for the Bayesian tests}}
\label{subsubsec:loss_function}
Recall that the ``$0-1-c$" loss function (see, for example, \ctn{Schervish95}, for details)
entails zero loss under the correct decisions, %regarding choices of $H_0$ and $H_1$, 
loss $1$ under false acceptance of the null hypothesis and loss $c~(c>0)$ under false rejection of the null. 
Thus, we accept $H_{0d^*}$ under the ``$0-1-c$" loss function if
\begin{equation}
P\left(d^*<\varepsilon|\mbox{Data}\right)\geq\frac{1}{1+c}.
\label{eq:posterior_H_0}
\end{equation}
%where $c>0$.
%for some reasonable choice of $\varepsilon>0$. 
%In fact, one can choose $\varepsilon$ such that
%the set $\{d^*<\varepsilon\}$ boils down to the 95\% highest posterior density credible region
%of the posterior of $d^*$.
Since usually there is no clear-cut way of specifying $c$, we shall generally select the default value
$c=1$, reducing the ``$0-1-c$" loss function to the well-known ``$0-1$" loss function.

However, for the tests associated with $d^*_E$ and $d^*_{E,\min}$, which are to be considered
only if $H_{0d^*}$ is accepted,
%for re-confirmation
%of $H_0$ using the Euclidean metric based significance test, 
we shall select $c$ much
larger than 1. This is because %as already pointed out in Section \ref{subsubsec:avoid_permutations}, 
it makes sense to provide greater protection to the null hypothesis $H_0$ given that the clustering test 
has already provided partial support to $H_0$, indicating that at least the clusterings are not
significantly different (see also Section S-6 of the supplement). 
%Secondly, the Euclidean distance may be large even if the two vectors
%are the same but differing only by a permutation; some protection against this issue can be achieved
%by setting $c$ to be reasonably large. 

Note that, under the ``$0-1$" loss, the null hypothesis is to be accepted if its posterior
probability exceeds $1/2$, while under the ``$0-1-c$" loss, the threshold  posterior probability
is $1/(1+c)$. For the hypotheses involving $d^*_E$ and $d^*_{E,\min}$ %In our applications with the ``$0-1-c$" loss,  
we shall set $c=19$ so that $1/(1+c)= 0.05$. This choice is motivated by
the $5\%$ level of significance of classical significance tests. Some other choices  
will also be briefly touched upon.

\subsubsection{{\bf Choice of $\varepsilon$}}
\label{subsubsec:choice_epsilon}
Choices of $\varepsilon$ are expected to be problem specific. 
In Section S-7 %S-4.1.2 
of the supplement, %\ref{subsubsec:threshold}, 
we discuss in detail the choices of $\varepsilon$ in our applications. Briefly,
we first consider an appropriate null model, for instance the same model as ours but
with $\bA$ and $\bSigma$ set to identity matrix to reflect the null hypotheses of 
``no interaction" and same mixture distributions  under cases and controls for each gene for no genetic effect.
We then generate case-control genotype data from the null model and fit our general Bayesian model
to the ``null data" and set $\varepsilon$ as the $55$-th percentile of the relevant posterior distribution.

\subsection{{\bf Bayesian tests for individual genetic effects when $H_0$ is rejected}}
\label{subsec:individual_tests}
If $H_0$ given by (\ref{eq:H_0}) is finally accepted 
%based on our clustering metric and re-confirmed by the significance test
%based on the Euclidean metric, 
then we may conclude that there is no significant evidence to 
claim that the genes, individually, or in interaction with the
other genes, are important factors in the case-control study.

On the other hand, if $H_0$ is rejected, then we check for significances of the individual genes 
by applying our Bayesian testing procedure on the hypotheses 
\begin{equation}
H_{0j}:~h_{0j}=h_{1j}~\mbox{versus}~H_{1j}:~h_{0j}\neq h_{1j},~\mbox{for}~j=1,\ldots,J.
\label{eq:H0j}
\end{equation}
For each $j=1,\ldots,J$, we adopt the same procedure for testing the hypothesis as 
for testing $H_0$ versus $H_1$; only $d^*$, $d^*_E$ and $d^*_{E,\min}$ are to be replaced with
$d_j$, $d_{E,j}$ and $d_{E,\min,j}$, respectively. Note that at each stage associated with 
$d^*$, $d^*_E$ and $d^*_{E,\min}$, the Bayesian hypotheses testing framework is equivalent to
a Bayesian multiple testing paradigm. Specifically, testing $H_{0d_j}:d_j<\varepsilon$ versus $H_{1d_j}:d_j\geq\varepsilon$
for $j=1,\ldots,J$ using our Bayesian methods is equivalent to minimizing the Bayes risk of the additive ``0-1" loss function,
and the subsequent Bayesian tests for the hypotheses 
$H_{0d_{E,j}}:d_{E,j}<\varepsilon$ versus $H_{1d_{E,j}}:d_{E,j}\geq\varepsilon$
and $H_{0d_{E,\min,j}}:d_{E,\min,j}<\varepsilon$ versus $H_{1d_{E,\min,j}}:d_{E,\min,j}\geq\varepsilon$, for relevant 
indices $j$, are Bayesian multiple testing procedures that minimize the Bayes risk of the additive ``0-1-c" loss function,
where we choose $c=19$.

%(we re-formulate the $j$-th hypothesis as 
%$H_{0j}:~\hat d\left(\bP_{Mjk=0},\bP_{Mjk=1}\right)<\varepsilon$ versus
% $H_{1j}:~\hat d\left(\bP_{Mjk=0},\bP_{Mjk=1}\right)\geq \varepsilon$ in conjunction with
%the Euclidean distance based significance test), then the $j$-th gene
%%individually, or in interaction with the other genes, 
%plays a significant role in the study.
%In order to detect if the $j$-th gene is individually important, we set the off-diagonal elements of $\bA$ 
%to be zero, so that
%there is no interaction between the genes, 
%%associated with the $j$-th gene, 
%and re-conduct the test $H_{0j}:~h_{0j}=h_{1j}$ versus 
%$H_{1j}:~h_{0j}\neq h_{1j}$. If $H_{0j}$ is rejected, then one may conclude that at least the $j$-th 
%gene is individually important. 
If $H_{0j}$ is accepted, then it is possible that the $j$-th gene is not
individually influential, but some interaction effect involving the $j$-th gene may be significant. To check which interactions
are significant (we may check this even if $H_{0j}$ is rejected, since the $j$-th gene may be marginally
significant as well as interactive with the other genes), one may conduct the tests 
$H_{0,j,j^*}:~\left|\bA_{jj^*}\right|<\varepsilon$ 
versus $H_{1,j,j^*}:~\left|\bA_{jj^*}\right|\geq\varepsilon$,
for $j^*\neq j$,  
%\begin{equation}
%\rho_{jj^*} = \frac{\bA_{jj^*}}{\sqrt{\bA_{jj}\bA_{j^*j^*}}};
%\label{eq:rho}
%\end{equation}
$\bA_{jj^*}$ being the $(j,j^*)$-th
element of $\bA$.
Acceptance of $H_{1,j,j^*}$ for some (or many) $j^*\neq j$, indicates which of the genes
interact with the $j$-th gene to contribute significantly to the underlying case-control study.

\section{{\bf Validation of our model and methodologies with biologically realistic simulated data sets}}
\label{sec:simstudy_briefing}
We evaluate our model and methodologies on data sets generated from the GENS2 software
designed by \ctn{Pinelli12}. In a nutshell, the software creates large, biologically realistic data sets
having realistic LD patterns,
where risks of complex diseases are influenced by known gene-gene and gene-environment interactions.
We consider two simulated data sets for our experiments -- in the first experiment, we generate a 
case-control data set under the effect of
gene-gene interaction, fit our model to the data set, and test the relevant hypotheses. We show that
our model and methodology successfully captures the relevant information regarding the effects of the
individual genes, gene-gene interaction, and the number of sub-populations. %Quite importantly, 
We also show that, in spite of LD, our model succeeds in capturing the close neighborhoods of the actual
disease predisposing loci (DPL) of the genes.

In the second experiment we generate a data set where
disease risk is devoid of any genetic effect and is influenced only by some environmental exposure. Application of
our model and methods to this data set again successfully captures the correct situation, clearly
indicating lack of genetic influence. 

For both the simulated data sets we perform simulation experiments by randomly permuting the labels of the loci
of each gene. For both cases we obtain results associated with the permuted labels that completely support those
obtained from the original simulated data sets.

Details are provided in Section S-7 of the supplement.

\section{{\bf Application of our model and methodologies to a real, case-control dataset on Myocardial Infarction}}
\label{sec:realdata}
Application of our ideas to a case-control dataset on early-onset of myocardial infarction (MI) from MI Gen study, obtained from the dbGaP database
({\bf http://www.ncbi.nlm.nih.gov/gap}), led to some interesting findings.

MI (more commonly, heart attack), is a complex disease and is a leading cause
of death and disability all over the world. Much investigation has been carried out for detecting the genetic causes
of myocardial infarction, all of which are based on the assumption that 
the main contributory factors for the disease are the mutations in the proteins associated with 
the pathophysiology of atherosclerosis (see \ctn{Erdmann10}).

Although the GWA studies have revealed a lot of genetic information regarding
MI (an overview of the main results can be found in \ctn{Erdmann10}), only a very few of the detected genes are related to traditional
risk factors (LDL-cholesterol, diabetes and LP[a] etc.), and the other genes increase the risk by pathogenetic mechanisms
that are not yet properly understood. Despite much success in deciphering the marginal effects of many SNPs, not much has been achieved in the gene-gene interaction front. According to \ctn{Musameh15}, burden of multiple testing renders the standard
GWAS samples underpowered to detect such effects, while \ctn{LucasG12} blame the complexity of the epistatic effects as a reason behind the difficulty in detecting them. 

\subsection{{\bf Data description}}
\label{subsec:myo_data}
%The MI Gen data obtained from dbGaP consists of observations on presence/absence of
%minor alleles at $727478$ SNP markers associated with 22 autosomes and the sex chromosomes of $2967$ cases of early-onset myocardial infarction, $3075$ age and sex matched controls. 
%The average age at the time of MI was 41 years among the male cases and 47 years among the female cases. 

The MI Gen data obtained from dbGaP broadly represents a mixture of four sub-populations: Caucasian, Han Chinese,
Japanese and Yoruban. Since the names of the genes were not provided in the dataset, SNPs were mapped on to the corresponding genes using the  Ensembl human genome database ({\bf http://www.ensembl.org/}). However, technical glitches prevented us from
obtaining information on the genes associated with all the markers. As such, we could categorize
$446765$ markers out of $727478$ with respect to $37233$ genes.

For our analysis, we considered a set of SNPs that are found to be individually associated with different cardiovascular end points like LDL cholesterol, smoking, blood pressure, body mass etc. in various GWA studies published in NHGRI catalogue and augmented this set further with another set of SNPs found to be marginally associated with MI in the MIGen study (see \ctn{LucasG12}). Our study also includes SNPs that are reported to be associated with MI in various other studies, see \ctn{Erdmann10}, \ctn{LuQi11} and \ctn{Wang04}. In all, we obtained 271 SNPs.
Unfortunately, only 33 of them turned out to be common to the SNPs of our original MI dataset on genotypes, which has been mapped on to the genes using the Ensembl human genome database.
However, we included in our study all the SNPs associated with the genes containing the 33 common SNPs. Specifically, our study involves the genotypic information on 32 genes covering 1251 loci, including
the 33  previously identified loci for all the $6042$ individuals available in our dataset.

Categorization of the case-control genotype data into the four sub-populations, each of which are likely
to represent several further and rather varied sub-populations genetically, implies that  
the maximum number of mixture components must be fixed at some value much higher than $4$. As before, we set $M=30$
and $\alpha_{jk}=10$ for every $(j,k)$, to facilitate data-driven inference. 
Interestingly, the distributions of the number of distinct components for $\alpha_{jk}=1.5$
(so that the prior mean and variance are approximately $5$) were not significantly different from
those of $\alpha_{jk}=10$, indicating prior robustness.

We chose a similar set-up for the null model. That is, we chose the same number of genes and the
same number of loci for each gene, the same number of cases and controls, the same value $M=30$, but
$\alpha_{jk}=1.5$ for every $(j,k)$, as in our simulation studies. 
%As in the simulation studies, this entails that about $5$ components are to be expected 
%{\it a priori} and {\it a posteriori} under the null model for each $(j,k)$ pair.
We use the same priors as in the real data set-up except that we set
$\bA$ and $\bSigma$ to be identity matrices to ensure that the genetic interaction is not present 
and set the same mixture distribution under cases and controls for each gene to ensure the absence of genetic effects. 
For details see Section 4.1.2 of the supplement.

\subsection{{\bf Remarks on model implementation}}
\label{subsec:myo_implementation}
We implemented our parallel MCMC algorithm detailed in Section S-4 of the supplement for posterior simulation on a 
VMware consisting of $60$ double-threaded, $64$-bit physical cores, each running at $2.5$ GHz; $50$ such cores
were available to us. The mixture components associated with $\left\{(j,k):j=1,\ldots,J;k=0,1\right\}$, with
$J=32$, are updated in parallel, on $64$ of the total $100$ available threads. This is followed
by updating the interaction parameters on a separate processor using a mixture of additive and
additive-multiplicative TMCMC (see Section S-4.1 of the supplement). However, in this problem, the
interaction matrix $\bA$ is of order $32\times 32=1024$, and the associated Cholesky decomposition (see Section S-4.1
of the supplement) then consists of $33\times 16=528$ parameters. Furthermore, here $\blambda$ is a $2J=64$-dimensional
vector, $\left\{(u_r,v_r):r=1,\ldots,L\right\}$, where $L=207$, consists of $2\times 207=414$ parameters
and $\bSigma$, with its Cholesky decomposition, consists of $3$ unknowns. 
Hence, in all, there are $1009$ interaction parameters to be updated.

Updating too many parameters in a single block, even with TMCMC, need not guarantee
automatic efficiency. Here we consider updating sub-blocks of parameters at a time using additive TMCMC. 
Specifically, we update $\bLambda$
by updating the $64$-dimensional $\bLambda_k=\left\{\lambda_{jk}:j=1,\ldots,J\right\}$ separately for $k=0,1$;
we also update the blocks $\left\{u_r:r=1,\ldots,L\right\}$ and $\left\{v_r:r=1,\ldots,L\right\}$ and $\bSigma$ separately.
Since $\bA$ consists of $528$ parameters, at each iteration we update only $32$ randomly chosen non-zero elements of the 
Cholesky factor of $\bA$ using additive TMCMC. The latter is certainly a valid TMCMC strategy, 
which is theoretically a mixture of TMCMC strategies (see, for example, \ctn{Tierney94} in the context of Metropolis-Hastings), 
and maintains very reasonable acceptance rate in our application.

The above parallel MCMC algorithm takes about $31$ hours to yield $30,000$ iterations in our aforementioned
VMware machine. We discard the first $10,000$ iterations as burn-in.
%Implementation of the null model took about $19$ hours and $21$ minutes to yield $30,000$ iterations in the same machine.
%The reduced time compared to the real data implementation is because of lesser number of mixture components
%under the null model.
% Recall that gene-gene interactions in the non-null model influence the posterior distributions
%of the number of components, while this is not the case for the null model which lacks gene-gene interaction.
%This issue seems to be responsible for the discrepancy between the posteriors of the number of distinct components
%in the null and non-null models.
%
Informal convergence diagnostics such as trace plots exhibited adequate mixing properties
of our parallel algorithm.
\subsection{{\bf Results of the real data analysis}}
\label{subsec:realdata_results}

\subsubsection{{\bf Influential genes obtained from our analysis}}
\label{subsubsec:influential_genes}
Our Bayesian hypotheses testing using both clustering metric and the Euclidean distance reveal
that there is very significant overall genetic influence on MI. Indeed, it turned out that 
$P\left(d^*<\epsilon_1|\mbox{Data}\right)\approx 0.0335$ and $P\left(d^*_E<\epsilon_2|\mbox{Data}\right)\approx 0$,
where $\epsilon_1$ and $\epsilon_2$ are the $55$th percentiles of the null distributions of $d^*$ and $d^*_E$.
Furthermore, testing for the effects of the genes individually using the clustering metric showed that apart from
only $5$ genes, namely, AP006216.10, AP006216.5, APOC1 and OR4A48P and AP00621.5, all other genes have 
significant effect on MI, while
with the Euclidean metric, all the genes considered for study turned out to be significant.
The posterior probabilities of the null hypotheses (of no significant genetic influence) are shown in 
Figure S-4 %\ref{fig:null_hypotheses} 
of the supplement.

Interestingly, with
respect to the Euclidean metric, all the five posterior probabilities of the null hypotheses associated
with the aforementioned 5 genes, %AP006216.10, AP006216.5, APOC1 and OR4A48P and AP006216.5, 
turned out to be empirically zero.
Thus, even though the clustering metric accepts 5 null hypotheses, the
confirmation tests with the Euclidean distances suggest rejection of all of them. We hence conclude
that all the genes considered in the study have significant effect on MI. This is in keeping with the fact that the genes considered in our study were found to be associated with different cardiovascular endpoints in various GWA studies or have been confirmed to play important roles in causing MI in earlier studies.

\subsubsection{{\bf Disease predisposing loci detected by our Bayesian analysis}}
\label{subsubsec:our_DPL}
%\ctn{Erdmann10} enlists $11$ SNPs which have been flagged by GWAS investigations as having significant effects
%on MI. Such reportedly important markers, which are available in our dataset, 
%are $rs599839$, $rs6725887$, $rs9818870$, $rs12526453$, $rs2048327$, $rs3127599$, $rs7767084$, 
%$rs10755578$, $rs1333049$ and $rs2259816$. In our notation, these
%correspond to Gene-1 to Gene-9, with number of loci 
%$6$, $5$, $17$, $177$, $39$, $15$, $19$, $21$ and $8$, respectively. Note that $rs7767084$ and 
%$rs10755578$ represent the same gene. 
%We now report the DPLs that we obtain by our analysis
%and show that % the existing ones %For brevity of exposition, we consider 15 genes containing 
%important SNPs, 
We now show that the most influential SNPs corresponding to the maximum Euclidean distance in each of the significant 
genes in our study, which we continue to refer to as the DPLs, are usually close to, 
and sometimes exactly the same as the SNPs, already flagged by the earlier studies as influential.

\begin{figure}%[htp]
\centering
\subfigure[DPL of $HLA-DRA$.]{ \label{fig:DPL_gene_10}
\includegraphics[width=6cm,height=6cm]{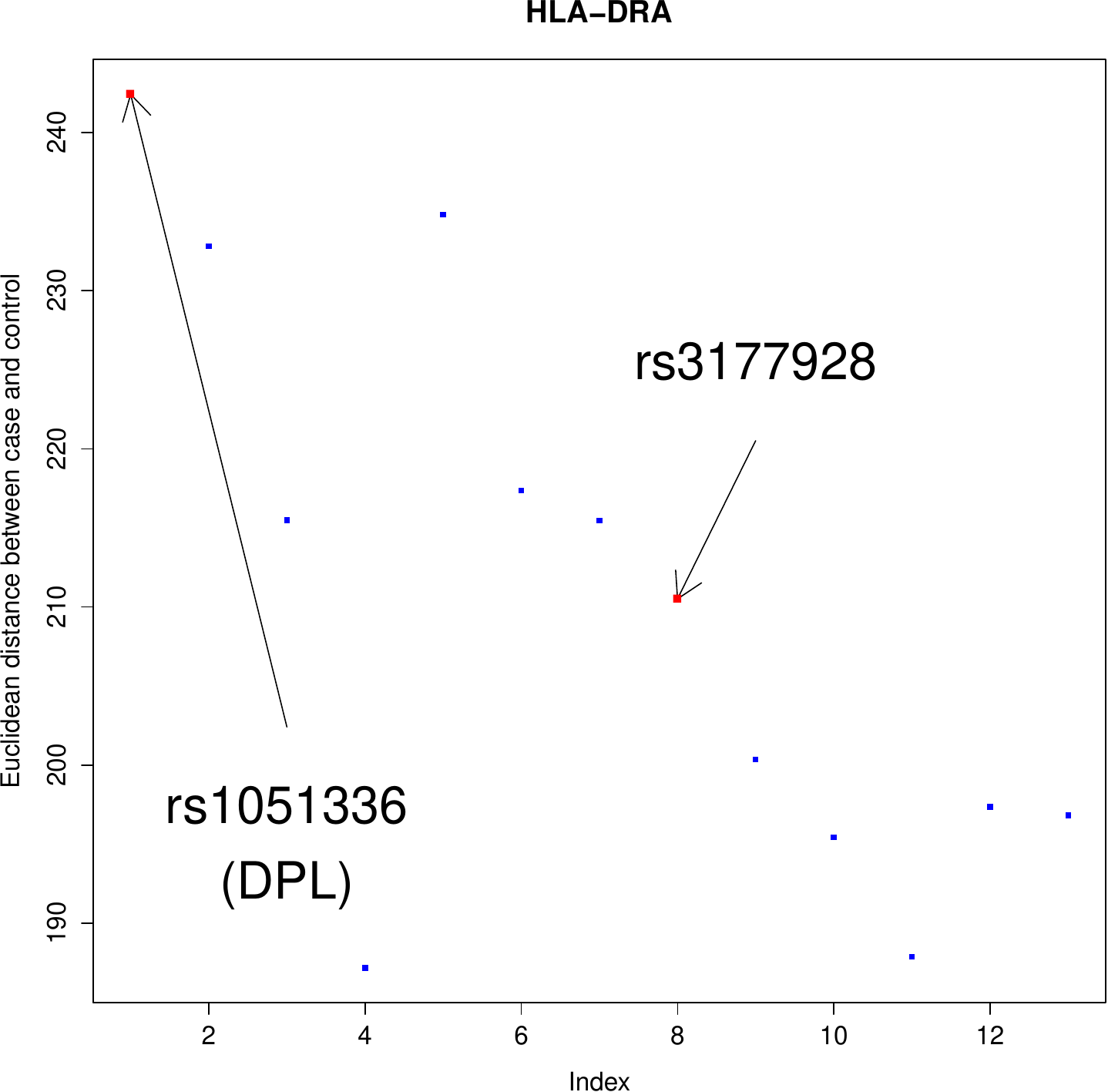}}
\hspace{2mm}
\subfigure[DPL of $RP11-306G20.1$.]{ \label{fig:DPL_gene_13}
\includegraphics[width=6cm,height=6cm]{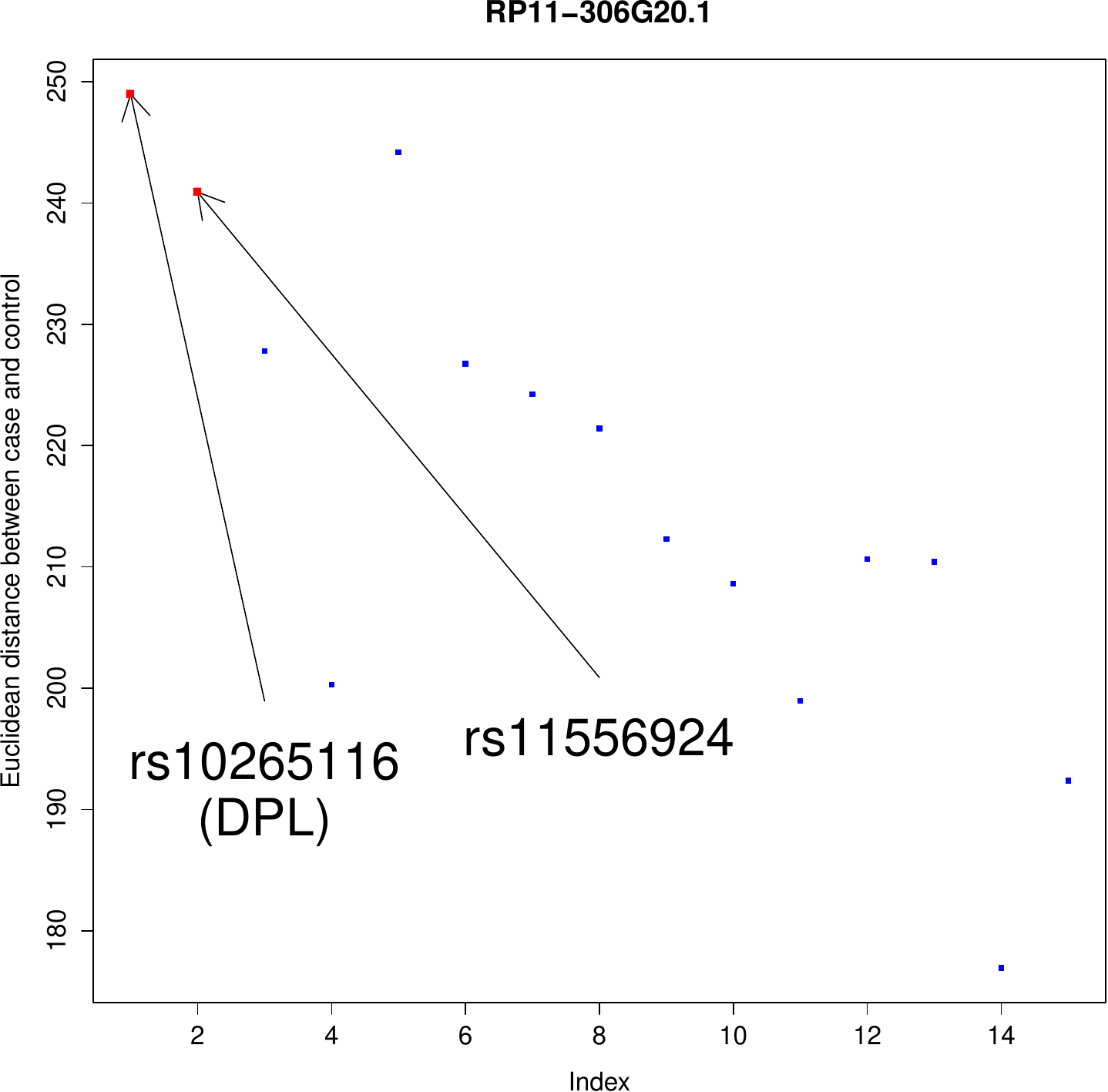}}\\
\vspace{2mm}
\subfigure[DPL of $PCSK9$.]{ \label{fig:DPL_gene_25}
\includegraphics[width=6cm,height=6cm]{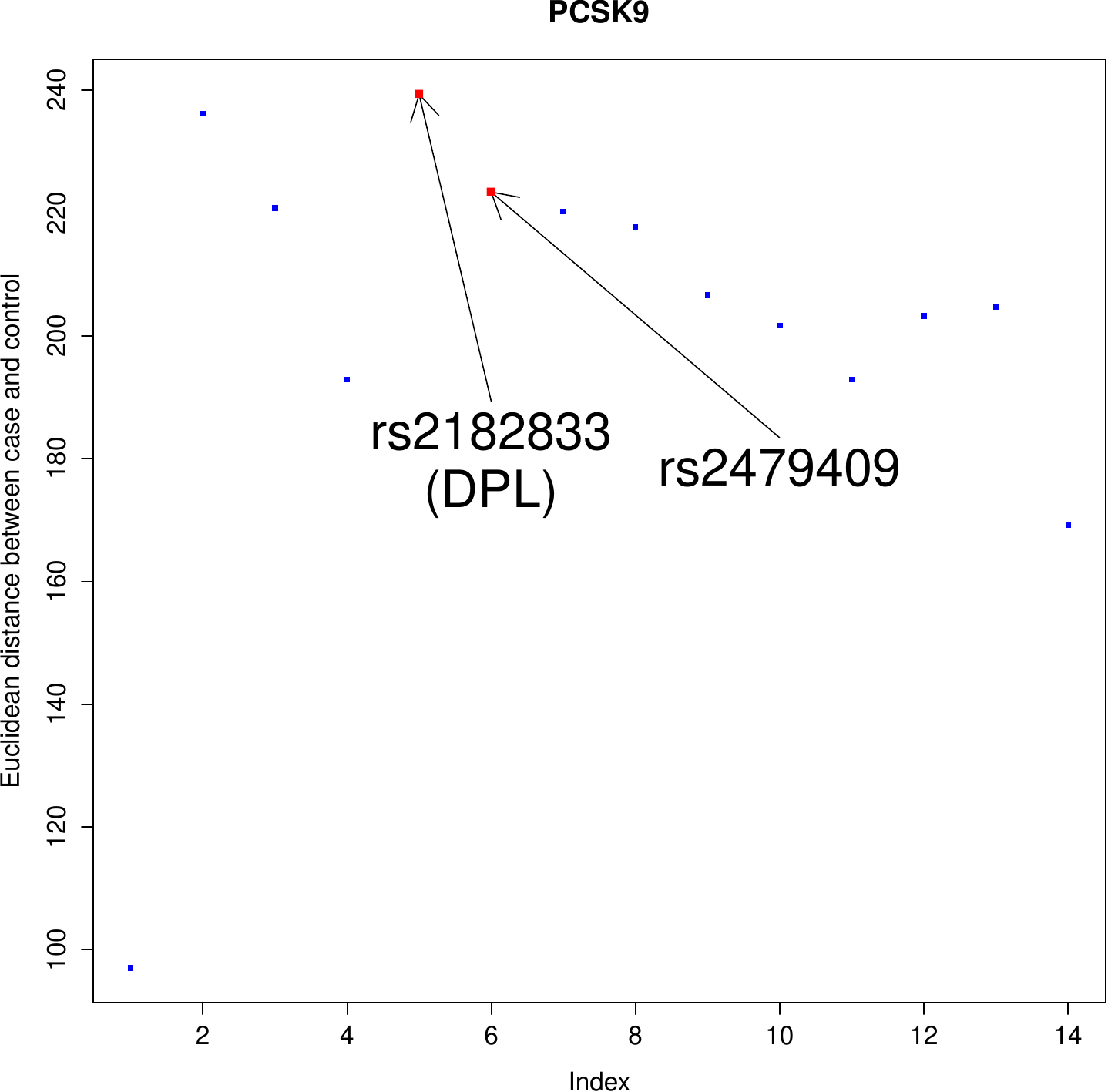}}
\hspace{2mm}
\subfigure[DPL of $ADCY5$.]{ \label{fig:DPL_gene_30}
\includegraphics[width=6cm,height=6cm]{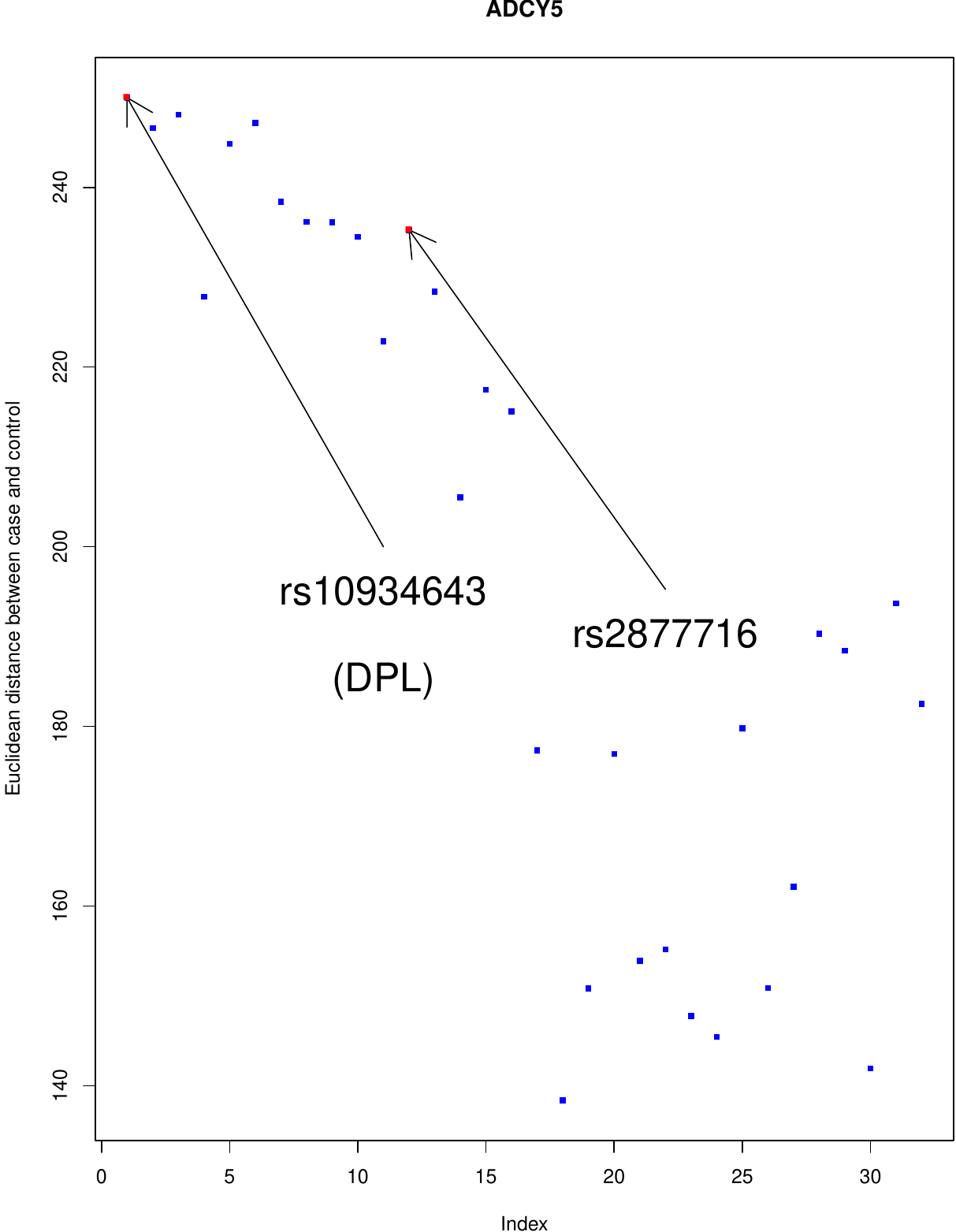}}
\caption{{\bf Disease predisposing loci of the genes influential with respect to the clustering metric:} 
Plots of the Euclidean distances 
$\left\{d^r_j\left(\mbox{logit}\bp^r_{jk=0},\mbox{logit}\bp^r_{jk=1}\right);~r=1,\ldots,L_j\right\}$
against the indices of the loci. In terms of the Euclidean distances, panels (b) and (c), and panel (d) to some extent, 
show adequate agreement of our DPLs with loci known to
be influential, while panel (a) shows disagreement of our obtained DPL
with the locus believed to be influential.}
\label{fig:DPL_genes_clustering}
\end{figure}

Figure S-5 of the supplement %\ref{fig:metric_medians} 
shows the index plots of the posterior medians of the 
clustering and Euclidean distances between case and control, with respect to the corresponding genes.
%\begin{figure}%[htp]
%\centering
%\subfigure[Clustering metric medians.]{ \label{fig:clustering_medians}
%\includegraphics[width=15cm,height=6cm]{plots_realdata/Clustering_medians-crop.pdf}}\\
%\vspace{4mm}
%\subfigure[Euclidean metric medians.]{ \label{fig:euclidean_medians} 
%\includegraphics[width=15cm,height=6cm]{plots_realdata/Euclidean_medians-crop.pdf}}
%\caption{{\bf Posterior medians of the Euclidean distances:} 
%Index plots of the posterior medians of the Euclidean distances
%with respect to the $32$ genes.}
%\label{fig:metric_medians}
%\end{figure}
In terms of the clustering metric, genes %numbered $10$, $13$, $25$ and $30$ 
$HLA-DRA$, $RP11-306G20.1$, $PCSK9$ and $ADCY5$ 
are associated with the largest medians of the clustering distances, ranging between $0.57$ and $0.68$. These genes consist
of number of loci $13$, $15$, $14$ and $32$, respectively. 
On computing the averaged Euclidean distances 
$\left\{d^r_j\left(\mbox{logit}\left(\bp^r_{jk=0}\right),\mbox{logit}\left(\bp^r_{jk=1}\right)\right);
~r=1,\ldots,L_j\right\}$, of the loci in 
each such Gene-$j$, where the averages are taken over the TMCMC samples, we 
found that loci $rs1051336$ in $HLA-DRA$, $rs10265116$ in $RP11-306G20.1$, $rs2182833$ in $PCSK9$ and $rs10934643$ in 
$ADCY5$ have the largest distances among all the loci of the $4$ respective genes.
These are depicted in Figure \ref{fig:DPL_genes_clustering}. Note that for the genes $RP11-306G20.1$, $PCSK9$, and $ADCY5$ to some extent, the significant SNPs from our study are not only close to the SNPs found significant in the existing studies (see \ctn{LucasG12}), with respect to the Euclidean distance, but also lie in their close neighborhoods, suggesting
relative agreement between the SNPs found significant in our study and the loci considered to be influential for MI in the literature.
On the other hand, $rs3177928$ on $HLA-DRA$ which has been reported by \ctn{Teslovich10} to be associated with LDL cholesterol and total cholesterol (TC) does not turn out to be a significant SNP for MI according to our analysis. 

%Figure \ref{fig:DPL_genes_euclidean} analyzes the 
We now focus attention to the genes that turned out to be more influential
than the remaining in the sense that the medians of the Euclidean distances exceed $100$. 
These are the genes $AP006216.10$, $APOC1$ and $OR4A48P$
%numbered $6$ $19$ and $27$, 
with corresponding median Euclidean distances $110.0097$, $108.4569$ and $163.8584$.
These three genes clearly stand out in Figure S-5 (panel (b)).
Each of them consists of a single locus, and are yet highly influential.

Figures S-6, S-7 and S-8 %\ref{fig:DPL_genes_other}, \ref{fig:DPL_others} and \ref{fig:DPL_others2} 
of the supplement analyze the SNPs of some of the other influential genes and point out the significant ones. 
Except for the genes $MIA3$ and $PHACTR1$, the SNPs found significant in our study closely agree
with the SNPs that are considered in the literature as influential.

%In a nutshell, our Bayesian analysis yields DPL which generally agree with the SNPs that are
%considered in the literature as influential. However, there are many cases where our findings are
%not in agreement with the literature. Figure \ref{fig:DPL_others2} shows that for genes $3$ and $5$,
%the DPL that we obtain do not match the corresponding literature-based influential SNPs.
%of genes $1$, $5$ and $9$ (panel (a) of Figure \ref{fig:DPL_genes_clustering} and
%panels (a) and (b) of Figure \ref{fig:DPL_genes_other}), but for the other genes, our findings do not
%match those reported in \ctn{Erdmann10}.
%10 13 25 30

%\begin{figure}%[htp]
%\centering
%\subfigure[DPL of $AP006216.10$.]{ \label{fig:DPL_gene_6}
%\includegraphics[width=6cm,height=6cm]{plots_realdata/plot_DPL_gene_6-crop.pdf}}
%\hspace{2mm}
%\subfigure[DPL of $APOC1$.]{ \label{fig:DPL_gene_19}
%\includegraphics[width=6cm,height=6cm]{plots_realdata/plot_DPL_gene_19-crop.pdf}}\\
%\vspace{2mm}
%\subfigure[DPL of $OR4A48P$.]{ \label{fig:DPL_gene_27}
%\includegraphics[width=6cm,height=6cm]{plots_realdata/plot_DPL_gene_27-crop.pdf}}
%\caption{{\bf Disease predisposing loci of genes influential with respect to the Euclidean metric:} 
%Plots of the Euclidean distances 
%$\left\{d^r_j\left(\mbox{logit}\bp^r_{jk=0},\mbox{logit}\bp^r_{jk=1}\right);~r=1,\ldots,L_j\right\}$
%against the indices of the loci; the genes consist of one locus each.}
%\label{fig:DPL_genes_euclidean}
%\end{figure}

In the next section we argue that gene-gene interaction plays a vital role in explaining the discrepancies
between our findings and the existing results based on previous studies.
\subsubsection{{\bf Roles of gene-gene and SNP-SNP interactions discriminating our gene findings with influential 
genes reported in the literature}}
\label{subsubsec:ggi_contrast_influential}

\begin{figure}%[htp]
\centering
%\subfigure[Colorplot of actual posterior gene-gene interaction.]{ \label{fig:gene_clustering_dendrogram} 
\includegraphics[width=16cm,height=16cm]{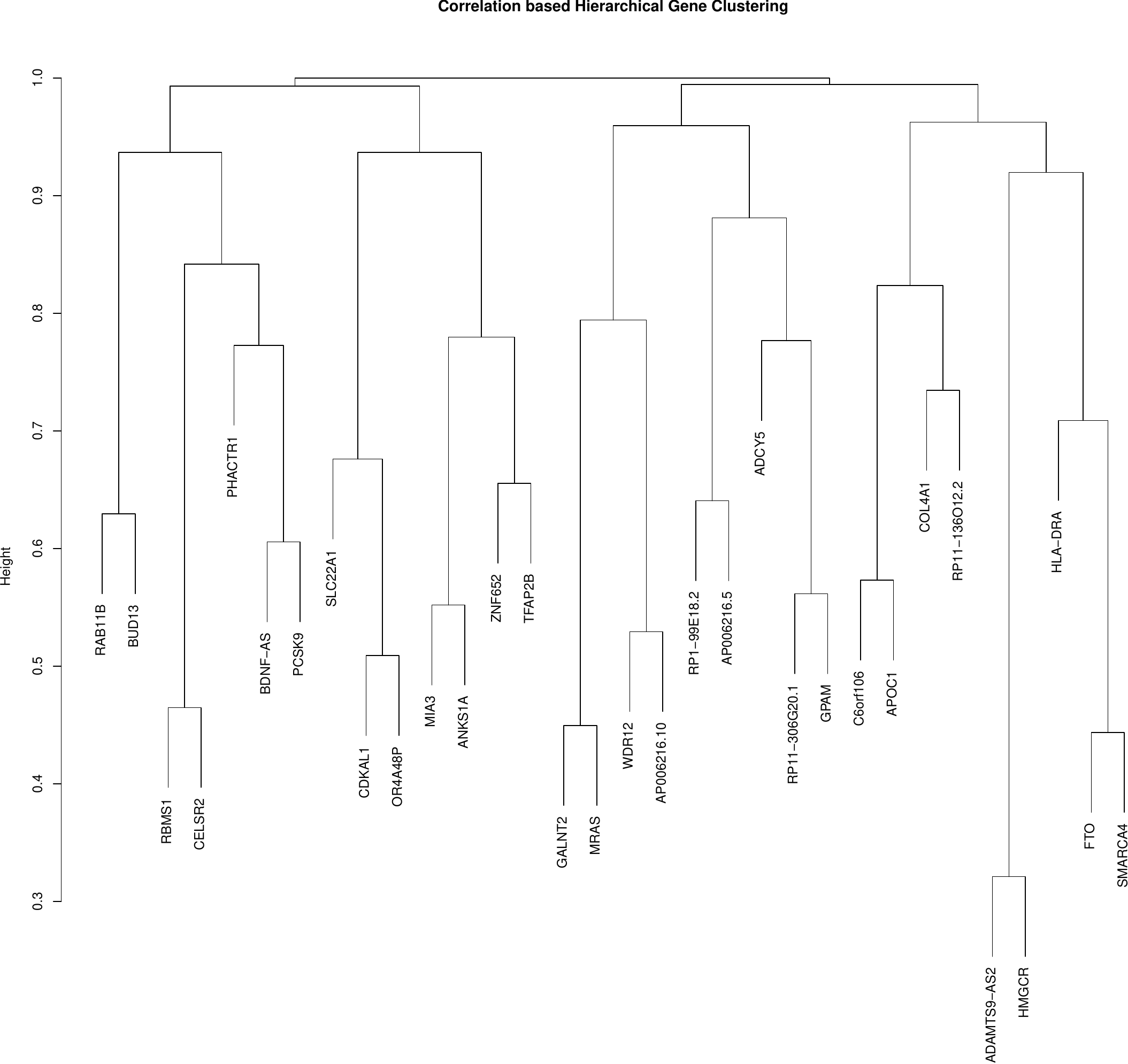}
\caption{{\bf Hierarchical gene clustering based on gene-gene interactions.}}
\label{fig:gene_clustering_dendrogram}
\end{figure}

%The results on interaction testing are depicted in Figure \ref{fig:ggi_plots1}. 
The actual gene-gene correlations based on medians of the posterior covariances,
are shown in Figure \ref{fig:ggi_plots2}, while Figure S-9 of the supplement depicts the results
on interaction testing. The color intensities correspond to the absolute values of the
correlations. Note that the correlation structure involves both positive and negative values where
negative correlations occur in more than 45\% of the cases. A hierarchical clustering of the genes based
on the absolute values of the correlations, are provided in Figure \ref{fig:gene_clustering_dendrogram}.
The vertical axis of the diagram represents $1-\left|\rho\right|$, $\rho$ standing for the correlations
shown in Figure \ref{fig:ggi_plots2}. In a nutshell, lower the order of the hierarchy, stronger are the correlations
between the genes. For instance, Figure \ref{fig:gene_clustering_dendrogram} shows that the correlation
between genes $ADAMTS9-AS2$ and $HMGCR$ is the strongest; moreover, the correlation between 
$ADAMTS9-AS2$ and $MIA3$, for example, is stronger than the correlation between $ADAMTS9-AS2$ and $PHACTR1$.
\begin{figure}%[htp]
\centering
\subfigure[Colorplot of actual posterior gene-gene interaction.]{ \label{fig:ggi_plot} 
\includegraphics[width=16cm,height=16cm]{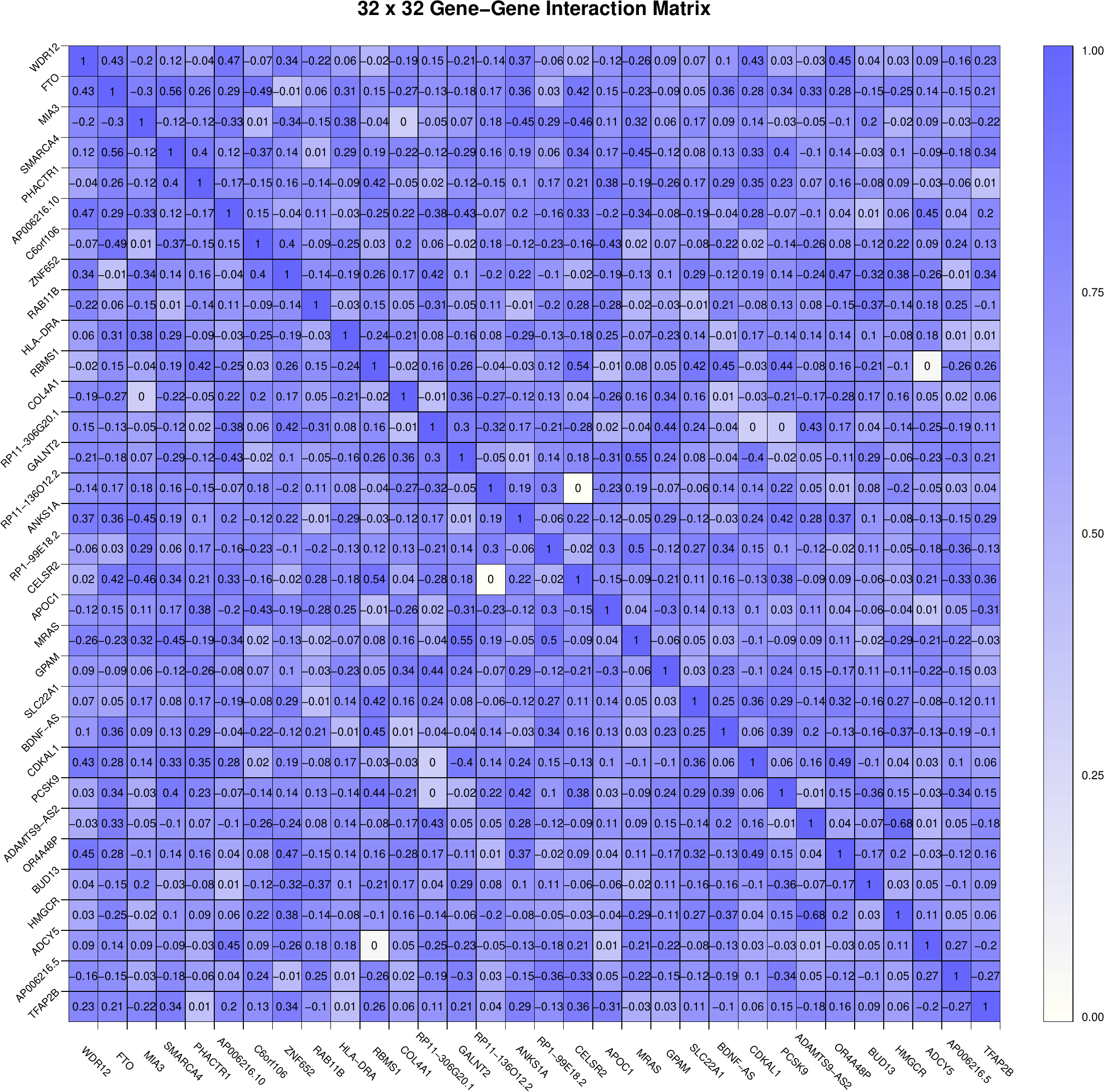}}
\caption{{\bf Gene-gene interaction plot:} Actual gene-gene interactions
based on medians of the absolute values of the posterior covariances.}
\label{fig:ggi_plots2}
\end{figure}

Figures \ref{fig:ggi_plots2} and \ref{fig:gene_clustering_dendrogram} depict complex interplay among the genes, many
of which negatively influence each other with respect to the Euclidean distances.
These gene-gene interactions also seem to influence the SNP-wise Euclidean distances between cases and controls, 
revealing the effect of some new SNPs that failed to make their presence felt in the previous studies due to lack
of adequate dependence structure. There is another important issue to point towards in this context. As will
be seen, the correlations with respect to our current dataset are usually of small magnitude. 
This is not because the covariances are of small magnitude; indeed, they are all significantly bounded away
from zero, but the variances are of very large order, so that the covariances scaled by the square roots of the variances,
are small. These correlations, depending upon positive or negative signs, and in conjunction with the very 
complex interplay among the genes and the SNPs, 
are instrumental in deciding whether a SNP appears as the most influential one among a set of SNPs in any gene. 
The issue is explained mathematically in Section S-8 of the supplement.

We elucidate the aforementioned issue with respect to the genes 
$HLA-DRA$, $RP11-306G20.1$, $PCSK9$ and $ADCY5$; see Figure \ref{fig:DPL_genes_clustering}. Indeed, among the SNPs of $HLA-DRA$,
the case-control Euclidean distance associated with $rs1051336$, which has turned out to be influential in our analysis has maximum negative correlation of $-0.02361464$ with
that of $rs3177928$, the SNP pointed significant by \ctn{Teslovich10}. Hence, it is not surprising that $rs3177928$ failed to be
close to $rs1051336$ in terms of the Euclidean distance between case and control. On the other hand, $rs10265116$
of $RP11-306G20.1$ is positively correlated with the literature-based important SNP $rs11556924$, the correlation
being $0.007653744$, supporting the closeness of the Euclidean distances between $rs11556924$ and $rs10265116$.
For gene $PCSK9$, the correlation between $rs2182833$, the influential SNP by our study and the literature-based $rs2479409$, is
$-0.005999625$. The consequence of this small, albeit negative correlation, is well-reflected in panel (c)
of Figure \ref{fig:DPL_genes_clustering}; the corresponding Euclidean distances are close but there are other
SNPs, positively correlated with $rs2182833$, that have larger Euclidean distances compared to that of $rs2479409$.  
For gene $ADCY5$, however, the influential SNP by our study, $rs10934643$ is positively correlated with the literature-based
SNP $rs2877716$, the correlation being $0.00880905$. The fact that in spite of the positive correlation
the two SNPs are not adequately close in terms of case-control Euclidean distances, begs some explanation, which
we provide next.

Interestingly, the key to this phenomenon lies in inter-genetic SNP-SNP interaction. Indeed, $OR4A48P$, the most influential
gene in terms of the maximum Euclidean distance, and consisting of the single locus $rs7395662$, 
exerts negative influence on $rs2877716$ through
the negative correlation $-0.02919905$, but has small and positive
correlation, $0.00298559$, with $rs10934643$. This gene, through its only SNP, also influences the other
genes via positive or negative correlations. For instance, its correlations with $rs1051336$ and $rs3177928$
of $HLA-DRA$ are $0.05828725$ and $-0.01058818$, respectively. In other words, the former receives lot more weight
compared to the latter, so that $rs1051336$ becomes influential with respect to our model. For 
gene $RP11-306G20.1$, its correlations with the relevant two loci $rs10265116$ and $rs11556924$, are $0.05306113$ and
$0.003207612$, respectively, so that both SNPs are relatively close but the former takes precedence, becoming DPL,
thanks to its larger correlation with $OR4A48P$. For gene $PCSK9$, the correlations of $rs2182833$ and $rs2479409$
with $OR4A48P$ are $-0.001009753$ and $-0.01482009$, respectively. Hence, the former locus becomes the DPL because
of its smaller negative correlation. Thus, there is a very complex interplay among the different genes, their SNPs
and among the SNPs of different genes. In general it is infeasible to keep track of these complex dependencies
and provide simple explanations for the different DPLs and their differences with the SNPs believed to be important
by the scientific community.

The above elucidations attempt to point out that unless gene-gene and SNP-SNP interactions are taken into account
through a sophisticated, nonparametric framework, such complex interaction effects might have been missed,
which would perhaps lead to declaration of some truly influential SNPs as non-significant,
and some non-influential SNPs as influential.

\subsubsection{{\bf Posteriors of the number of distinct mixture components distinguishing 
important sets of genes}}
\label{subsubsec:no_of_components}
%The discrepancies between the sets of genes %$\{10,13,25,30\}$, $\{6,19,27\}$ 
%$\{HLA-DRA, RP11-306G20.1, PCSK9, ADCY5\}$,
%$\{AP 006216.10, APOC1, OR4A48P\}$ 
%and the remaining genes
%are also reflected in the
%posterior distributions of the number of mixture components associated with them.
%For the remaining genes, here we consider only %$\{1,2,4,8\}$, 
%$\{WDR12, FTO, SMARCA4, ZNF652\}$
%for the sake of brevity.
Figures %\ref{fig:ggi_comp_realdata1}, \ref{fig:ggi_comp_realdata2} and \ref{fig:ggi_comp_realdata3} 
S-10, S-11 and S-12 show the posteriors of the number of distinct components associated with genes from the sets
$\{WDR12$, $FTO$, $SMARCA4$, $ZNF652\}$, $\{HLA-DRA$, $RP11-306G20.1$, $PCSK9$, $ADCY5\}$
and $\{AP006216.10$, $APOC1$, $OR4A48P\}$, respectively.
The posteriors confirm our expectation that the four broad sub-populations composed of Caucasians, Han Chinese,
Japanese and Yoruban admit further sub-divisions in general. Indeed, although for genes $FTO$, $SMARCA4$ and $ZNF652$, the number
of subpopulations turned out to be less than $5$ with high posterior probabilities, for the other genes the number of subpopulations have 
exceeded $5$ with almost full posterior probabilities.   
The shown posteriors are negligibly different for case and control, for all the three sets of genes, which is
to be expected because of the high positive correlations between $\lambda_{j0}$ and $\lambda_{j1}$.

It is interesting to observe that the posteriors of the number of components of the first
set of genes  $\{WDR12$, $FTO$, $SMARCA4$, $ZNF652\}$ are roughly stochastically
dominated by the second set $\{HLA-DRA$, $RP11-306G20.1$, $PCSK9$, $ADCY5\}$, which, in turn, are dominated by those of the set
$\{AP006216.10$, $APOC1$, $OR4A48P\}$. The implication is that the third set consists of more genetic variations, followed
by the second set, while the first set consists of least genetic variations. Thus, the last set of genes,
consisting of only one locus each, seems to be most likely to affect the disease, while the first set 
seems to be the least influential on MI.

\subsection{{\bf Discussion of our Bayesian methods and GWAS in light of our findings}}
\label{subsec:discussion}
Our Bayesian analysis yielded results that are broadly in agreement with those obtained by GWA investigations
reported in the literature. 
However, the fact that some of the SNPs which are flagged by the literature as important, 
did not show up as the most significant ones, deserves attention. The main issue that emerged
in our investigation is that the gene-gene interactions are responsible for suppression 
of the so-called important SNPs via implicit induction of negative correlations among Euclidean
distances between cases and controls for the associated genes. 
Had there been no such negative correlations, it is plausible that these SNPs would turn out to be the most
influential ones. 
%This is supported by the observation that Gene-5, consisting of the citedly influential SNP, $rs2048327$, turned out to be 
%one of the most significant genes in our analysis, and the clustering metric for this gene is positively correlated with 
%that of Gene-33, the most significant gene in our analysis. Recalling that the metric for 
%Gene-33 is significantly negatively correlated with those of the other GWAS-based genes, it seems clear that
%this positive interaction with Gene-33 granted importance to Gene-5 and negative interaction of the other
%GWAS-based genes with Gene-33 suppressed their relevance in our study. Nevertheless, the DPL
%that we obtained from genes $1$, $5$ and $9$ (all of them emerged significant with respect
%to our Bayesian hypotheses testng) closely matched those obtained by GWAS.

Apart from a few agreements, the literature based SNPs are different from the SNPs detected significant in our analysis, many of which lie in the intronic regions and have not been thoroughly explored and hence need further investigation.
As per our investigation, sophisticated, nonparametric modeling of 
gene-gene interactions plays a very crucial role in imparting significance
to the overall effect of the individual genes. Since the GWAS did not incorporate the complex intra and inter-genetic interactions into the model, it is perhaps not 
very unreasonable to question if the same genes would emerge as significant if realistic modeling of gene-gene  
interactions is taken into account.

For the current MI study, we summarize our findings in Tables \ref{table:summary} and \ref{table:summary2}, where we present the 32 genes 
ordered with respect to the median case-control Euclidean distances, the SNPs flagged by the literature as significant, 
the corresponding SNPs detected by our Bayesian model and methods, and the phenotypes of the reportedly 
significant SNPs and our SNPs. 

%\begin{comment}

%\newpage

%\begin{table}[h!]
\begin{table}
\centering
\caption{Summary of the MI data investigation, where genes are ranked in order of their decreasing case-control 
based Euclidean distance.}
\label{table:summary}
\begin{tabular}{|c|c|c|c|c|c|}
\hline
            & Chr              & Literature  & Reported     & Bayesian  &  Reported\\
Genes    & No.                               & based SNPs   &         Phenotype    &  SNPs& Phenotype             \\ 
         &         &                  &                    &           &                 \\ 
\hline
OR4A48P & 11  & rs7395662 & LDL, HDL, & rs7395662 &  LDL, HDL,  \\
 & & &   tryglycerides & &  tryglycerides\\
AP006216.10 & 11  & rs964184 & triglycerides, LDL, & rs964184 &triglycerides, LDL,\\
 & & &  HDL cholesterol& & HDL cholesterol\\
APOC1 & 19  & rs4420638 &  LDL, HDL& rs4420638 &  LDL, HDL\\
 & & &  cholesterol & & cholesterol\\
TFAP2B & 6  & rs987237 & BMI  & rs2011201 & \\
AP006216.5 & 11  & rs7396835 &Body weight, BMI  & rs1263172 & \\
 & & &  triglyceride& &\\
RAB11B & 19  & rs2967605 & Carotid Artery  & rs2913973 & \\
 & & &  heart disease& & \\
BUD13 & 11  & rs28927680 & HDL, cholesterol & rs10488699 & LDL, cholesterol\\
 & & &  triglycerides&  & HDL cholesterol\\
CELSR2 & 1  & rs599839 & CHD, CAD, &  rs14000 & CHD\\
 & & &   LDL cholesterol& & \\
RP1-99E18.2 & 6  & rs6922269 & CHD  & rs11155760 & \\
WDR12 & 2  & rs6725887 & CHD, CAD, MI & rs10205697 & \\
C6orf106 & 6  & rs2814944 &HDL, LDL, BMI  & rs1201872 & \\
 & & &  cholesterol& &\\
HLA-DRA & 6  & rs3177928 &cholesterol, LDL  & rs1051336 & \\
RP11-306G20.1 & 7 & rs11556924 &  & rs10265116 & \\

\hline
\end{tabular}
\end{table}

\begin{table}
\centering
\caption{Continuation of Table \ref{table:summary}.}
%Summary of the MI data investigation, where genes are ranked in order of their decreasing case-control 
%based Euclidean distance.}
\label{table:summary2}
\begin{tabular}{|c|c|c|c|c|c|}
\hline
            & Chr              & Literature  & Reported     & Bayesian  &  Reported\\
Genes    & No.                               & based SNPs   &         Phenotype    &  SNPs& Phenotype             \\ 
         &         &                  &                    &           &                 \\ 
\hline
PCSK9 & 1  & rs2479409 & cholesterol,  & rs2182833 & cholesterol\\
 & & &  LDL cholesterol& &\\
ADCY5 & 3  & rs2877716 & Carbohydrate metabolism& rs10934643 & \\
HMGCR & 5  & rs3846662 & LDL cholesterol & rs12654264 & \\
GALNT2 & 1  & rs4846914 & cholesterol, HDL & rs1474925 & \\
 & & &  Triglycerides& & \\
MRAS & 3  & rs9818870 &  CAD & rs1199335 & \\
ZNF652 & 17 & rs16948048 & Diastolic blood  & rs12940887 & \\
 & & &  pressure& & \\
ANKS1A & 6  & rs17609940 & CAD & rs17647222 & \\
SLC22A1 & 6  & rs1564348 & LDL & rs1564348 & \\
RP11-136O12.2 & 8 & rs17321515 & Triglycerides & rs16900615 & \\
GPAM & 10  & rs1129555 &  LDL& rs10885315 & \\
RBMS1 & 2  & rs7593730 &  Type 2 diabetes& rs11694165 & \\
BDNF-AS & 11  & rs1013442 & Smoking & rs1013442 & \\
SMARCA4 & 19  & rs1122608 & MI(early onset) & rs10415811 & \\
FTO & 16  & rs1121980 & BMI & rs10521303 & \\
MIA3 & 1  & rs17465637 & MI(early onset) & rs17163303 & \\
CDKAL1 & 6  & rs10946398 &  Type 2 diabetes& rs1012625 & \\ 
ADAMTS9-AS2 & 3  & rs4607103 & Type 2 diabetes & rs10510917 & \\
COL4A1 & 13  & rs3742207 & Arterial stiffness & rs1000989 & \\
PHACTR1 & 6  & rs12526453 &  MI (early onset)& rs1014342 & \\
\hline
\end{tabular}
\end{table}

%\end{comment}

%\newpage
%\section{{\bf Summary and conclusion}}
\section{{\bf Concluding remarks}}
\label{sec:conclusion}

In this work we have focused exclusively on gene-gene interaction. Recently, \ctn{Bhattacharya17a} and \ctn{Bhattacharya17b} 
have extended this model to incorporate gene-environment interactions in our model, and developed tests for
the effects of gene-environment interactions as well as gene-gene interactions on case-control. 
They have successfully applied the ideas to various simulated datasets generated from the GENS2 software,
and to the MI dataset, considering sex as the environmental variable.
The results they obtained are broadly in agreement with the results on this MI dataset already existing in the literature. 

In this paper, we were compelled to consider a small part of the available real dataset consisting of SNPs cited 
in the literature as important. This small dataset, however, has the added advantage of alleviating computational
burden. Indeed, since only $50$ two-threaded cores were available to us for implementation of our ideas, 
it is anyway imperative for us to confine attention to a (relatively small) subset of the available dataset. 
We are, however, expecting to expand our current parallel computing infrastructure, which would be of immense help in analysing
the complete dataset, which is our actual goal.

\section*{{\bf Acknowledgment}}

We are sinceely grateful to the two reviewers whose encouraging and constructive comments have led to significant improvement of our manuscript.
We are also grateful to Dr. Arunabha Majumdar for providing useful feedback on an earlier
version of our manuscript.

\newpage

\renewcommand\thefigure{S-\arabic{figure}}
\renewcommand\thetable{S-\arabic{table}}
\renewcommand\thesection{S-\arabic{section}}

\setcounter{section}{0}
\setcounter{figure}{0}
\setcounter{table}{0}

\begin{center}
{\bf \Large Supplementary Material}
\end{center}

\section{{\bf A rule of thumb for choosing $M$ and $\alpha$}}
\label{sec:thumb_rule}
Following \ctn{Majumdar13}, \ctn{Sabya12}, \ctn{Sabya11}, \ctn{Bhattacharya08}, we set $M=30$ in our applications.
It follows from \ctn{Antoniak74} that the mean and variance of the distinct parameter vectors in the set
$\bp_{1jk},\bp_{2jk},\ldots,\bp_{Mjk}$ are both given by approximately $\alpha_{jk}\log\left(1+\frac{M}{\alpha_{jk}}\right)$.
When prior information regarding the true number of mixture components is lacking, 
it may be reasonable to specify the expected number of distinct components
to be close to half of the maximum number of components possible, namely, close to $M/2$. 
With $M=30$, we fix $\alpha_{jk}=10$, so that about $14$ distinct mixture components
%in (\ref{eq:mixture1}) 
are to be expected {\it a priori}. Apart from this choice, we also considered
the possibilities $\alpha_{jk}=1$, $\alpha_{jk}\sim\mbox{Gamma}\left(0.1,0.1\right)$, that is, the gamma
distribution with mean $1$ and variance $10$, and $\alpha_{jk}\sim\mbox{Gamma}\left(1,0.1\right)$
(so that the mean and variance are $10$ and $100$, respectively); however, the choice $\alpha_{jk}=10$
for all $(j,k)$ outperformed the other choices with regard to capturing the true number of mixture components.
Hence, in this work, we report all our results associated with $M=30$ and $\alpha_{jk}=10$.
According to this specification, the prior mean and variance of the number of distinct components are approximately
$14$. Thus, compared to smaller values of $\alpha_{jk}$, this choice ensures greater variability
so that data-driven inference on the number of components receives greater weight. 

\section{{\bf Choices of $\bA_0$ and $\bSigma_0$}}
\label{subsec:iw3}

For $k = 0, 1$; $i = 1,\ldots,N_k$ and $j = 1,\ldots,J$, here we denote
by $w^r_{ijk}$ the count of the minor allele at the $r$-th locus of the $j$-th gene and $k$-th case-control status.
In other words, $w^r_{ijk}=x^1_{ijkr}+x^2_{ijkr}$.
With this notation we define
\begin{equation}
\bar{w}_{ijk}=\frac{1}{L_j}\sum_{r=1}^{L_j}{w^r_{ijk}}.
\label{eq:w_bar1}
\end{equation}
Also let
\begin{equation}
\bar{w}_{\cdot j\cdot}=\frac{1}{N_0+N_1}\sum_{k=0}^1\sum_{i=1}^{N_k}\bar{w}_{ijk}.
\label{eq:w_bar2}
\end{equation}
With these, we specify the $(j_1,j_2)$-th element of $\bA_0$ as
\begin{equation}
a_{0,j_1j_2}=\frac{1}{N_0+N_1}
\sum_{k=0}^1\sum_{i=1}^{N_k}\left(\bar{w}_{ij_1k}-\bar{w}_{\cdot j_1\cdot}\right)
\left(\bar{w}_{ij_2k}-\bar{w}_{\cdot j_2\cdot}\right).
\label{eq:w_bar3}
\end{equation}
For the specification of $\bSigma_0$, we first consider
\begin{equation}
\bar{w}_{\cdot \cdot k}=\frac{1}{N_kJ}\sum_{i=1}^{N_k}\sum_{j=1}^J\bar{w}_{ijk}.
\end{equation}
Then, letting $N=\min\{N_0,N_1\}$, we specify the $(k_1,k_2)$-th element of $\bSigma$ as 
\begin{equation}
\sigma_{0,k_1k_2}=\frac{1}{NJ}
\sum_{i=1}^{N}\sum_{j=1}^J\left(\bar{w}_{ijk_1}-\bar{w}_{\cdot \cdot k_1}\right)
\left(\bar{w}_{ijk_2}-\bar{w}_{\cdot \cdot k_2}\right).
\label{eq:w_bar4}
\end{equation}

\section{{\bf Elucidation that the $r$-th loci of any two different genes can be independent with positive
probability}}
\label{sec:independent_loci}
Our assumption that $u_r$ and $v_r$ of each locus $r$ is shared by all the genes does not
imply that the labels of the loci of the genes are not exchangeable. Indeed, the $r$-th loci of two 
different genes may be independent, given the data. 
To understand this, note that the $r$-th loci of any gene $j$ 
does not only have the effect $u_r$ and $v_r$, but also $\lambda_{jk}$, for given $k\in\{0,1\}$.
In other words, all the loci of gene $j$ share the common effect $\lambda_{jk}$. Hence, for any two genes
denoted by $j_1$ and $j_2$, and $k_1,k_2\in\{0,1\}$, and considering only $u_r$, the $r$-th loci have the effects 
$u_r+\lambda_{j_1k_1}$ and $u_r+\lambda_{j_2k_2}$. By our distributional assumptions it follows that
the covariance between these effects is $1+a_{j_1j_2}\sigma_{k_1k_2}$, given $a_{j_1j_2}$ and $\sigma_{k_1k_2}$, 
which is equal to zero if $a_{j_1j_2}=-\sigma^{-1}_{k_1k_2}$. Independence follows due to normality of our 
specified distributions. 
Now note that by our inverse-Wishart priors on $\bA$ and $\bSigma$, the event
$\left|a_{j_1j_2}+\sigma^{-1}_{k_1k_2}\right|<\epsilon$ gets positive probability for any $\epsilon>0$, so that
the covariance can be in any neighborhood of zero with positive probability, if connoted by the data.

\section{{\bf A parallel MCMC algorithm for model fitting}}
\label{sec:computation}

Recall that the mixtures associated with gene $j\in\{1,\ldots,J\}$ and case-control status $k\in\{0,1\}$
are conditionally independent of each other, given the interaction parameters.
This allows us to update the mixture components in separate parallel processors, conditionally on the
interaction parameters. Once the mixture components are updated, we update the interaction parameters 
using a specialized form of TMCMC, in a single processor. The details of updating the mixture components 
in parallel are as follows.

\begin{itemize}

%\subsection{{\bf Updating the allocation variables using Gibbs steps}}
%\label{subsec:fullcond_z}

\item[(1)] Split the pairs $\left\{(j,k):~j=1,\ldots,J;~k=0,1\right\}$ in the available
parallel processors.

\item[(2)] During each MCMC iteration, for each $(j,k)$ in each available parallel processor, do the following 
\begin{enumerate}
%\item[{\bf Update allocation variables}]
\item[(i)]
For $i=1,\ldots,N_k$, update the allocation variables $z_{ijk}$ by simulating
from the full conditional distribution of $z_{ijk}$, given by
\begin{equation}
[z_{ijk}=m|\cdots]\propto\pi_{mjk}\prod_{r=1}^{L_j}f\left(\bx_{ijkr}|p_{mjkr}\right);
%{\sum_{m'=1}^M\pi_{m'jk}\prod_{r=1}^{L_j}f\left(\bx_{ijkr}|p_{m'jkr}\right)}
\label{eq:fullcond_z}
\end{equation}
for $m=1,\ldots,M$.

%\subsection{{\bf Updating the configuration indicators associated with the mixtures using Gibbs steps}}
%\label{subsec:fullcond_c}

%\item[{\bf Updating the configuration indicators associated with the mixtures using Gibbs steps}]
\item[(ii)]

%For each $(j,k)$, 
Let $\left\{\bp^*_{1jk},\ldots,\bp^*_{\tau_{jk} jk}\right\}$ denote the distinct elements
in $\bP_{Mjk}=\left\{\bp_{1jk},\ldots,\bp_{Mjk}\right\}$. Also let $\bC_{jk}=\left\{c_{1jk},\ldots,c_{Mjk}\right\}$
denote the configuration vector, where $c_{mjk}=\ell$ if and only if $\bp_{mjk}=\bp^*_{\ell jk}$.

Now let $\tau^{(m)}_{jk}$ denote the number of distinct elements in $\bP_{-Mjkm}=\bP\backslash\left\{\bp_{mjk}\right\}$ 
and let ${\bp^m}^*_{\ell}=\left\{{p^{m}}^*_{\ell jkr};~r=1,\ldots,L_j\right\};~\ell=1,\ldots,\tau^{(m)}_{jk}$ 
denote the distinct parameter vectors. Further, let ${\bp^m}^*_{\ell}$
occur $M_{\ell m}$ times. 

Then update $c_{mjk}$ using Gibbs steps, where the full conditional distribution of $c_{mjk}$ is given by
\begin{equation}
[c_{mjk}=\ell|\cdots]\propto\left\{\begin{array}{ccc}q^*_{\ell, mjk} & \mbox{if} & \ell=1,\ldots,\tau^{(m)}_{jk};\\
q_{0,mjk} & \mbox{if} & \ell=\tau^{(m)}_{jk}+1,\end{array}\right.
\label{eq:fullcond_c}
\end{equation}
where
\begin{align}
q_{0,mjk} &=\alpha_{jk}\prod_{r=1}^{L_j}\frac{\beta\left(n_{1mjr}+\nu_{1jkr},n_{2mjr}+\nu_{2jkr}\right)}
{\beta\left(\nu_{1jkr},\nu_{2jkr}\right)};
\label{eq:q0}\\
q^*_{\ell, mjk} &=M_{\ell m}\prod_{r=1}^{L_j}\left\{{p^{m}}^*_{\ell jkr}\right\}^{n_{1mjr}}
\left\{1-{p^{m}}^*_{\ell jkr}\right\}^{n_{2mjr}}.
\label{eq:q1}
\end{align}
In (\ref{eq:q0}) and (\ref{eq:q1}), $n_{1mjr}$ and $n_{2mjr}$ denote the number of $``a"$ and $``A"$ alleles,
respectively, at the $r$-th locus of the $j$-th gene associated with the $m$-th mixture component.
In other words, $n_{1mjr}=\sum_{i:z_{ijk}=m}\left(x^1_{ijkr}+x^2_{ijkr}\right)$ and
$n_{2mjr}=\sum_{i:z_{ijk}=m}\left\{2-\left(x^1_{ijkr}+x^2_{ijkr}\right)\right\}$.
The function $\beta(\cdot,\cdot)$ in the above equations is the Beta function such that for any $s_1>0, s_2>0$,
$\beta(s_1,s_2)=\frac{\Gamma(s_1)\Gamma(s_2)}{\Gamma(s_1+s_2)}$; $\Gamma(\cdot)$ being the Gamma function.

%\subsection{{\bf Updating the parameters of the distinct components associated with the mixtures using Gibbs steps}}
%\label{subsec:fullcond_distinct}

%\item[{\bf Update the parameters of the distinct components associated with the mixtures using Gibbs steps}]
\item[(iii)]

Let ${n}^*_{1\ell jr}=\sum_{m:c_{mjk}=\ell}n_{1mjr}$ and  ${n}^*_{2\ell jr}=\sum_{m:c_{mjk}=\ell}n_{2mjr}$.
Then, for $\ell=1,\ldots,\tau_{jk}$; $r=1,\ldots,L_j$; $j=1,\ldots,J$ and $k=0,1$, 
update ${p}^*_{\ell jkr}$ by simulating from its full conditional distribution,
given by
\begin{equation}
[{p}^*_{\ell jkr}|\cdots]\sim \mbox{Beta}\left({n}^*_{1j\ell r}+\nu_{1jkr},{n}^*_{2j\ell r}+\nu_{2jkr}\right).
\label{eq:fullcond_p}
\end{equation}
\end{enumerate}
\item[(3)] During each MCMC iteration, update the interaction parameters $\left\{(u_{r'},v_{r'});~r'=1,\ldots,L\right\}$,
$\bLambda$, $\bA$ and $\bSigma$ in a single processor using TMCMC, conditionally on the remaining 
parameters. The details of updating the interaction parameters are provided in Section \ref{subsec:interaction_update}.
\end{itemize}

\subsection{{\bf Updating the interaction parameters using a mixture of additive and additive-multiplicative TMCMC}}
\label{subsec:interaction_update}

We now provide details on updating the parameters $\left\{(u_{r'},v_{r'});~r'=1,\ldots,L\right\}$, 
$\bLambda$, $\bA$ and $\bSigma$. %using a mixture of additive and additive-multiplicative TMCMC. 
Note, however, that since $\bA$ and $\bSigma$ are positive definite matrices, directly updating
these matrices is not straightforward, since the MCMC proposals need not preserve positive definiteness
and checking positive definiteness, which is required while evaluating the acceptance ratio, is not straightforward
for high dimensional matrices.
Therefore, we resort to Cholesky decompositions, $\bA=\bC_1\bC_1'$ and $\bSigma=\bC_2\bC'_2$,
where $\bC_1$ and $\bC_2$ are lower triangular matrices. Thus, instead of updating $\bA$ and $\bSigma$
directly, we can update the elements of $\bC_1$ and $\bC_2$, with the only constraint that the diagonal
elements are positive. 

Before we provide the problem-specific details, let us first recall the main ideas of additive, 
multiplicative, and additive-multiplicative TMCMC; for details see \ctn{Dutta14} and \ctn{Dey14}.

\subsubsection{{\bf Additive TMCMC}}
\label{subsubsec:additive_tmcmc}

%Note that our moves at each step were symmetric and the magnitude of the jump would depend on the choice of the proposal density $q$. 
%Again, $q$ must have its support as $\mathbb{R}^{+}$, and from now onwards, we shall assume that it is a $N(0,1)$ distribution truncated at $0$. 
%Note that at each step , we sample only one $\epsilon$ from this proposal distribution and updates all the co-ordinates at one go. 
%The notion of symmetrical transitions at each step can be expressed from a mathematical point of view as follows. 
Suppose that we are simulating from a $d$ dimensional space (usually $\mathbb{R}^{d}$), and suppose we are currently at a point 
$x= (x_{1}, \ldots, x_{d})$.
Let us define $d$ random variables $b_{1}, \ldots, b_{d}$, such that, for $i=1,\ldots,d$, 
\begin{equation}
b_{i} =\left\{\begin{array}{ccc} +1 & \mbox{with probability} & p_i; \\
 -1 & \mbox{with probability} & 1-p_i.
 \end{array}\right.
 \label{eq:b_add}
\end{equation}
The additive TMCMC uses moves of the following type: 
\begin{equation*}
(x_{1}, \ldots, x_{d}) \rightarrow (x_{1}+ b_{1}\epsilon, \ldots, x_{d}+b_{d}\epsilon), 
\end{equation*}
where $\epsilon\sim g^{(1)}=q^{(1)}(\cdot)I_{\{\epsilon>0\}}$. Here $q^{(1)}(\cdot)$ is an arbitrary density with support $\mathbb R_+$, the
positive part of the real line, and
for any set $A$, $I_{A}$ denotes the indicator function of $A$. We define $T^{(1)}_b(x,\epsilon)=(x_1+b_1\epsilon,\ldots,x_d+b_d\epsilon)$ 
to be the additive transformation of $x$ corresponding to the `move-type' $b$.
In our applications,
we shall assume that $p_i=1/2$ for $i=1,\ldots,d$.
%, and $q(\epsilon)I_{\{\epsilon>0\}}\equiv N(0,\frac{\ell^2}{d})I_{\{\epsilon>0\}}$.
%Note that, for each $i$, $b_i\epsilon\sim N(0,\frac{\ell^2}{d})$, but even though $b_i\epsilon$ are pairwise uncorrelated
%($E(b_i\epsilon\times b_j\epsilon)=0$ for $i\neq j$), they are not independent since all of them involve the same $\epsilon$. 
%Also observe that
%$b_i\epsilon+b_j\epsilon=0$ with probability $1/2$ for $i\neq j$, showing that the linear combinations of $b_i\epsilon$ need
%not be normal. In other words, the joint distribution of $(b_1\epsilon,\ldots,b_d\epsilon)$ is not normal, even though the marginal
%distributions are normal and the components are pairwise uncorrelated. This also shows that $b_i\epsilon$ are not independent,
%because independence would imply joint normality of the components.

Thus, a single $\epsilon$ is simulated from $q^{(1)}(\cdot)I_{\{\epsilon>0\}}$, which is then either added to, or subracted
from each of the $d$ co-ordinates of $x$ with probability $1/2$. Assuming that the target distribution is proportional to 
$\pi$, the new move $T^{(1)}_b(x,\epsilon)$, corresponding to the move-type $b$,
is accepted with probability
\begin{equation}
\alpha=\min\left\{1,\frac{\pi(T^{(1)}_b(x,\epsilon))}{\pi(x)}\right\}.
\label{eq:acc_additive}
\end{equation}

\subsubsection{{\bf Multiplicative TMCMC}}
\label{subsubsec:multiplicative_tmcmc}

%Note that our moves at each step were symmetric and the magnitude of the jump would depend on the choice of the proposal density $q$. 
%Again, $q$ must have its support as $\mathbb{R}^{+}$, and from now onwards, we shall assume that it is a $N(0,1)$ distribution truncated at $0$. 
%Note that at each step , we sample only one $\epsilon$ from this proposal distribution and updates all the co-ordinates at one go. 
%The notion of symmetrical transitions at each step can be expressed from a mathematical point of view as follows. 
Again suppose that we are simulating from a $d$ dimensional space (say, $\mathbb R^d$), and that we are currently at a point 
$x= (x_{1}, \ldots, x_{d})$.
Let us now modify the definition of the random variables $b_{1}, \ldots, b_{d}$, such that, for $i=1,\ldots,d$, 
\begin{equation}
b_{i} =\left\{\begin{array}{ccc} +1 & \mbox{with probability} & p_i; \\
0 & \mbox{with probability} & q_i;\\
 -1 & \mbox{with probability} & 1-p_i-q_i.
 \end{array}\right.
 \label{eq:b_mult}
\end{equation}

Let $\epsilon\sim g^{(2)}=q^{(2)}(\cdot)I_{\{|\epsilon|\leq 1\}}$. If $b_i=+1$, then $x_i\rightarrow x_i\epsilon$, if $b_i=-1$,
then $x_i\rightarrow x_i/\epsilon$ and if $b_i=0$, then $x_i\rightarrow x_i$, that is, $x_i$ remains unchanged.
Let the transformed co-ordinate be denoted by $x^*_i$.
Also, let $J(b,\epsilon)$ denote the Jacobian of the transformation $(x,\epsilon)\mapsto (x^*,\epsilon)$. 
We denote $x^*$ by $T^{(2)}_b(x,\epsilon)$, the multiplicative transformation 
$(x,\epsilon)\mapsto (x^*,\epsilon)$ associated with the move-type $b$. 

For example, if $d=2$, then for $b=(1,1)$, $T^{(2)}_b(x,\epsilon)=(x_1\epsilon,x_2\epsilon)$ and the Jacobian is $\epsilon^2$, 
for $b=(-1,-1)$, $T^{(2)}_b(x,\epsilon)=(x_1/\epsilon,x_2/\epsilon)$ and $|J(b,\epsilon)|=\epsilon^{-2}$. 
For $b=(1,-1)$, $b=(-1,1)$, and $b=(0,0)$, $T^{(2)}_b(x,\epsilon)=(x_1\epsilon,x_2/\epsilon)$, $(x_1/\epsilon,x_2\epsilon)$, 
and $(x_1,x_2)$, respectively, and in all these three instances, 
$|J(b,\epsilon)|=1$. 
For $b=(1,0)$ and $b=(0,1)$, 
$T^{(2)}_b(x,\epsilon)=(x_1\epsilon,x_2)$ and $T^{(2)}_b(x,\epsilon)=(x_1,x_2\epsilon)$, respectively, and in both these cases
$|J(b,\epsilon)|=|\epsilon|$. For $b=(-1,0)$ or $b=(0,-1)$, $T^{(2)}_b(x,\epsilon)=(x_1/\epsilon,x_2)$ and
$(x_1,x_2/\epsilon)$, respectively, and the Jacobian is $|\epsilon|^{-1}$ in both these cases.
In general, the Jacobian for multiplicative TMCMC is given by $|\epsilon |^{\sum_{i=1}^d b_i}$. 

For our purpose, we assume that $p_i=q_i=1/3;~i=1,\ldots,d$.
Then assuming that the target distribution is proportional to 
$\pi$, the new move $T^{(2)}_b(x,\epsilon)$ is accepted
with probability
\begin{equation}
\alpha=\min\left\{1,\frac{\pi(T^{(2)}_b(x,\epsilon))}{\pi(x)}|J(b,\epsilon)|\right\}.
\label{eq:acc_multiplicative}
\end{equation}

\subsubsection{{\bf Additive-Multiplicative TMCMC}}
\label{subsubsec:add_mult_tmcmc}
\ctn{Dutta14} described another TMCMC algorithm that uses the additive transformation
for some co-ordinates of $x$ and the multiplicative transformation for the remaining co-ordinates. 
\ctn{Dutta14} refer to this as additive-multiplicative TMCMC. Let the target density $\pi$ be
supported on $\mathbb R^d$. Then, if the additive transformation is used for the $i$-th co-ordinate, 
we update $x_i$
to $x_i+b_i\epsilon_1$, where $b_i$ is defined by (\ref{eq:b_add}), and $\epsilon\sim g^{(1)}$.
On the other hand, if for any co-ordinate $x_j$, the multiplicative transformation is used,
then we simulate $b_j$ following (\ref{eq:b_mult}), simulate $\epsilon_2\sim g^{(2)}$,
and update $x_j$ to either $x_j\epsilon_2$ or $x_j/\epsilon_2$ accordingly as $b_j=+1$ or $-1$.
If $b_j=0$, then we leave $x_j$ unchanged. The new proposal is accepted with probability 
having the same form as (\ref{eq:acc_multiplicative}). Note that unlike the cases of additive
TMCMC and multiplicative TMCMC, which use a single $\epsilon$ to update all the $d$ co-ordinates
of $x$, here we need two $\epsilon$'s: $\epsilon_1$ and $\epsilon_2$, to update the $d$ co-ordinates.

\subsubsection{{\bf Mixture of additive and additive-multiplicative TMCMC for updating the interaction parameters}}
\label{subsubsec:mixture_tmcmc}

Note that additive TMCMC is expected to make shorter jumps, which maintain high acceptance rate, while
multiplicative TMCMC makes longer jumps on the average, which improves mixing behaviour of the underlying
Markov chain. Hence, it is expected that a mixture of additive and multiplicative TMCMC should outperform
the two individual TMCMC strategies. \ctn{Dey14} demonstrate with simulation studies that this is indeed the case.

For our purpose, we consider a mixture of additive and additive-multiplicative TMCMC, giving equal weight to 
both, for updating the interaction parameters. In the additive-multiplicative TMCMC we update 
$\left\{\left(u_{r'},v_{r'}\right);~r'=1,\ldots,L\right\}$, $\bLambda$, and the
diagonal elements of the lower triangular matrices $\bC_1$ and $\bC_2$, 
using the additive transformation,
while using the multiplicative transformation to update the off-diagonal elements of $\bC_1$ and $\bC_2$. 

Implementation of mixture TMCMC with equal mixing weights involves, for each iteration
of TMCMC, simulating a random number $R\sim U(0,1)$;
if $R<1/2$, additive TMCMC is to be employed. Otherwise, additive-multiplicative TMCMC must be implemented. 
The acceptance ratio (without the Jacobian) is obtained by evaluating 
\[
\pi(\bLambda)\pi(\bC_1\bC'_1)\pi(\bC_2\bC'_2)\times\prod_{\ell=1}^d\prod_{j=1}^J\prod_{k=0}^1\prod_{r=1}^{L_j}
\left\{{p}^*_{\ell jkr}\right\}^{{n}^*_{1j\ell r}+\exp\left(u_r+\lambda_{jk}\right)}
\left\{1-{p}^*_{\ell jkr}\right\}^{{n}^*_{2j\ell r}+\exp\left(v_r+\lambda_{jk}\right)}
\]
at the proposed and the old values of the interaction parameters, conditionally on the remaining parameters.
In the above, $\pi(\bLambda)$, $\pi(\bC_1\bC'_1)$ and $\pi(\bC_2\bC'_2)$ are given by
%(\ref{eq:pi_Lambda}), (\ref{eq:pi_A}) and (\ref{eq:pi_Sigma}), respectively.
(3.15), (3.21) and (3.23) of our main manuscript.

In our applications we chose $\epsilon\sim g^{(1)}\equiv N(0,1)I_{\{\epsilon>0\}}$ for additive
transformations and $\eta\sim g^{(2)}\equiv N(0,1)I_{\left\{\left|\eta\right|<1\right\}}$ for multiplicative transformations.
It is also important to mention that in our applications of additive transformation, we considered
the positive scaling factors $\varphi_1,\ldots,\varphi_d$, so that the transformation takes the form
\begin{equation*}
(x_{1}, \ldots, x_{d}) \rightarrow (x_{1}+ b_{1}\varphi_1\epsilon, \ldots, x_{d}+b_{d}\varphi_d\epsilon). 
\end{equation*}
For $\left\{(u_{r'},v_{r'});~r'=1,\ldots,L\right\}$ and $\bLambda$, choosing all the scale factors to be 
$0.01$ and choosing the relevant scale factors to be $0.05$ in the cases of $\bC_1$ and $\bC_2$ yielded
reasonable convergence.

\section{{\bf Hellinger distance for hypothesis testing and associated computational challenge}}
\label{sec:hellinger}

An appropriate divergence measure 
between any two probability distributions $f_1$ and $f_2$ over the same domain $\mathcal \bY$ is the
Hellinger distance %associated with the Bhattacharyya coefficient (\ctn{Bhattacharyya43}) 
given by
\begin{equation}
d(f_1,f_2)=\sqrt{1-BC(f_1,f_2)},
\label{eq:Bmetric}
\end{equation}
where $BC(f_1,f_2)$ is the Bhattacharyya coefficient (\ctn{Bhattacharyya43}), given, in the discrete case, by
\begin{equation}
BC(f_1,f_2)=\sum_{\by\in\mathcal \bY}\sqrt{f_1(\by)f_2(\by)}.
\label{eq:BC}
\end{equation}
It is well-known that $d(f_1,f_2)$ defined as (\ref{eq:Bmetric}), is a metric.

In our situation, $h_{0j}$ and $h_{1j}$ are distributions of $L_j$-variate binary random variables, so that
the support is $\left\{0,1\right\}^{L_j}$. For large $L_j$, this renders the $BC$ coefficient (\ref{eq:BC})
infeasible to compute. Indeed, in our applications, $L_j$ is of the order of thousands, and this compels
us to seek alternatives to the Hellinger metric.

\section{{\bf Clustering and Euclidean metrics}}
\label{sec:clustering_euclidean}

\subsection{{\bf Computationally efficient alternative based on clustering ideas}}
\label{subsec:clustering}

Ideas on clusterings of the mixture distributions $h_{0j}$ and $h_{1j}$ provides us with a novel
and computationally efficient procedure for testing $H_0$. 
Briefly, we assess discrepancies between the two mixture distributions
%implied by $k=0$ and $k=1$ 
$h_{0j}$ and $h_{1j}$
by studying the divergence between the two clusterings 
of $\bP_{Mjk=0}=\left\{ \bp_{1jk=0},\bp_{2jk=0},\ldots,\bp_{Mjk=0}\right\}$
and $\bP_{Mjk=1}=\left\{ \bp_{1jk=1},\bp_{2jk=1},\ldots,\bp_{Mjk=1}\right\}$, for $j=1,\ldots,J$.
Significantly large divergence between the two clusterings for some $j=1,\ldots,J$ clearly leads to rejection of $H_0$.
An appropriate metric for studying divergence between clusterings is described next. %in Section \ref{subsec:metric}.

\subsubsection{{\bf Choice of the clustering metric}}
\label{subsubsec:metric}
To avoid computational burden we work with the following clustering metric suggested by 
\ctn{Sabya11} as an approximation to the one coined by \ctn{Ghosh08}:
\begin{equation}
\hat d(I,II)=\max\left\{\bar d(I,II),\bar d(II,I)\right\},
\label{eq:approx_final}
\end{equation}
where
\begin{eqnarray}
\bar d(I,II)&=&\left\{\tilde n_{00}-\sum_{{i}=1}^{K_1}
\max_{1\leq j\leq K_2}\tilde n_{{i}{j}}\right\}\bigg/{\tilde n}_{00}\label{eq:approx1}\\
&=&1-\frac{\sum_{{i}=1}^{K_1}\underset{1\leq j\leq K_2}{\max}
{\tilde n}_{{i}{j}}}{{\tilde n}_{00}}.\label{eq:approx2}
\end{eqnarray}

\subsubsection{{\bf Shortcoming of the clustering metric for hypothesis testing}}
\label{subsubsec:clustering_shortcoming}
Significantly large divergence between clusterings of $\bP_{Mjk=0}$ and $\bP_{Mjk=1}$ indicate significant
difference between the mixture densities $h_{0j}$ and $h_{1j}$.
However, insignificant clustering distance between $\bP_{Mjk=0}$ and $\bP_{Mjk=1}$ need not necessarily 
imply insignificant difference between the above mixture densities. As a simple example, let us consider
two different parameter vectors $\{\theta_1,\theta_1,\theta_2\}$ and $\{\vartheta_1,\vartheta_1,\vartheta_2\}$.
Although these two vectors have the same clustering $\{\{1,2\},\{3\}\}$, the parameter vectors themselves may be
significantly different. Therefore, whenever the clustering distance is insignificant, it is important to check 
whether or not the parameter vectors being compared, are significantly different. We next propose a divergence
based on the Euclidean distance between two vectors for this purpose.

\subsection{{\bf Divergence based on Euclidean metric in conjunction with the clustering metric 
for hypothesis testing}}
\label{subsec:euclidean_clustering}
Note that when two clusterings are the same, %(at least not significantly different under $H_0$), 
minimizing
the Euclidean distance over all possible permutations of the clusters, provides a sensible measure
of divergence. 
In other words, for any two vectors $\bv^{(1)}=\left(v^{(1)}_1,\ldots,v^{(1)}_K\right)$ and 
$\bv^{(2)}=\left(v^{(2)}_1,\ldots,v^{(2)}_K\right)$ 
in $K$-dimensional Euclidean space, where $K>1$,
we propose the following divergence measure:
\begin{equation}
d_{E,\mbox{min}}\left(\bv^{(1)},\bv^{(2)}\right)=\min_{j_1,\ldots,j_K}\sqrt{\sum_{i=1}^K\left(v^{(1)}_i-v^{(2)}_{j_i}\right)^2}, 
\label{eq:d_min}
\end{equation}
the minimization being over all possible permutations $(j_1,j_2,\ldots,j_K)$ of $(1,2,\ldots,K)$.

Note that the maximum or the average over all possible permutations
is not appropriate -- even when the two vectors being compared are the same, taking maximum
or average over the permutations results in non-zero divergence.
The above divergence is non-negative, symmetric in that 
$d_{E,\mbox{min}}\left(\bv^{(1)},\bv^{(2)}\right)=d_{E,\mbox{min}}\left(\bv^{(2)},\bv^{(1)}\right)$, 
satisfies the property
$d_{E,\mbox{min}}\left(\bv^{(1)},\bv^{(2)}\right)=0$ if and only if $\bv^{(1)}=\bv^{(2)}$,
and is 
invariant with respect to permutations of the clusters. However, we refer to $d_{E,\mbox{min}}$ as a pseudo-metric as the divergence measure does not satisfy 
the triangular inequality. 
Failure of the triangle inequality is not unusual,  
a very well-known instance being the Kullback-Leibler divergence. See also \ctn{Basu11} for the
general class of divergence measures which do not satisfy the triangular inequality.
Hence, we do not perceive $d_{E,\mbox{min}}$ as suffering from any serious drawback.

%It is useful to remark that this Euclidean based divergence is itself not sufficient for the testing purpose
%and must be used in conjunction with the clustering metric.
%Indeed, two parameter vectors, one of which is just
%a permutation of the other, yields the same magnitude of the mixture density, so that even if
%the Euclidean based divergence between the parameter vectors is significantly large, this need not imply significant
%difference between the mixture densities associated with the parameter vectors.
%However, if the clusterings associated with the mixture densities are known to be of insignificant difference,
%then insignificant divergence between the parameter vectors does imply insignificant difference between the mixture
%densities.

\subsubsection{{\bf Strategy for avoiding minimization over permutations}}
\label{subsubsec:avoid_permutations}

Since the number of possible permutations can be quite large, computation 
of $d_{E,\mbox{min}}$ can be burdensome in the extreme for large number of MCMC iterations.
%for any iteration of our MCMC algorithm, we approximate the permutation-minimized Euclidean 
%distance with the simple Euclidean distance 
%for the particular permutations of the components of the two vectors obtained in the specific MCMC iteration.  
%
Hence, we consider the following strategy for actual testing of hypothesis using $d_{E,\mbox{min}}$ when 
the null hypothesis has been accepted by the clustering based test. 

We first test the hypothesis using the simple Euclidean metric $d_E$ after attaching
significant weight to the null hypothesis. 
Since $d_E\geq d_{E,\mbox{min}}$,
acceptance of the null hypothesis with respect to $d_E$ implies acceptance of the null with respect to $d_{E,\mbox{min}}$.
The strategy of providing preference to the null is justifiable on the ground that the
clustering metric has already provided partial evidence in favour of the null that at least the clusterings
are not significantly different. 
%In Section \ref{subsec:testing} we provide details on the method
%of imposing larger weightage to the null.

If the null hypothesis is accepted with respect to $d_E$, then we have clearly been able to
avoid minimization over permutations.
If, on the other hand, the null is rejected when tested with $d_E$, then one must re-test the null 
using $d_{E,\mbox{min}}$, which would involve dealing with permutations.
%Such an event is not very likely since the null, after being accepted with respect to the clustering metric, has significant weightage.

\subsubsection{{\bf Computation of the simple Euclidean metric in our case after logit transformation}}
\label{subsubsec:euclidean_logit}
In our case, in order to compute the simple Euclidean distance, we first compute the averages 
$\bar{p}_{mjk}=\sum_{r=1}^{L_j}p_{m,jkr}/L_j$, then consider their logit transformations
$\mbox{logit}\left(\bar{p}_{mjk}\right)=\log\left\{\bar{p}_{mjk}/(1-\bar{p}_{mjk})\right\}$. 
Then, we compute the Euclidean distance between the vectors
$$\mbox{logit}\left(\bar{\bP}_{Mjk=0}\right)=\left\{\mbox{logit}\left(\bar{p}_{1jk=0}\right),
\mbox{logit}\left(\bar{p}_{2jk=0}\right),
\ldots, \mbox{logit}\left(\bar{p}_{Mjk=0}\right)\right\}$$
and 
$$\mbox{logit}\left(\bar{\bP}_{Mjk=1}\right)=\left\{\mbox{logit}\left(\bar{p}_{1jk=1}\right),\mbox{logit}\left(\bar{p}_{2jk=1}\right),
\ldots, \mbox{logit}\left(\bar{p}_{Mjk=1}\right)\right\}.$$
We denote the Euclidean distance associated with the $j$-th gene by\\ 
$d_{E,j}=d_{E,j}\left(\mbox{logit}\left(\bar{\bP}_{Mjk=0}\right),
\mbox{logit}\left(\bar{\bP}_{Mjk=1}\right)\right)$, and denote $\underset{1\leq j\leq J}{\max}~d_{E,j}$ by $d^*_E$.

\section{{\bf Simulation studies}}
\label{sec:simulation_study}

\subsection{{\bf First simulation study: gene-gene interaction}}
\label{subsec:first_simulation_study}

\subsubsection{{\bf Data description}}
\label{subsubsec:data_description}
In the first simulation study we simulated 5 case-control type data sets associated with 5 different sub-populations in the context
of gene-gene interaction associated with two genetic factors. The
data sets consist of disease status, gender, environmental exposures and genotypes for each individual.
Two genes have been considered, one with 1084 SNPs and another with 1206 SNPs, with 
one DPL at each gene. Each of the 5 data sets consists of 113 individuals.
From the 5 data sets, we selected a total of 100 individuals without replacement with probabilities 
assigned to the 5 data sets being $(0.1, 0.4, 0.2, 0.15, 0.15)$. That is, we chose one of the 
5 data sets with these probabilities and selected a row randomly from the chosen data set; we repeated
this procedure 100 times without replacing the rows. In our final data set thus obtained, there
were 41 cases and 59 controls arising out of 5 different sub-populations.

\subsubsection{{\bf Specifications of the thresholds $\varepsilon$'s using null distributions}}
\label{subsubsec:threshold}
Before testing the relevant hypotheses, it is important to discuss how to choose the thresholds $\varepsilon$'s
associated with the hypotheses. Our idea is to study the null distribution of the distance measures in connection with the clusterings of the parameter vectors of the mixture distributions associated with a gene in cases and controls,
using which we specify the thresholds. In more details, 
we simulate a genotype data set using our own Bayesian semiparametric model, considering two genes,
the genes consisting of $L_1=1084$ and $L_2=1206$ SNPs, respectively, as in the original data set obtained
from GENS2. We also set $N_1+N_2=100$. 
To guarantee that there is no interaction between the genes, we set $\bA$ to be the identity matrix.
We also set $\bSigma$ to be the identity matrix.
For each gene $j$, and for control status $k=0$, we simulate
$\bP_{Mjk=0}$ using the Polya urn scheme, and set $\bP_{Mjk=1}=\bP_{Mjk=0}$, independently
for $j=1,2$; this ensures that 
for each gene, case and control are associated with exactly the same mixture, and that the genes are
unrelated to each other. 
Fitting our model to the data generated from the GENS2 software showed that about $5$ distinct
mixture components are highly probable for each $(j,k)$. Since our past research
on our Dirichlet process based mixture model (\ctn{Bhattacharya08}, \ctn{Sabya11}, \ctn{Sabya12},
\ctn{Majumdar13}) revealed that it is a reliable representative
of the true number of components, we assume that approximately $5$ components are to be expected for each $(j,k)$.
As such, we set $\alpha_{jk}=1.5$ so that $\alpha_{jk}\log\left(1+\frac{M}{\alpha_{jk}}\right)\approx 5$
is (approximately) the expected number of components. We set $\alpha_{jk}=1.5$ for generating the data from our model 
as well as for fitting our model to this generated data. Thus, about $5$ components are expected both
{\it a priori} and {\it a posteriori}.
%Recall that for the non-null model we had set $M=30$ and $\alpha_{jk}=10$ in contrast with the null model;
%the choice facilitates data-driven inference, which had the effect of capturing the correct
%number of distinct components by the non-null posterior.
%In general, however, there is no reason to expect the same distribution of the number of components under
%the null and non-null models.
%Indeed, under the non-null model gene-gene interactions directly affect $\bG_{0,jk}$, 
%which play the role of controlling the number of components through the Polya urn; see (S-1.3) of the supplement.
%This is not the case for the null model where gene-gene interactions are absent.

It is also important to note that, although we set $\bA$ and $\bSigma$ to be identity matrices while generating the data,
we fit our model to the data using the general set-up described in Sections 3.3.4 and 3.3.5 %and 3.3.6 
of our main manuscript.
%\ref{subsubsec:matrix_normal},
%\ref{subsubsec:other_priors} and \ref{subsubsec:iw3}.
%with $x^s_{ijkr}=1$ and $0$ with probabilities $1/2$ each, independently for $i=1,\ldots,N_k$; $k=0,1$;
%$r=1,\ldots,L_j$; $j=1,2$; $s=0,1$, where $L_1=1084$ and $L_2=1206$, and $N_1+N_2=100$. Thus, this data set
%is homogeneous for every $(i,j,k,r,s)$, although it has the same size and loci as the original simulated 
%data set. 

Hence, fitting our model to the data set generated under the absence of genetic and interaction effects 
are expected to yield posterior distributions of the relevant quantities 
which can serve as benchmark distributions under the null hypotheses. 
We generate posterior samples using the same parallel MCMC algorithm detailed in Section \ref{sec:computation}.
of the supplement.

We specify $\varepsilon$'s as $F^{-1}\left(0.55\right)$, where
$F$ is the distribution function of the relevant benchmark posterior distribution.
The reason for choosing $F^{-1}\left(0.55\right)$ instead of the median is to ensure that the
correct null hypothesis is accepted under the ``$0-1$" loss. Indeed, for the median, the posterior
probability of the true null is $0.5$, while under the ``$0-1$" loss, the true null will be accepted
if its posterior probability exceeds $1/2$.

%Note that alternatively, one could choose $\varepsilon=F^{-1}\left(0.01\right)$, for example, to ensure
%small left tail probability below $\varepsilon$, and choose $c$ of the ``$0-1-c$" loss to be large enough
%so that the null hypothesis is accepted at some appropriately small level of significance. However,
%in complex MCMC based analyses of high-dimensional posteriors such as ours, the extreme tails of the marginal 
%posteriors may be missed
%by the finite-size MCMC sample, yielding very poor estimates of the relevant tail probabilities, which
%can seriously affect the decisions regarding acceptance or rejection of the null hypotheses.
%Hence, we recommend setting $\varepsilon$ slightly greater than the median of the respective null posterior, which
%is also quite appropriate under the ``$0-1$" loss as discussed above.

%It is important to note that as the number of clusters within the clusterings increase, our clustering
%metric increases as well. Since under the null hypotheses we expect the clustering metric to concentrate
%around as small values as possible, for the appropriate null distribution we set $\alpha_{jk}=0.2$,
%so that the expected number of clusters, with $M=30$, is $\alpha_{jk}\log\left(1+\frac{M}{\alpha_{jk}}\right)\approx 1$,
%the minimum number of clusters permissible.

\subsubsection{{\bf Results of fitting our model}}
\label{subsubsec:results_first_simulation_study}

We implemented our parallel MCMC algorithm on a machine with i7 processors, splitting the mixture
updating mechanisms in 4 parallel processors, and updating the interaction parameters in a single processor.
Our code is written in C in conjunction with the Message Passing Interface (MPI) protocol for parallelisation.

The total time taken to implement $30,000$ MCMC iterations, where the first $10,000$ are discarded as burn-in,
is just about an hour. Informal convergence assessment with trace plots indicated reasonably good mixing.

\begin{figure}%[htp]
\centering
\subfigure[Posterior of $d^*$.]{ \label{fig:ggi_max_metric}
\includegraphics[width=6cm,height=6cm]{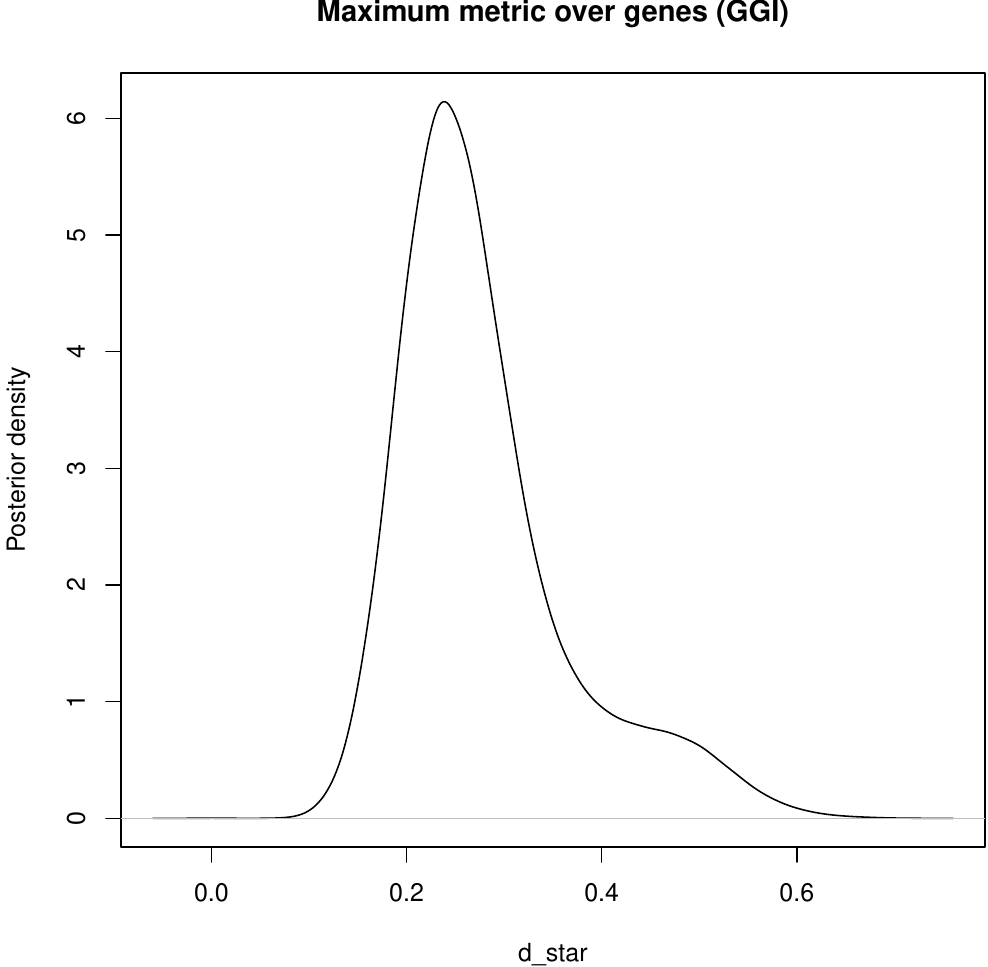}}
\hspace{2mm}
\subfigure[Posterior of $\hat d_1$.]{ \label{fig:ggi_metric_gene1} 
\includegraphics[width=6cm,height=6cm]{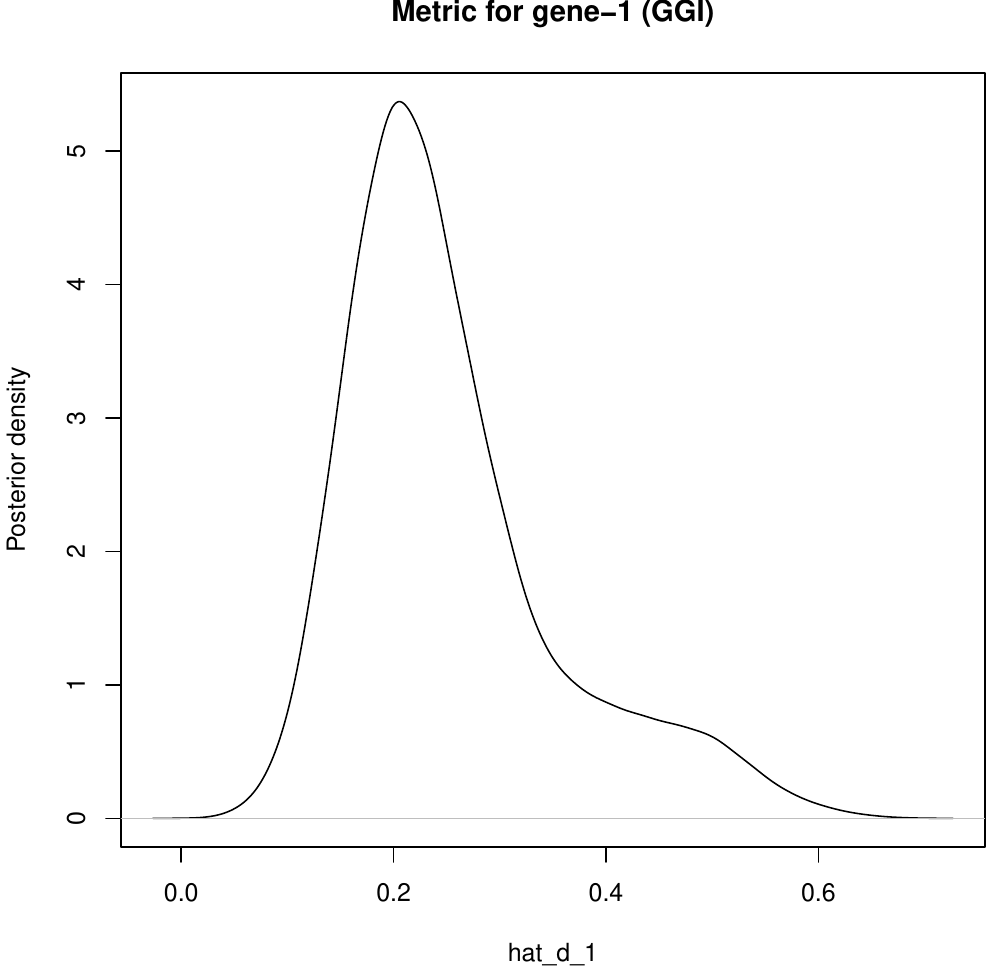}}\\
\vspace{4mm}
\subfigure[Posterior of $\hat d_2$.]{ \label{fig:ggi_metric_gene2} 
\includegraphics[width=6cm,height=6cm]{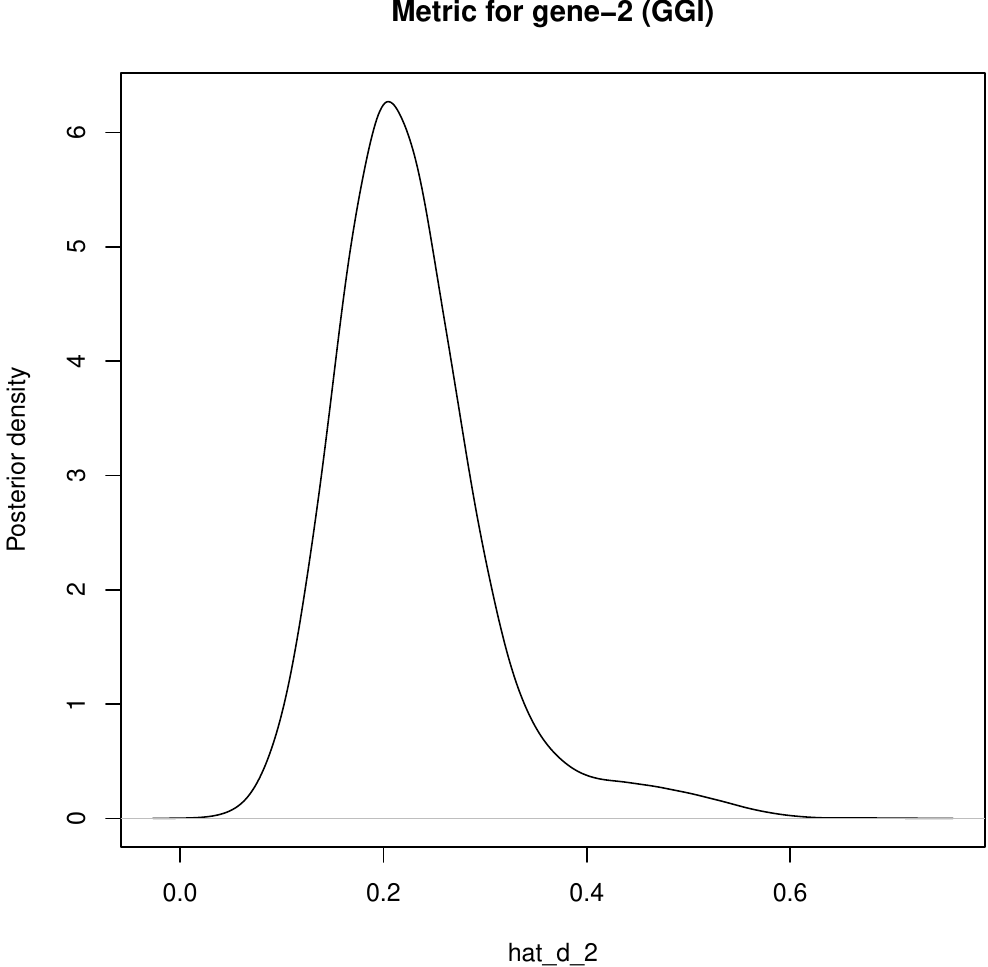}}
\caption{{\bf Gene-Gene Interaction:} Posterior distributions of 
$d^*=\underset{j=1,2}{\max}\hat d\left(\bP_{Mjk=0},\bP_{Mjk=1}\right)$ (panel (a)),  
$\hat d_1=\hat d\left(\bP_{Mjk=0},\bP_{Mjk=1}\right)$, with $j=1$ (panel (b)) and 
$\hat d_2=\hat d\left(\bP_{Mjk=0},\bP_{Mjk=1}\right)$, with $j=2$ (panel (c)).
In all the cases, $M=30$.}
\label{fig:ggi_metric_plots}
\end{figure}

Figure \ref{fig:ggi_metric_plots} displays the posterior distributions of 
$d^*=\underset{j=1,2}{\max}~\hat d\left(\bP_{30,j,0},\bP_{30,j,1}\right)$, 
$\hat d_1=\hat d\left(\bP_{30,1,0},\bP_{30,1,1}\right)$ and 
$\hat d_2=\hat d\left(\bP_{30,2,0},\bP_{30,2,1}\right)$, respectively.
The diagrams show that in all the three cases, regions that are significantly bounded away from zero
have high posterior probabilities compared to those closer to zero.
%the region to the right of $0.2$ has significantly higher posterior 
%probability compared to the left of $0.2$. 
%, while the maximum value is around $0.6$. 
For the purpose of formal Bayesian
hypothesis, following the discussion in Section \ref{subsubsec:threshold}, we
set $\varepsilon=0.233$. 
%to be $1/3$ of the maximum value $0.6$, that is,
%we specify $\varepsilon=0.2$. 
Then the posterior probability
$P\left(d^*<\varepsilon|\mbox{Data}\right)$, 
empirically obtained from $20,000$ MCMC samples,
turned out to be $0.230$. 
With $c=1$ in the ``$0-1-c$" loss (so that the popular ``$0-1$" loss is obtained),
this is far less than the threshold posterior probability $1/2$.
That is, under the ``$0-1$" loss, our Bayesian test of hypothesis clearly suggests significant overall genetic influence.
%Also note that, choosing $\varepsilon<0.233$ would have further substantially decreased 
%the posterior probability. 

It now remains to investigate individual and interaction effects of the genes.
The empirical posterior probabilities $P\left(\hat d_1<\varepsilon|\mbox{Data}\right)$
and $P\left(\hat d_2<\varepsilon|\mbox{Data}\right)$ turned out to be
$0.242$ and $0.280$, respectively, where we obtained $\varepsilon=0.2$. %suggesting significant individual genetic effects. 
Under the ``$0-1$" loss, our tests thus suggest significant individual genetic effects. 

Using the procedure detailed in Section \ref{subsubsec:threshold}, we obtain $\varepsilon=0.166$. %0.09.
The relevant empirical posterior probability is given by 
%$\mbox{P}\left(\left|\bA_{12}\right|<0.09|\mbox{Data}\right)\approx 0.169$, clearly pointing
$\mbox{P}\left(\left|\bA_{12}\right|<0.166|\mbox{Data}\right)\approx 0.326$, clearly pointing
towards significant gene-gene interaction under the ``$0-1$" loss.

%``no interaction" hypothesis.
%To make this more precise, we compute the posterior distribution of $\rho_{12}$, which may be 
%interpreted as the correlation
%between the genes. As seen in Figure \ref{fig:ggi_corr}, the posterior of $\rho_{12}$ gives high density
%around $-1$, relatively lesser density around $+1$, and quite small density to neighborhoods of $0$.
%Formal Bayesian tests with reasonable values of $\varepsilon$ will clearly reject the hypothesis
%of no interaction; see the first two columns of Table \ref{table:table1}.
%Thus, our results are in keeping with the genetic information used for generating the data, which is
%highly encouraging.

\begin{figure}%[htp]
\centering
\subfigure[Posterior of $\tau_{10}$.]{ \label{fig:ggi_comp_prob1}
\includegraphics[width=6cm,height=6cm]{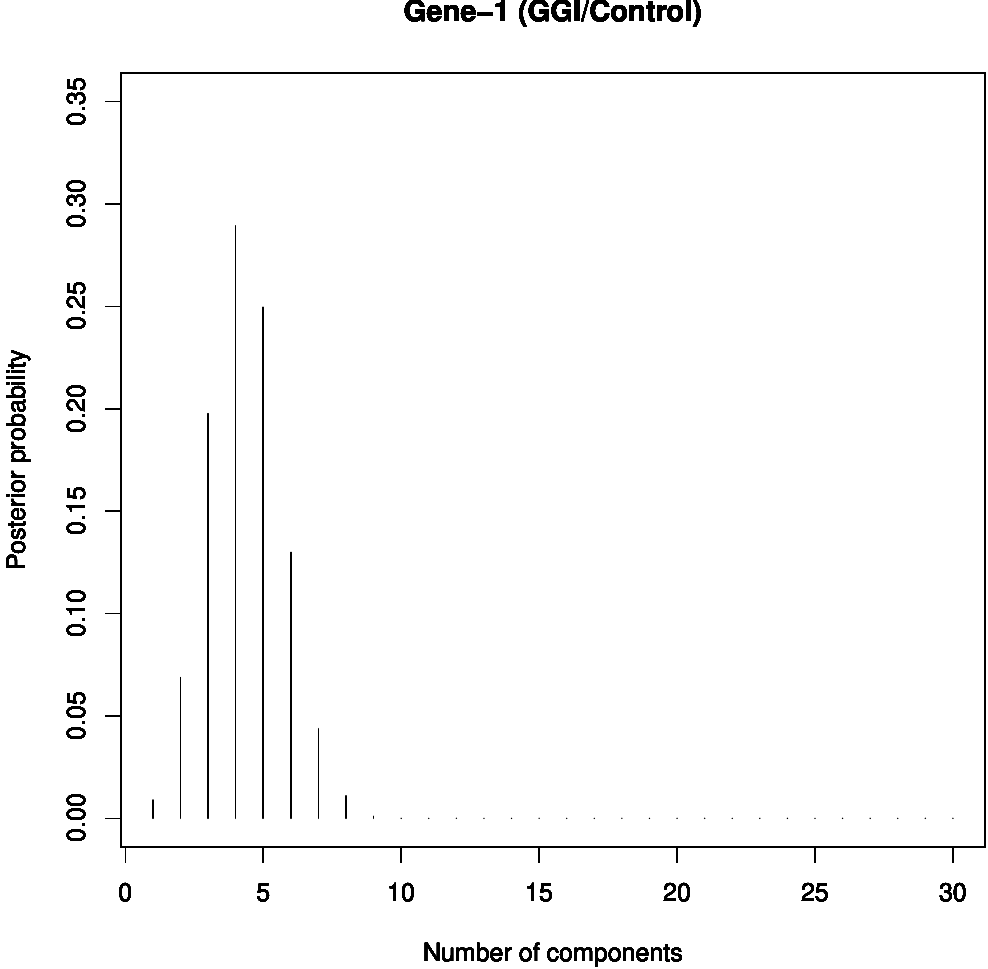}}
\hspace{2mm}
\subfigure[Posterior of $\tau_{11}$.]{ \label{fig:ggi_comp_prob2}
\includegraphics[width=6cm,height=6cm]{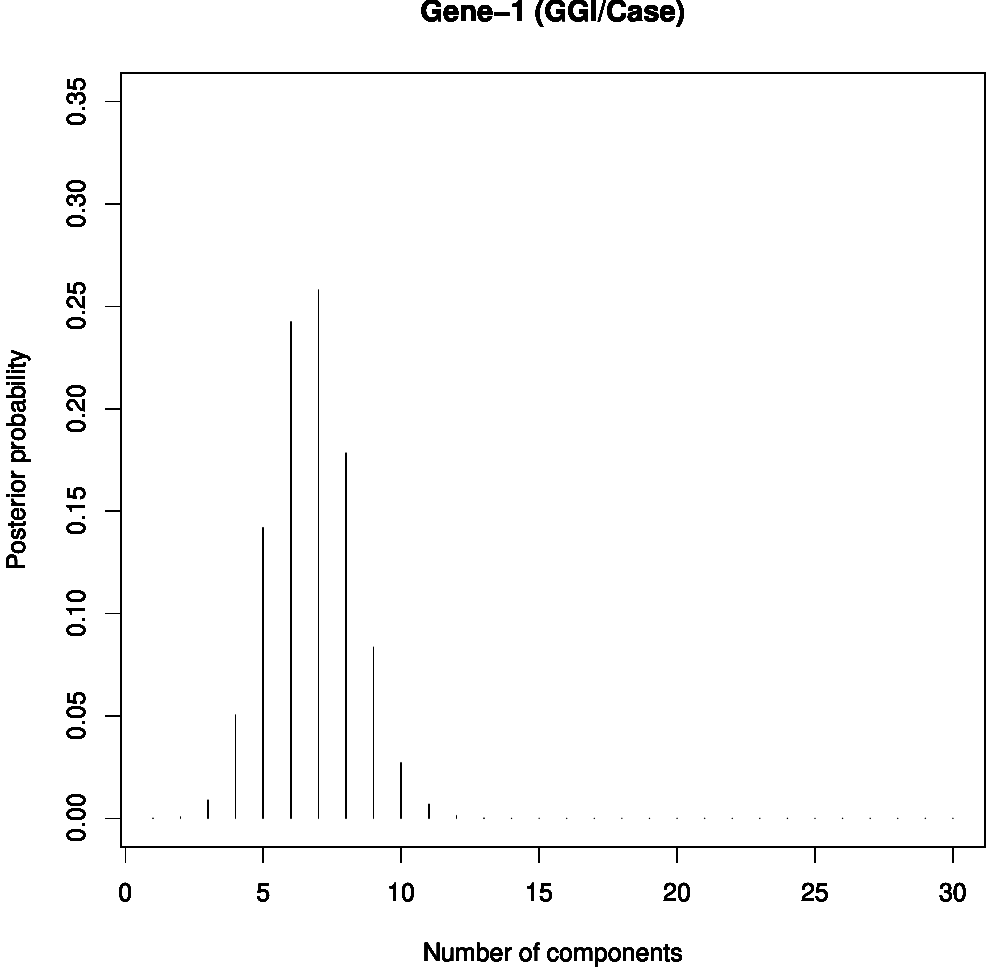}}
\hspace{2mm}
\subfigure[Posterior of $\tau_{20}$.]{ \label{fig:ggi_comp_prob3}
\includegraphics[width=6cm,height=6cm]{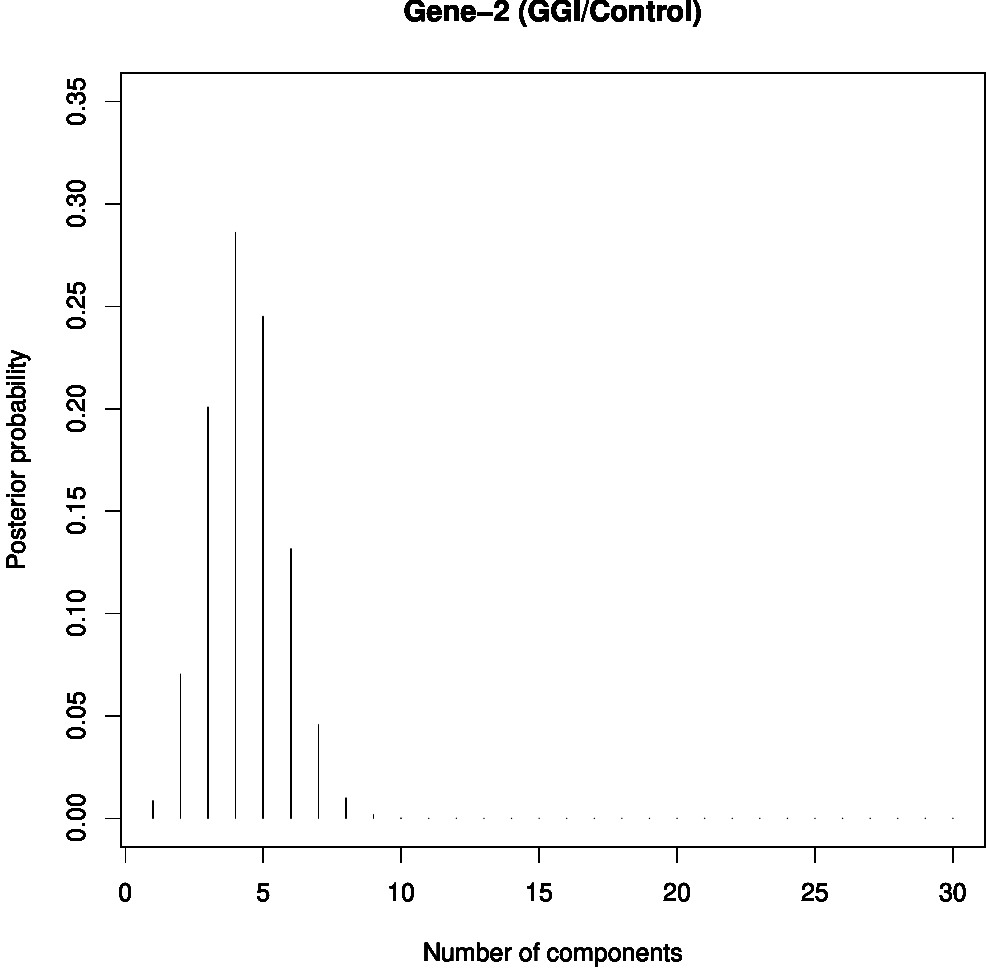}}
\hspace{2mm}
\subfigure[Posterior of $\tau_{21}$.]{ \label{fig:ggi_comp_prob4}
\includegraphics[width=6cm,height=6cm]{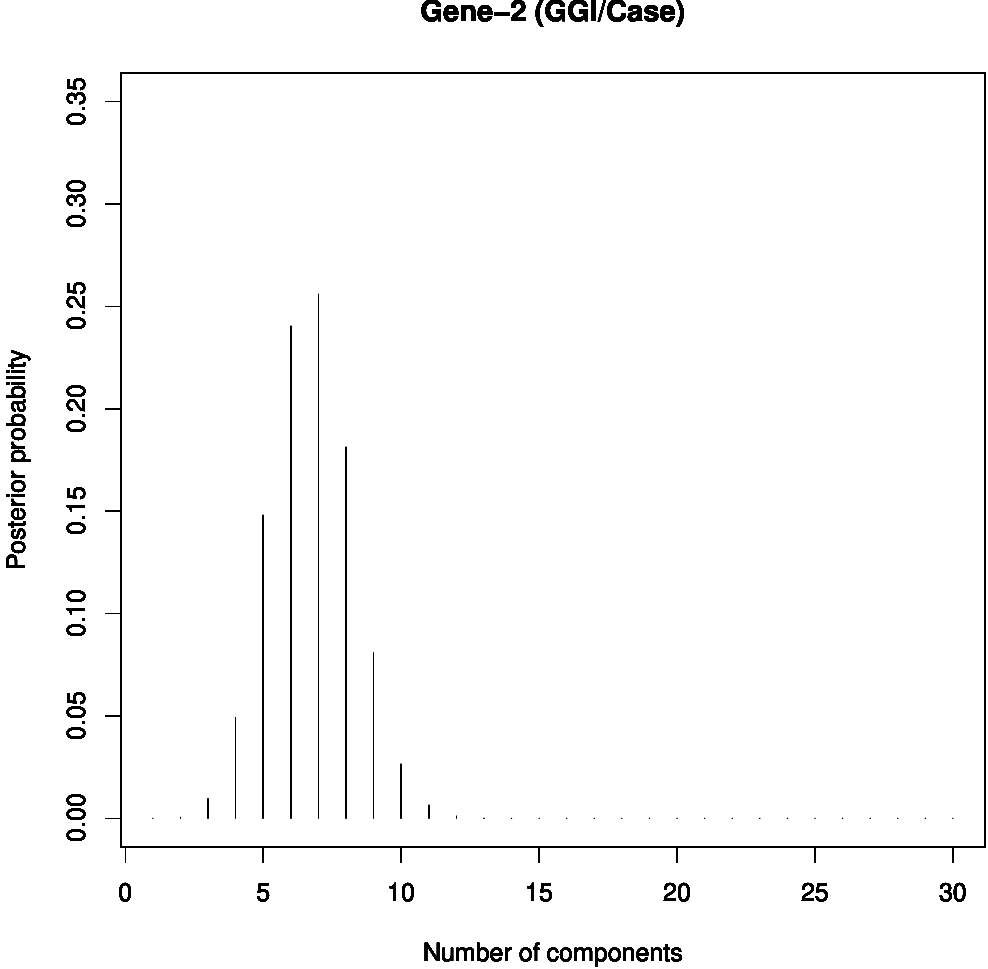}}
\caption{{\bf Gene-Gene Interaction:} Posterior distributions of the number of distinct components $\tau_{jk}$
for each pair ($j,k$); $j=1,2$; $k=0,1$.}
\label{fig:ggi_comp}
\end{figure}

Finally,the true numbers of sub-populations have been correctly captured by our
model and methodologies. Figure \ref{fig:ggi_comp} shows that although we started out with 
a maximum of $M=30$ components for each $(j,k)$; $j=1,2$; $k=0,1$, the posterior
distribution of the number of components in all the four pairs of $(j,k)$ have correctly concentrated
around 5, the true number of components. Once again, this is highly encouraging.

%\begin{figure}%[htp]
%\centering
%\subfigure[Posterior of $\bA_{11}$.]{ \label{fig:ggi_A11}
%\includegraphics[width=7cm,height=7cm]{plots/ggi_A11-crop.pdf}}
%\hspace{2mm}
%\subfigure[Posterior of $\bA_{12}$.]{ \label{fig:ggi_A12}
%\includegraphics[width=7cm,height=7cm]{plots/ggi_A12-crop.pdf}}\\
%\vspace{4mm}
%\subfigure[Posterior of $\bA_{21}$.]{ \label{fig:ggi_A21}
%\includegraphics[width=7cm,height=7cm]{plots/ggi_A21-crop.pdf}}
%\hspace{2mm}
%\subfigure[Posterior of $\bA_{22}$.]{ \label{fig:ggi_A22}
%\includegraphics[width=7cm,height=7cm]{plots/ggi_A22-crop.pdf}}\\
%\caption{{\bf Gene-Gene Interaction:} Posterior distributions of the elements of the 
%gene-gene interaction matrix $\bA$.}
%\label{fig:ggi_A}
%\end{figure}

%\begin{figure}%[htp]
%\centering
%\includegraphics[width=8cm,height=8cm]{plots/ggi_corr-crop.pdf}
%\caption{{\bf Gene-Gene Interaction:} Posterior distribution of $\rho_{12}$, correlation between the two genes.}
%\label{fig:ggi_corr}
%\end{figure}

\subsubsection{{\bf Detection of DPL}}
\label{subsubsec:dpl}

The case-control data simulated by the GENS2 software has one DPL in each of the two genes with the positions given by rs13266634 and rs7903146, for the first and second gene respectively.
However, as both the genes contain thousands of loci along with one DPL in each, with realistic patterns of Linkage Disequilibrium existing between them (see \ctn{Pinelli12}), We propose a graphical method to single out the influential SNPs. The details are as follows.

%Let $\bp^r_{jk}=\left\{p_{1jkr},p_{2jkr},\ldots,p_{Mjkr}\right\}$.
To check if the $r$-th locus of the $j$-th gene is disease producing, we assess if the Euclidean distance
$d^r_j\left(\bp^r_{jk=0},\bp^r_{jk=1}\right)$, between $\bp^r_{jk=0}$ and $\bp^r_{jk=1}$, is significantly larger than 
$d^{r'}_j\left(\bp^{r'}_{jk=0},\bp^{r'}_{jk=1}\right)$; for $r'\neq r$. Although
formal tests of significance for each locus is also possible, 
%particularly, in a Bayesian multiple testing
%framework (a general Bayesian multiple testing framework for dependent decisions
%is being developed by \ctn{Chandra14}), 
such tests can be computationally burdensome for large number of loci
such as ours. Hence, here we adopt a graphical approach based on index plots, in the 
spirit of detecting influential points in linear regression analysis. Such informal plots
are often advocated in statistics, see, for example, \ctn{Chatterjee06} and the references therein.

\begin{figure}%[htp]
\centering
\subfigure[Index plot for the first gene]{ \label{fig:index_plot_gene1}
\includegraphics[width=10cm,height=9cm]{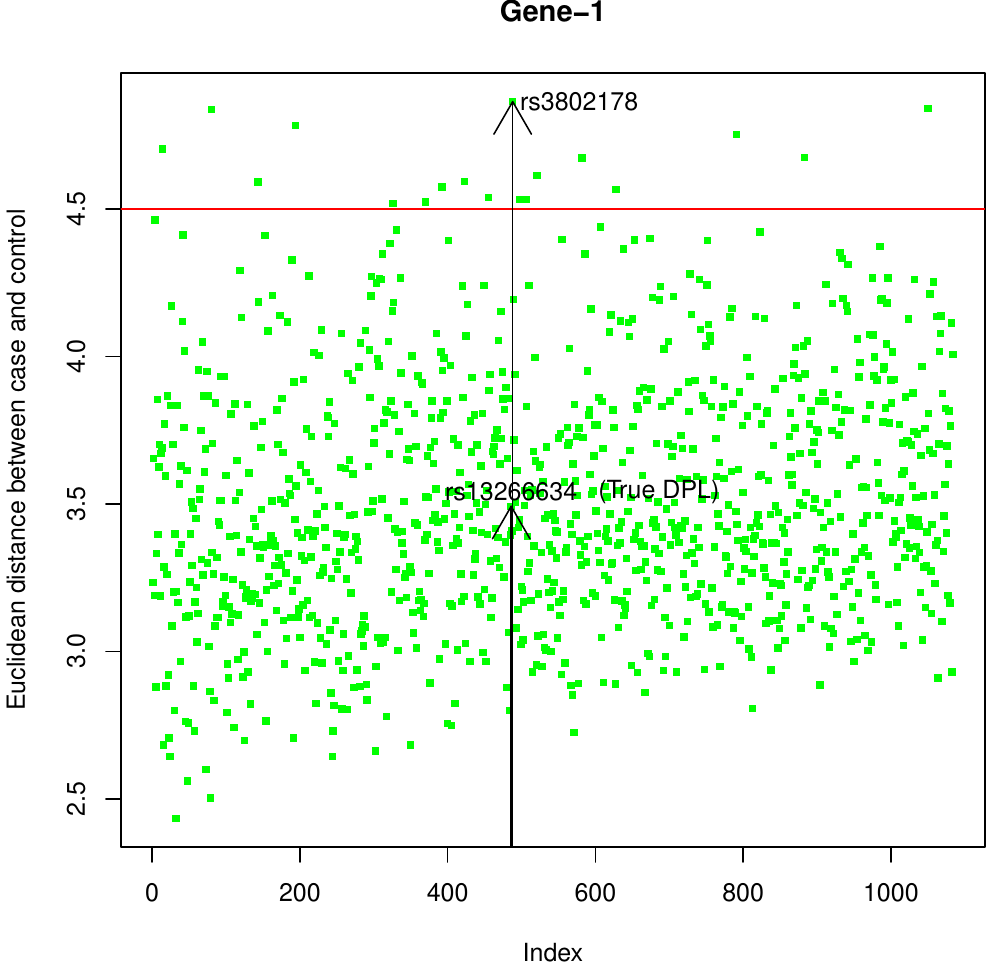}}
\vspace{2mm}
\subfigure[Index plot for the second gene.]{ \label{fig:index_plot_gene2}
\includegraphics[width=10cm,height=9cm]{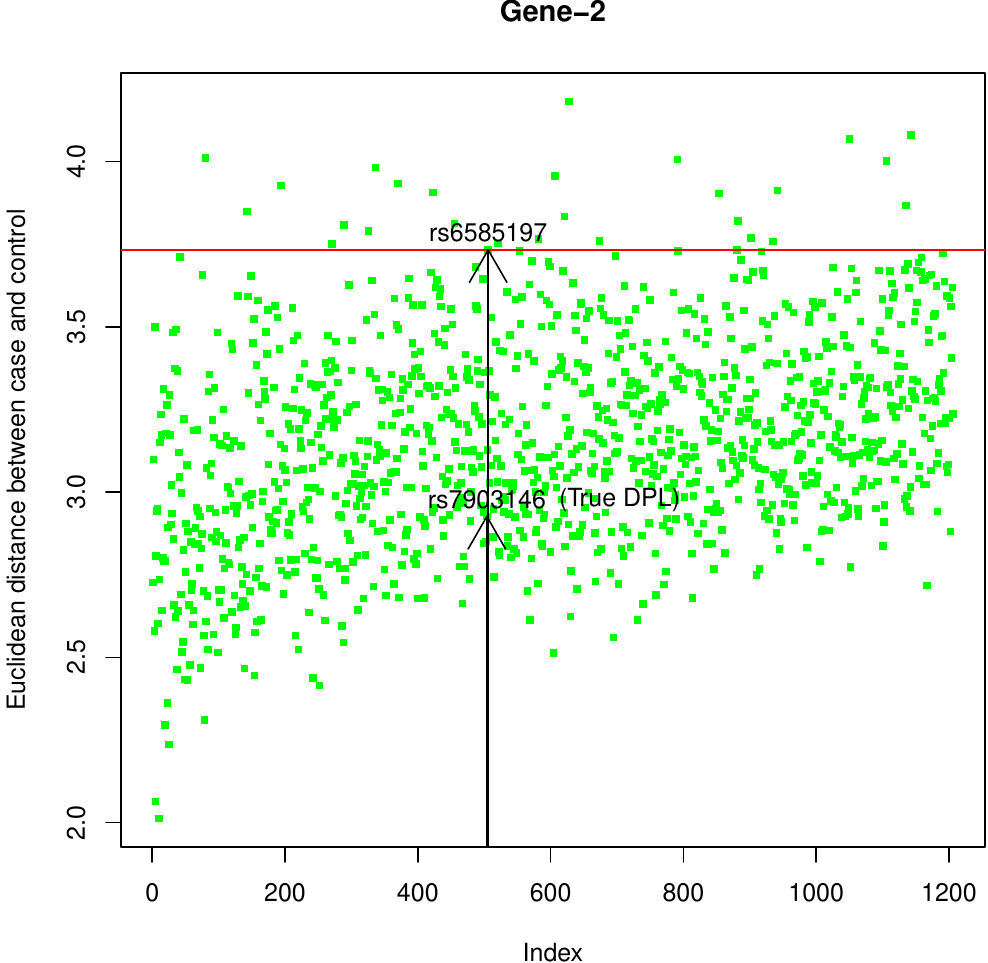}}
\caption{{\bf Index plots:} Plots of the Euclidean distances 
$\left\{d^r_j\left(\bp^r_{jk=0},\bp^r_{jk=1}\right);~r=1,\ldots,L_j\right\}$
against the indices of the loci, for $j=1$ (panel (a)) and $j=2$ (panel (b)).}
\label{fig:index_plots}
\end{figure}

In our case, for each gene $j$, 
we analyse the index plot of the Euclidean distances\\ 
$\left\{d^r_j\left(\mbox{logit}\left(\bp^r_{jk=0}\right),\mbox{logit}\left(\bp^r_{jk=1}\right)\right);~r=1,\ldots,L_j\right\}$. 
The plots, for the first and the second gene, are displayed in panels (a) and (b) of Figure 
\ref{fig:index_plots}.

The red, horizontal lines in the diagrams represent the cut-off value such that the points above the
horizontal line are those with the highest $2\%$ Euclidean distances. In panel (a) of Figure \ref{fig:index_plots},
the flagged point above the cut-off line, which is also associated with the maximum Euclidean distance, corresponds to SNP
position rs3802178. The figure also shows that the actual DPL rs13266634 is a very close neighbor of rs3802178. 
In panel (b) of Figure \ref{fig:index_plots}, the actual DPL rs7903146 is found to be lying very close to rs6585197, 
a SNP detected as influential by our method.
That is, our set of suspicious loci in the second gene again contains a very close neighbor of the true DPL. 
Realistically, it is appropriate to further investigate all the SNPs with Euclidean distances 
on or above the red, horizontal line, along with their close neighbors, as possible influential SNPs.

In spite of starting the investigation with simultaneous consideration of a large number of SNPs under the existence of realistic patterns of LD and a stratified population structure, our model and methodologies have not only detected the genetic and interaction effects correctly, but has also narrowed down the search for DPLs to a few influential SNPs, lying in the close neighborhoods of the actual DPL. Given that we assumed no knowledge of the true
model while fitting the data, this is highly encouraging.

\subsubsection{{\bf Results obtained after randomly permuting the labels of the loci in the dataset}}
\label{subsubsec:permutation1}
We conducted a further simulation study after randomly permuting the labels of the loci of the dataset.
In this case, we obtained exactly the same thresholds $\varepsilon$ as in the original simulation study, and the
probabilities $P\left(d^*<\varepsilon|\mbox{Data}\right)$, $P\left(\hat d_1<\varepsilon|\mbox{Data}\right)$
and $P\left(\hat d_2<\varepsilon|\mbox{Data}\right)$ are given, approximately, by $0.055$, $0.128$ and $0.123$,
respectively, strongly suggesting significant overall and marginal genetic effects. 
For gene-gene interaction effect, we obtained the threshold to be $\varepsilon=0.137$, and 
$P\left(|\bA_{12}|<\varepsilon|\mbox{Data}\right)\approx 0.132$, strongly suggesting significant gene-gene
interaction.

Thus, the results associated with randomly permuted labels of the loci are very much in keeping with
the results of the original simulation study.

\subsection{{\bf Second simulation study: no genetic effect}}
\label{subsec:second_simulation_study}

In this study, exactly in the same way as in the first simulation study, we simulated 
a mixture data set consisting of 5 sub-populations with
mixing proportions $(0.1, 0.4, 0.2, 0.15, 0.15)$;
the only difference with the first simulation study being the absence of any genetic effect and presence of the effect due environmental factor only. In a total of $100$ individuals simulated, there were 49 cases and 51 controls. 
For specification of the thresholds $\varepsilon$'s, we employ the same method proposed in Section 
\ref{subsubsec:threshold}. %but here we also set $\bA_{01}=\bA_{10}=0$.

We implement our model with the parallel MCMC algorithm in exactly the same way as in the first simulation
study, and obtained $30,000$ iterations with the first $10,000$ discarded as burn-in.
The posterior empirical probabilities $P\left(d^*<\varepsilon_1|\mbox{Data}\right)$, 
$P\left(d_1<\varepsilon_2|\mbox{Data}\right)$ and $P\left(d_2<\varepsilon_3|\mbox{Data}\right)$, where
$\varepsilon_1=0.233$, $\varepsilon_2=\varepsilon_3=0.2$, turned out to be
$0.554$, $0.511$, and $0.502$, respectively.
%all of which 
%provide strong evidence against the hypothesis of genetic influence under the ``$0-1$" loss, in terms
%of the clustering metric.
Note that, under the ``$0-1$" loss, the evidence associated with $d^*$ favours
the hypothesis of no genetic effect; the evidence is particularly strong because the posterior probability 
$P\left(d^*<\varepsilon_1|\mbox{Data}\right)$ almost exactly matches the corresponding 
posterior probability under the true null hypothesis of no genetic effect. 
The same argument clarifies that the other two posterior probabilities $P\left(d_1<\varepsilon_2|\mbox{Data}\right)$ and 
$P\left(d_2<\varepsilon_3|\mbox{Data}\right)$ also provide reasonably strong evidence 
against the hypothesis of genetic influence.

%> sum(ggni_max_eu<15.159)/20000
%[1] 0.0258
%> quantile(ggi_null_eu[FROM:TO,1],prob=c(0.025,0.5,0.975))
%     2.5%       50%     97.5% 
%   7.904776 13.747736 18.475205 
%> sum(ggni_eu[FROM:TO,1]<13.748)/20000
%      [1] 0.03605
%> quantile(ggi_null_eu[FROM:TO,2],prob=c(0.025,0.5,0.975))
%     2.5%       50%     97.5% 
%   8.048412 14.006883 18.818716 
%> sum(ggni_eu[FROM:TO,2]<14.007)/20000
%[1] 0.0869

%> sum(ggi_max_eu<15.159)/20000
%[1] 0.0115
%> sum(ggi_eu[FROM:TO,1]<13.748)/20000
%[1] 0.0248
%> sum(ggi_eu[FROM:TO,2]<14.007)/20000
%[1] 0.05625

To re-confirm the null hypotheses, we now resort to our tests based on the Euclidean metric.
%
%The posterior of $\rho_{12}$, displayed in Figure \ref{fig:ggni_corr}, assigns much higher probabilities
%to neighborhoods of zero
%compared to the gene-gene interaction scenario exemplified in the first simulation study.
%Table \ref{table:table1} compares the posterior probabilities 
%$\mbox{P}\left(\left|\rho_{12}\right|<\varepsilon|\mbox{Data}\right)$ for various choices of $\varepsilon$, 
%under situations when genetic influence
%is present and absent.
%For each choice of $\varepsilon$, the posterior probability is significantly larger when genetic effect
%is absent, providing substantial evidence against gene-gene interaction.
%With the Euclidean metric, as before $\varepsilon_E$ turned out to be $9$. 
Following the method proposed in Section \ref{subsubsec:threshold} we obtained the thresholds
%$\varepsilon_E=15.159$, $\varepsilon_{E,1}=13.748$ and $\varepsilon_{E,2}=14.007$, for evaluating
$\varepsilon_E=17.410$, $\varepsilon_{E,1}=16.250$ and $\varepsilon_{E,2}=16.307$, for evaluating
the relevant posterior probabilities, $P\left(d^*_E<\varepsilon_E|\mbox{Data}\right)$,
$P\left(d_{E,1}<\varepsilon_{E,1}|\mbox{Data}\right)$ and $P\left(d_{E,1}<\varepsilon_{E,2}|\mbox{Data}\right)$.
%These probabilities are evaluated to be approximately $0.026$, $0.036$, and $0.087$, respectively. 
These probabilities are evaluated to be approximately $0.118$, $0.156$, and $0.256$, respectively. 
%The choice $c=39$ is motivated by 2.5\% level of significance in the classical context.
For $c=19$ associated with the ``$0-1-c$" loss, so that $1/(1+c)= 0.05$, 
the above hypotheses are clearly accepted at $5\%$ level of significance, ensuring that the genes are
not responsible for the case-control status.
In fact, more generally, for $c\geq 9$, implying that $1/(1+c)\leq 0.1$, the
above posterior probabilities ensure
acceptance of the hypotheses at levels of significances not exceeding $10\%$.
Moreover, as in the first simulation study, even in this case 
the true number of sub-populations has been well-captured by our model (figures not shown).

Hence, all our results, under both the simulation studies, are very much in keeping with the underlying 
true genetic information used for generating the data sets.

\subsubsection{{\bf Results obtained after randomly permuting the labels of the loci in the dataset}}
\label{subsubsec:permutation2}
In this case we obtained
$P\left(d^*<\varepsilon|\mbox{Data}\right)\approx 0.439$, $P\left(\hat d_1<\varepsilon|\mbox{Data}\right)\approx 0.519$
and $P\left(\hat d_2<\varepsilon|\mbox{Data}\right)\approx 0.585$. Although $P\left(d^*<\varepsilon|\mbox{Data}\right)$
did not cross $0.5$, it is quite substantial, and in conjunction with the above marginal posterior probabilities suggest
no genetic effect. This is strongly confirmed by the tests based on the Euclidean metric, as
$P\left(d^*_E<\varepsilon_E|\mbox{Data}\right)$,
$P\left(d_{E,1}<\varepsilon_{E,1}|\mbox{Data}\right)$ and $P\left(d_{E,1}<\varepsilon_{E,2}|\mbox{Data}\right)$
are approximately $0.794$, $0.672$ and $0.671$, respectively, with respect to the thresholds 
$\varepsilon_E=18.195$, $\varepsilon_{E,1}=16.502$ and $\varepsilon_{E,2}=16.510$.
Even $P\left(|\bA_{12}|<\varepsilon|\mbox{Data}\right)\approx 0.582$, strongly suggesting insignificant gene-gene interaction.

In other words, again the results associated with random permutation of the labels of the loci of the genes
are consistent with the original simulation study and the truth.

%\section{{\bf Posterior probabilities of no genetic effect in the real data analysis}}
%\label{sec:postprob_realdata}
%\addtocounter{figure}{3}
\begin{comment}

\begin{figure}%[htp]
\centering
\subfigure[Real Data Analysis: Posterior probability of no genetic effect with respect to clustering metric.]
{ \label{fig:clustering_hypotheses}
\includegraphics[width=15cm,height=6cm]{plots_realdata/postprob_clustering-crop.pdf}}\\
\vspace{4mm}
\subfigure[Real Data Analysis: Posterior probability of no genetic effect with respect to Euclidean metric.]
{ \label{fig:euclidean_hypotheses} 
\includegraphics[width=15cm,height=6cm]{plots_realdata/postprob_euclidean-crop.pdf}}
\caption{{\bf Posterior probabilities of no individual genetic influence:} 
Index plots of the posterior probabilities of the null hypotheses for (a) clustering metric
and (b) Euclidean metric, for the $32$ genes.}
\label{fig:null_hypotheses}
\end{figure}

\begin{figure}%[htp]
\centering
\subfigure[Clustering metric medians.]{ \label{fig:clustering_medians}
\includegraphics[width=15cm,height=6cm]{plots_realdata/Clustering_medians-crop.pdf}}\\
\vspace{4mm}
\subfigure[Euclidean metric medians.]{ \label{fig:euclidean_medians} 
\includegraphics[width=15cm,height=6cm]{plots_realdata/Euclidean_medians-crop.pdf}}
\caption{{\bf Posterior medians of the Euclidean distances:} 
Index plots of the posterior medians of the clustering metric and the Euclidean distance
with respect to the $32$ genes.}
\label{fig:metric_medians}
\end{figure}

\end{comment}

\section{{\bf Explanation of the issue that even small correlations between SNP-wise case-control Euclidean
distances determine the DPL}}
\label{sec:small_corr_DPL}
Let us consider the following example where $(X,Y)\sim N_2\left(\mu_X,\mu_Y,\sigma^2_X,\sigma^2_Y,\rho\right)$, that is, 
$(X,Y)$ are distributed as bivariate normal with means $\mu_X$, $\mu_Y$; variances $\sigma^2_X$, $\sigma^2_Y$, 
and correlation $\rho$.
Then, the conditional expectation of $Y$ given $X=x$ is given by $E[Y|X=x,\rho]=\mu_Y+\frac{\sigma_Y}{\sigma_X}\rho x$.

Now, for any positive integer $n$, suppose that we wish to find the maximum among $\left\{\hat Y_1,\ldots,\hat Y_n\right\}$, 
where, for $i=1,\ldots,n$, $\hat Y_i=E[Y|X=x,\rho_i]=\mu_Y+\frac{\sigma_Y}{\sigma_X}\rho_i x$.
Clearly, the maximum will be $\hat Y_{i^*}$, where $i^*=\underset{i\in\{1,\ldots,n\}}{\arg\max}\rho_i$. 
In other words, irrespective of how small  
the values of $\left\{\rho_1,\ldots,\rho_n\right\}$ are, the maximum correlation among them dictates which value among
$\left\{\hat Y_1,\ldots,\hat Y_n\right\}$ will be the maximum. 

In our case, the quantile-quantile plots indicate that the SNP-wise Euclidean distances are quite close to normality, and since
our goal is to find the maximum among the SNP-wise expectations of the Euclidean distances (approximated
by averaging over the TMCMC samples), the above argument explains that indeed the correlations between the SNP-wise
Euclidean distances, however small in magnitude, dictate which SNP will be the DPL.

\section{{\bf Some other significant SNPs in the real data analysis}}
\label{sec:compare_DPL}
Figures %\ref{fig:DPL_genes_other1}, 
\ref{fig:DPL_genes_other2}, \ref{fig:DPL_others} and \ref{fig:DPL_others2} 
clearly indicate that in almost all the cases excepting the genes $MIA3$ and $PHACTR1$, 
the Euclidean distances of the significant SNP by our method agree quite closely with the SNPs 
detected as significant in the earlier studies (see \ctn{LucasG12} and the references therein). 
Note that for all the genes, including $MIA3$ and $PHACTR1$, the SNPs found significant in other studies 
lie in close neighbourhood of our most significant SNP with respect to the Euclidean distance; 
highlighting once again the need for close investigation of the SNPs lying in the close neighbourhood 
of those with highest Euclidean distance.

%\newpage

\addtocounter{figure}{3}
\begin{figure}%[htp]
\centering
\subfigure[Real Data Analysis: Posterior probability of no genetic effect with respect to clustering metric.]
{ \label{fig:clustering_hypotheses}
\includegraphics[width=15cm,height=6cm]{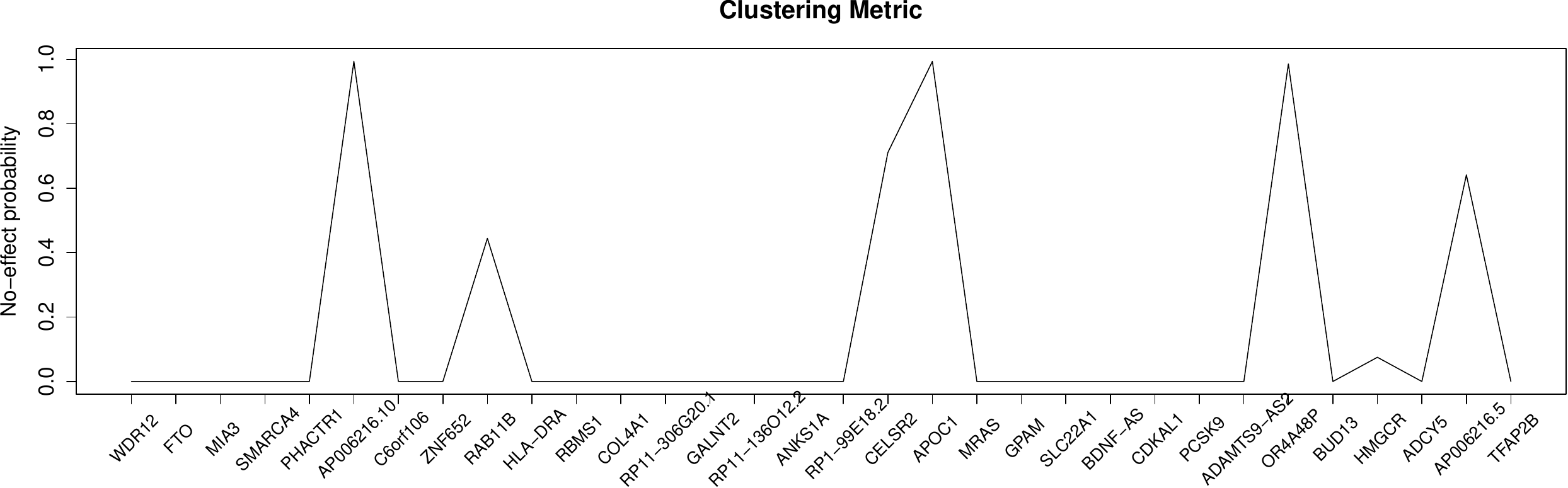}}\\
\vspace{4mm}
\subfigure[Real Data Analysis: Posterior probability of no genetic effect with respect to Euclidean metric.]
{ \label{fig:euclidean_hypotheses} 
\includegraphics[width=15cm,height=6cm]{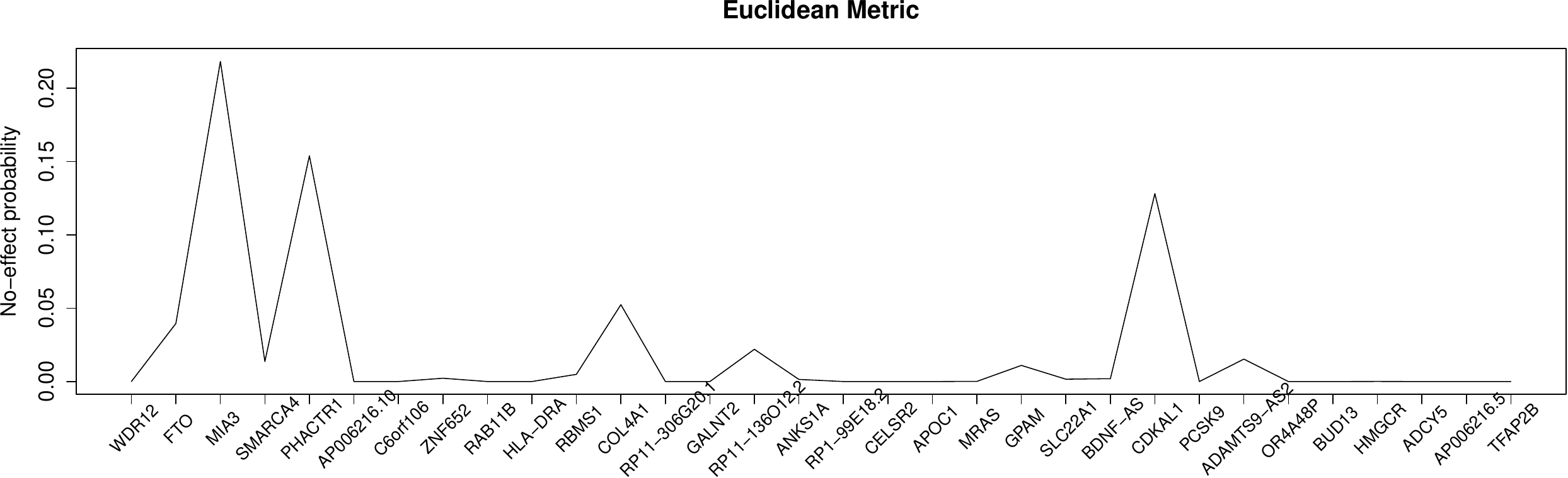}}
\caption{{\bf Posterior probabilities of no individual genetic influence:} 
Index plots of the posterior probabilities of the null hypotheses for (a) clustering metric
and (b) Euclidean metric, for the $32$ genes.}
\label{fig:null_hypotheses}
\end{figure}

\begin{figure}%[htp]
\centering
\subfigure[Clustering metric medians.]{ \label{fig:clustering_medians}
\includegraphics[width=15cm,height=6cm]{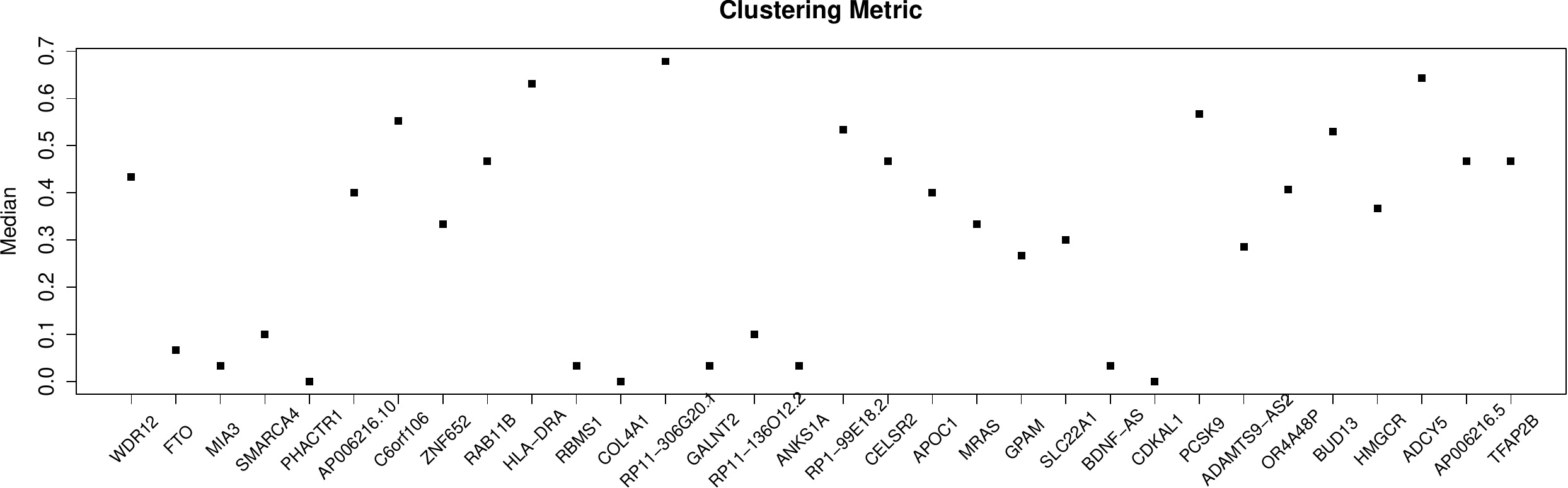}}\\
\vspace{4mm}
\subfigure[Euclidean metric medians.]{ \label{fig:euclidean_medians} 
\includegraphics[width=15cm,height=6cm]{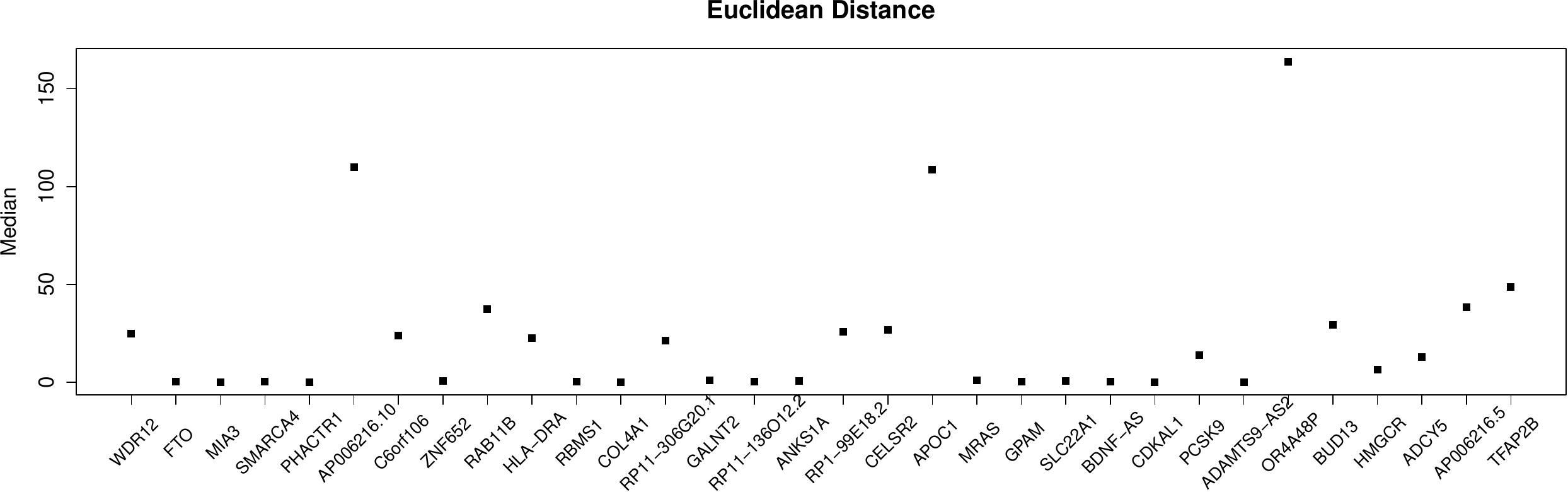}}
\caption{{\bf Posterior medians of the Euclidean distances:} 
Index plots of the posterior medians of the clustering metric and the Euclidean distance
with respect to the $32$ genes.}
\label{fig:metric_medians}
\end{figure}

\begin{comment}
\addtocounter{figure}{3}
\begin{figure}%[htp]
\centering
\subfigure[DPL of $WDR12$.]{ \label{fig:DPL_gene_1}
\includegraphics[width=6cm,height=6cm]{plots_realdata/plot_DPL_gene_1-crop.pdf}}
\hspace{2mm}
\subfigure[DPL of $FTO$.]{ \label{fig:DPL_gene_2}
\includegraphics[width=6cm,height=6cm]{plots_realdata/plot_DPL_gene_2-crop.pdf}}\\
\caption{{\bf Disease predisposing loci of other influential genes:} 
Plots of the Euclidean distances 
against the indices of the loci of genes $WDR12$ and $FTO$. %, $MIA3$, $SMARCA4$, $PHACTR1$, $C6orf106$. 
%The DPL that we obtained are close in terms of Euclidean distance to those loci which 
%are considered influential in the literature.
}
\label{fig:DPL_genes_other1}
\end{figure}
\end{comment}

%\pagebreak
%\addtocounter{figure}{3}
\begin{figure}%[htp]
\centering
%\vspace{2mm}
\subfigure[DPL of $WDR12$.]{ \label{fig:DPL_gene_1}
\includegraphics[width=5cm,height=5cm]{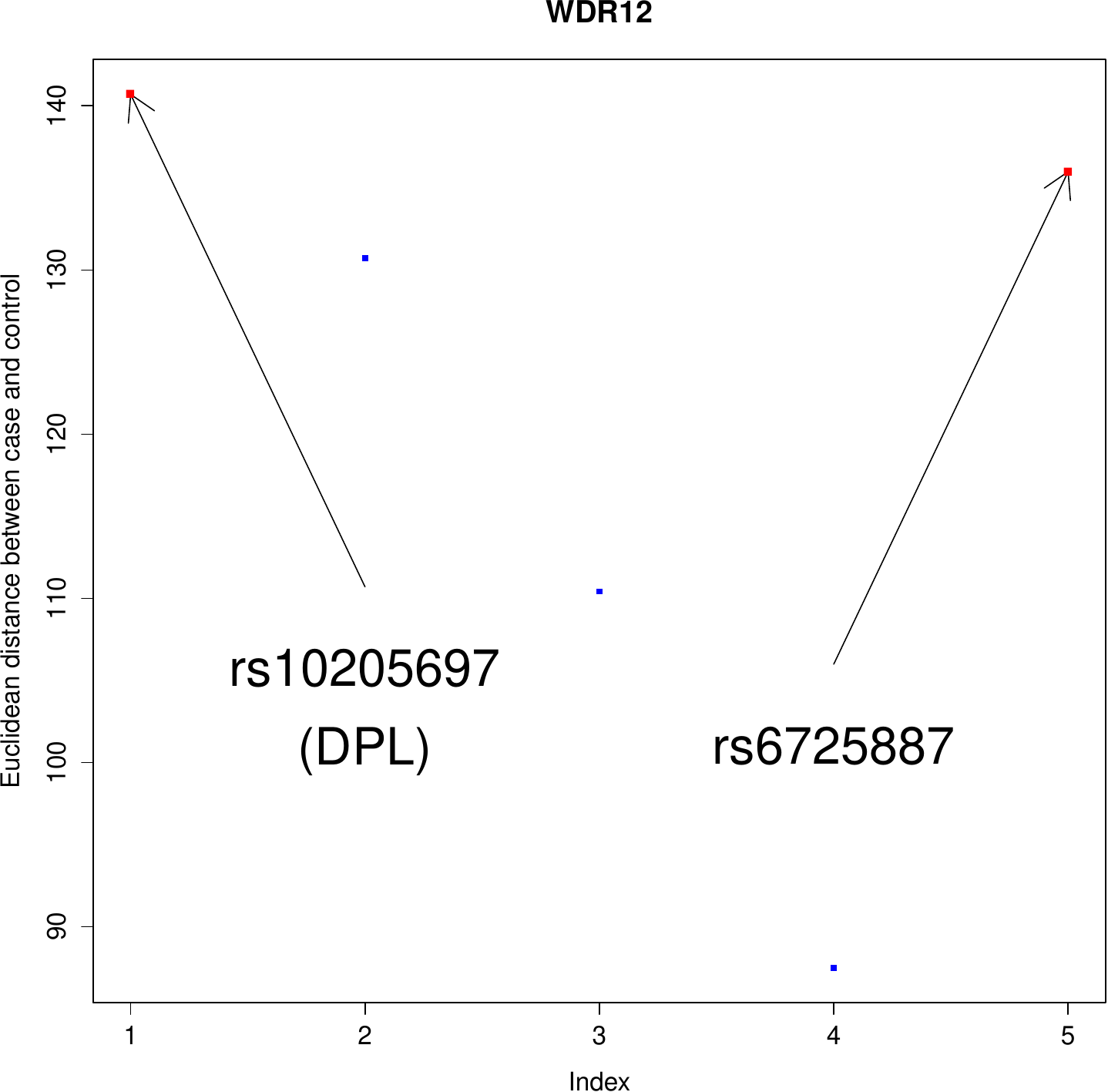}}
\hspace{2mm}
\subfigure[DPL of $FTO$.]{ \label{fig:DPL_gene_2}
\includegraphics[width=5cm,height=5cm]{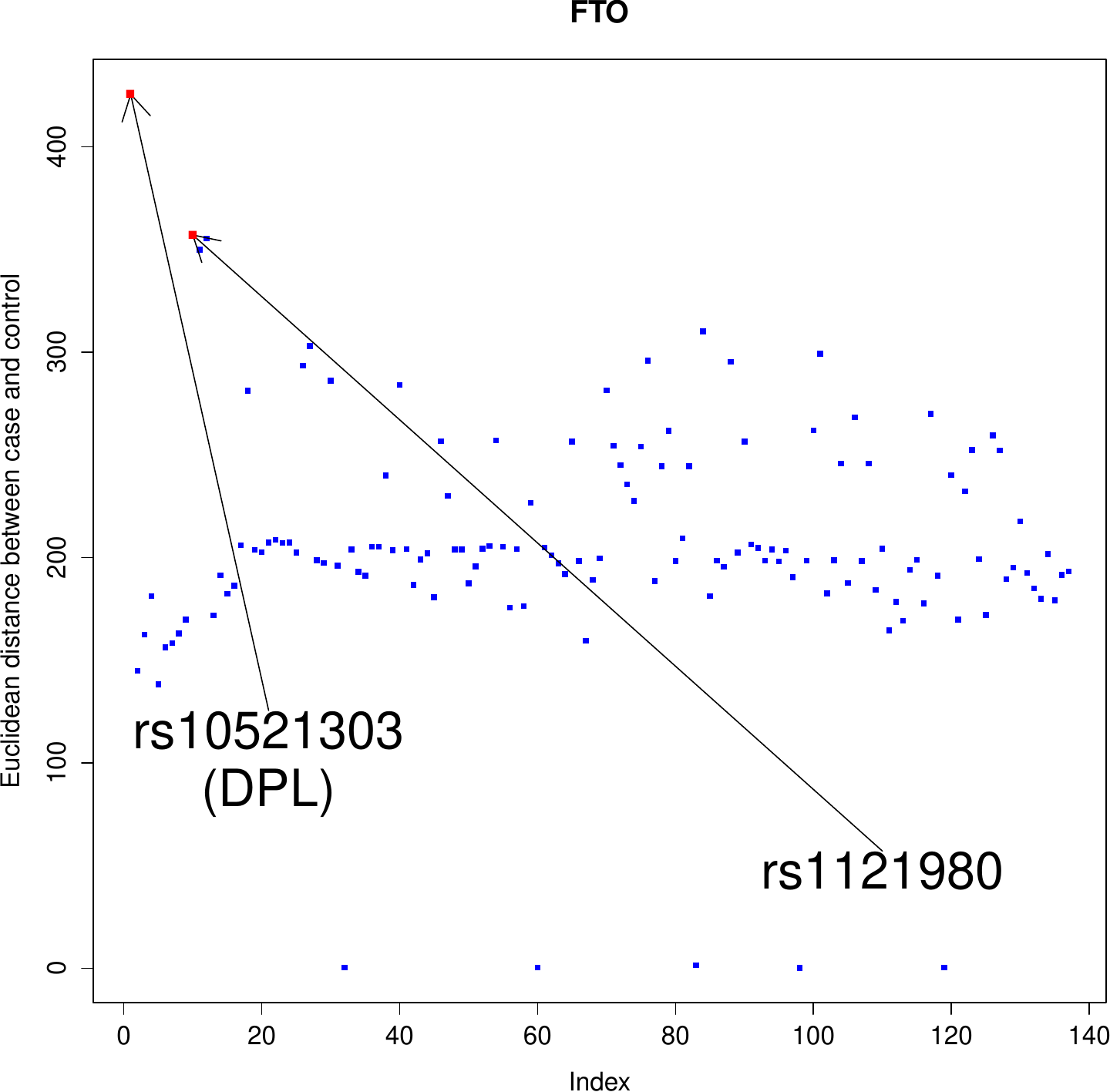}}\\
\vspace{2mm}
\subfigure[DPL of $MIA3$.]{ \label{fig:DPL_gene_3}
\includegraphics[width=5cm,height=5cm]{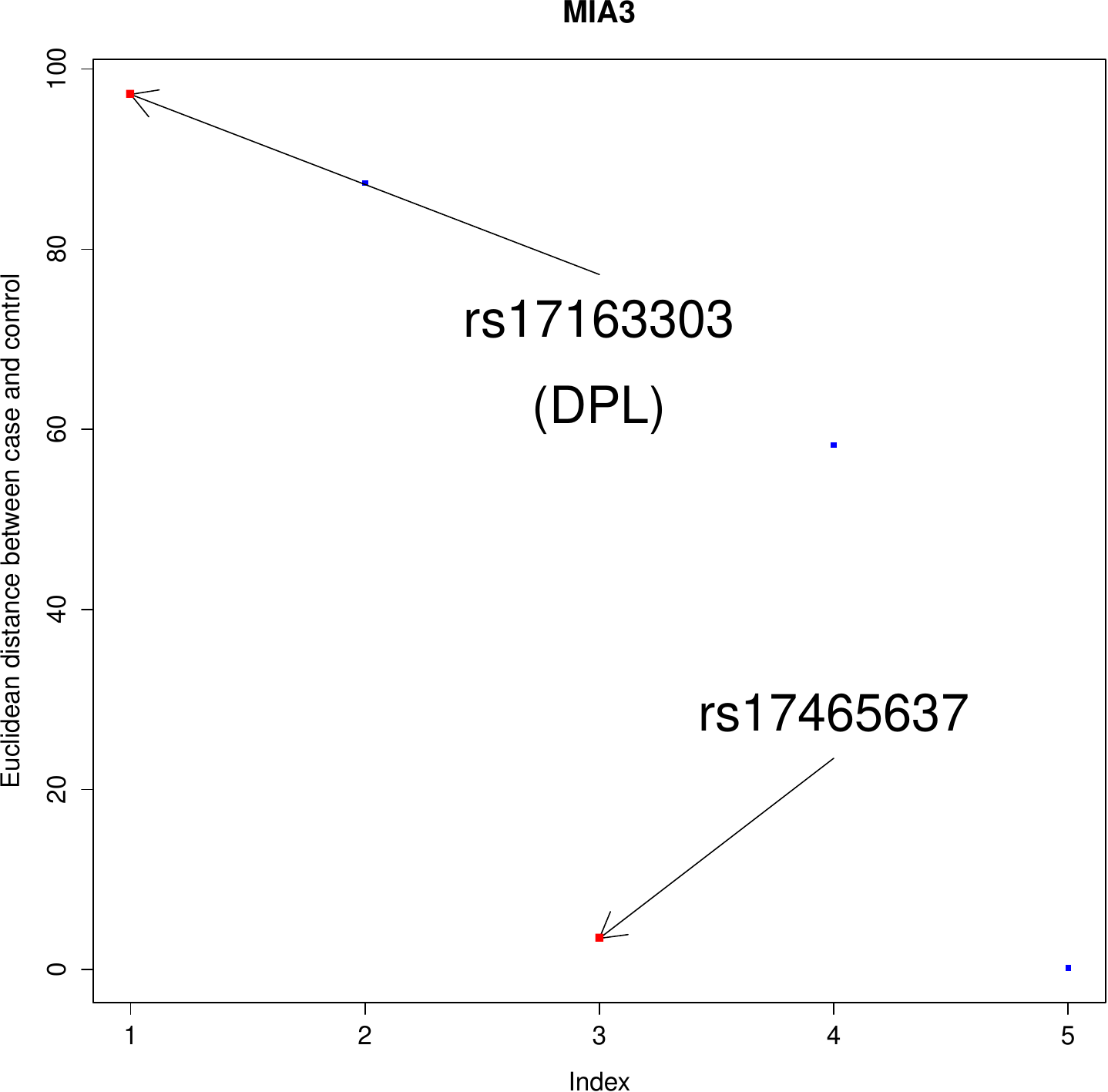}}
\vspace{2mm}
\subfigure[DPL of $SMARCA4$.]{ \label{fig:DPL_gene_4}
\includegraphics[width=5cm,height=5cm]{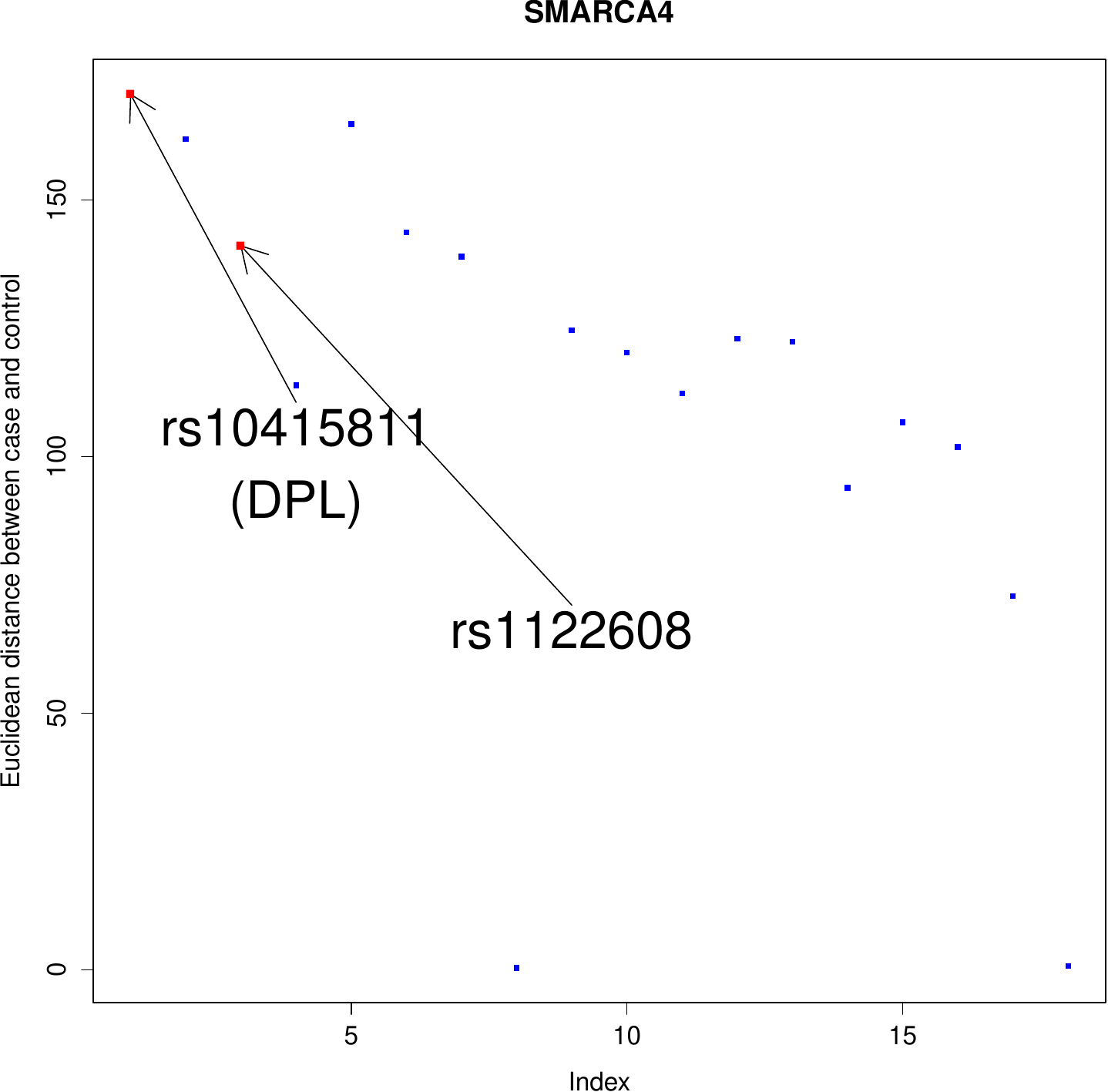}}\\
\vspace{2mm}
\subfigure[DPL of $PHACTR1$.]{ \label{fig:DPL_gene_5}
\includegraphics[width=5cm,height=5cm]{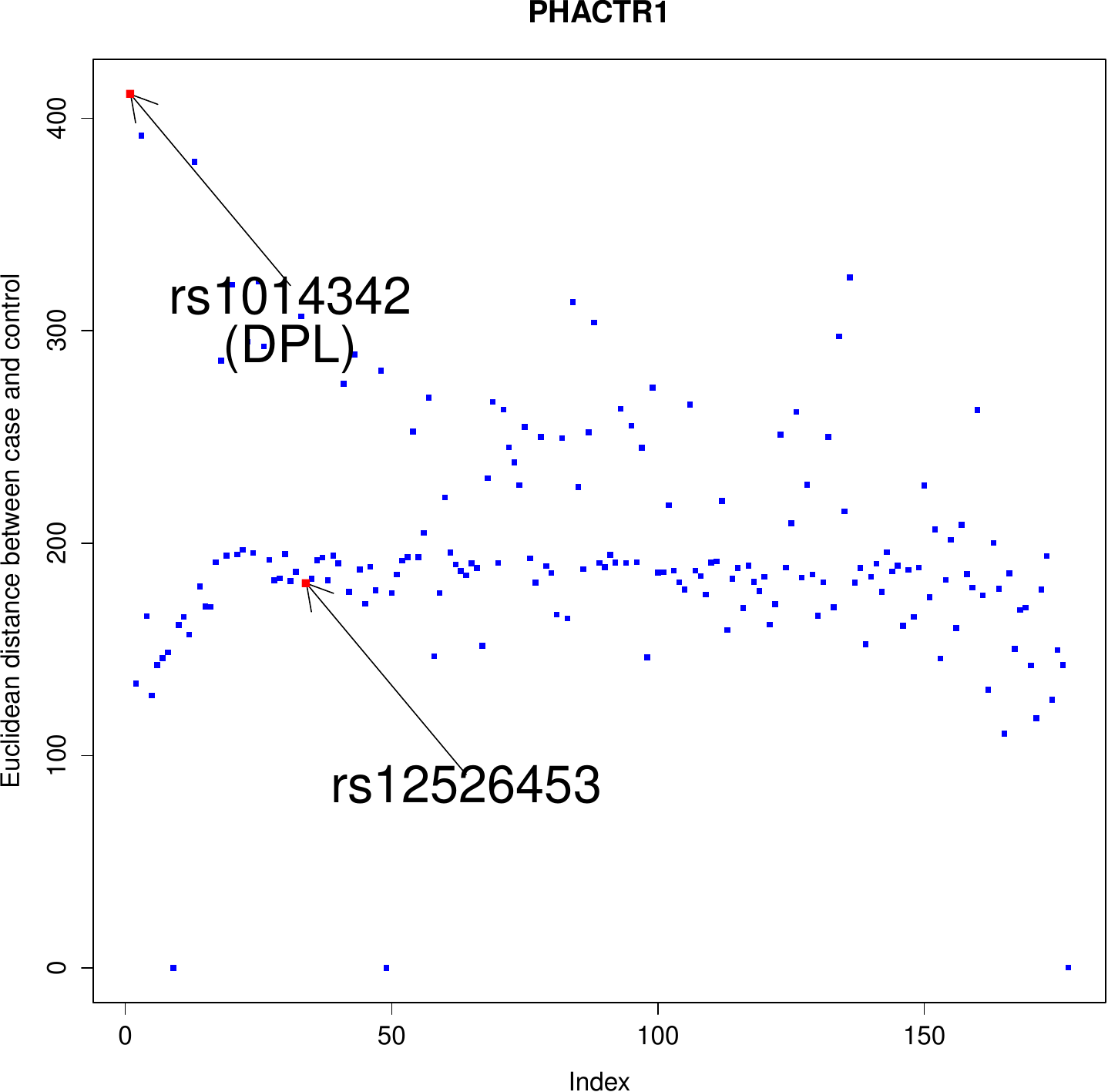}}
\hspace{2mm}
\subfigure[DPL of $C6orf106$.]{ \label{fig:DPL_gene_7}
\includegraphics[width=5cm,height=5cm]{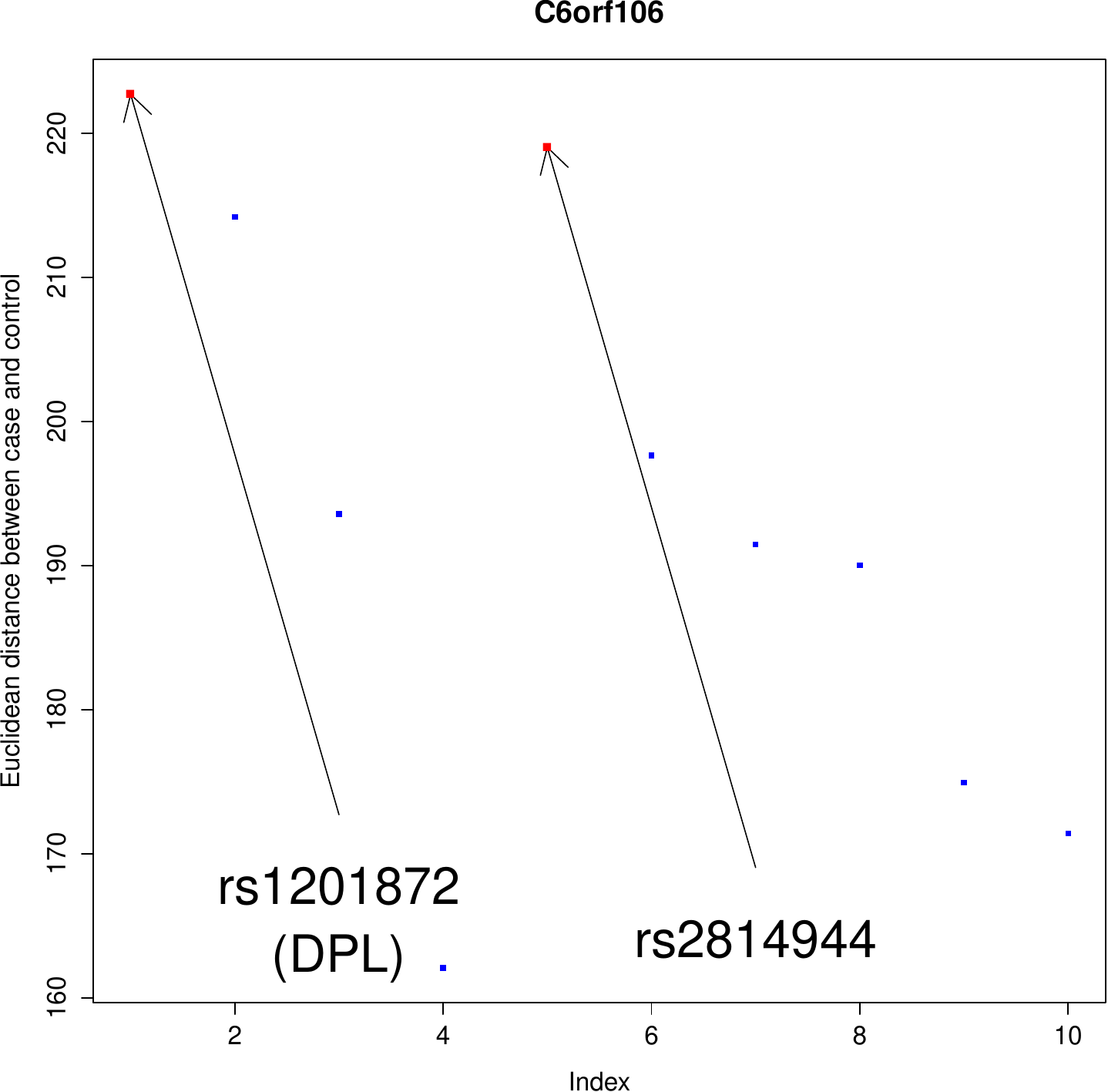}}
\caption{{\bf Disease predisposing loci of other influential genes:} 
Plots of the Euclidean distances 
against the indices of the loci of genes $WDR12$, $FTO$, 
$MIA3$, $SMARCA4$, $PHACTR1$, $C6orf106$. 
%The DPL that we obtained are close in terms of Euclidean distance to those loci which 
%are considered influential in the literature.
}
\label{fig:DPL_genes_other2}
\end{figure}

\begin{figure}%[htp]
\centering
\subfigure[DPL of $ZNF652$.]{ \label{fig:DPL_gene_8}
\includegraphics[width=5cm,height=5cm]{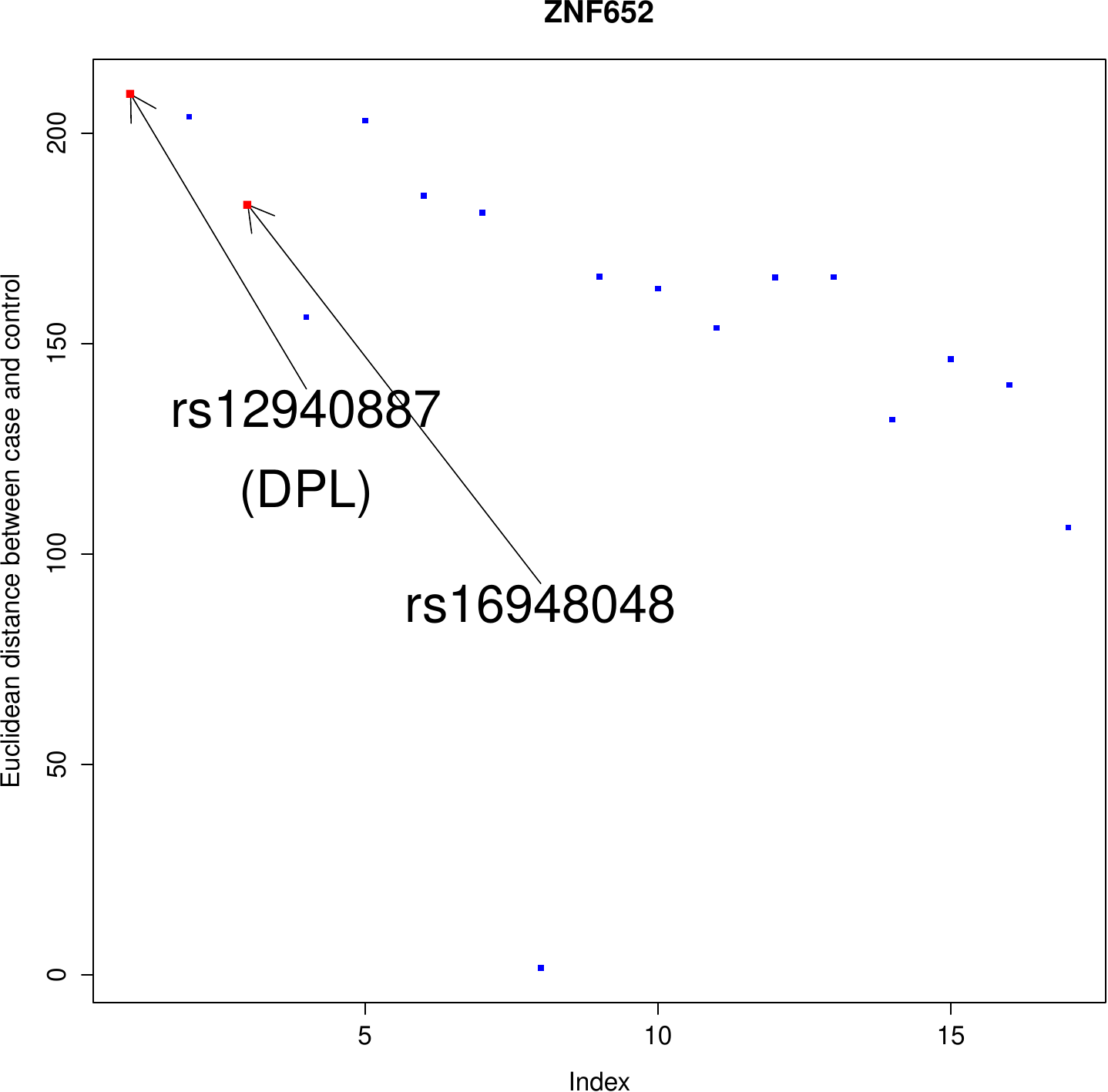}}
\hspace{2mm}
\subfigure[DPL of $RAB11B$.]{ \label{fig:DPL_gene_9}
\includegraphics[width=5cm,height=5cm]{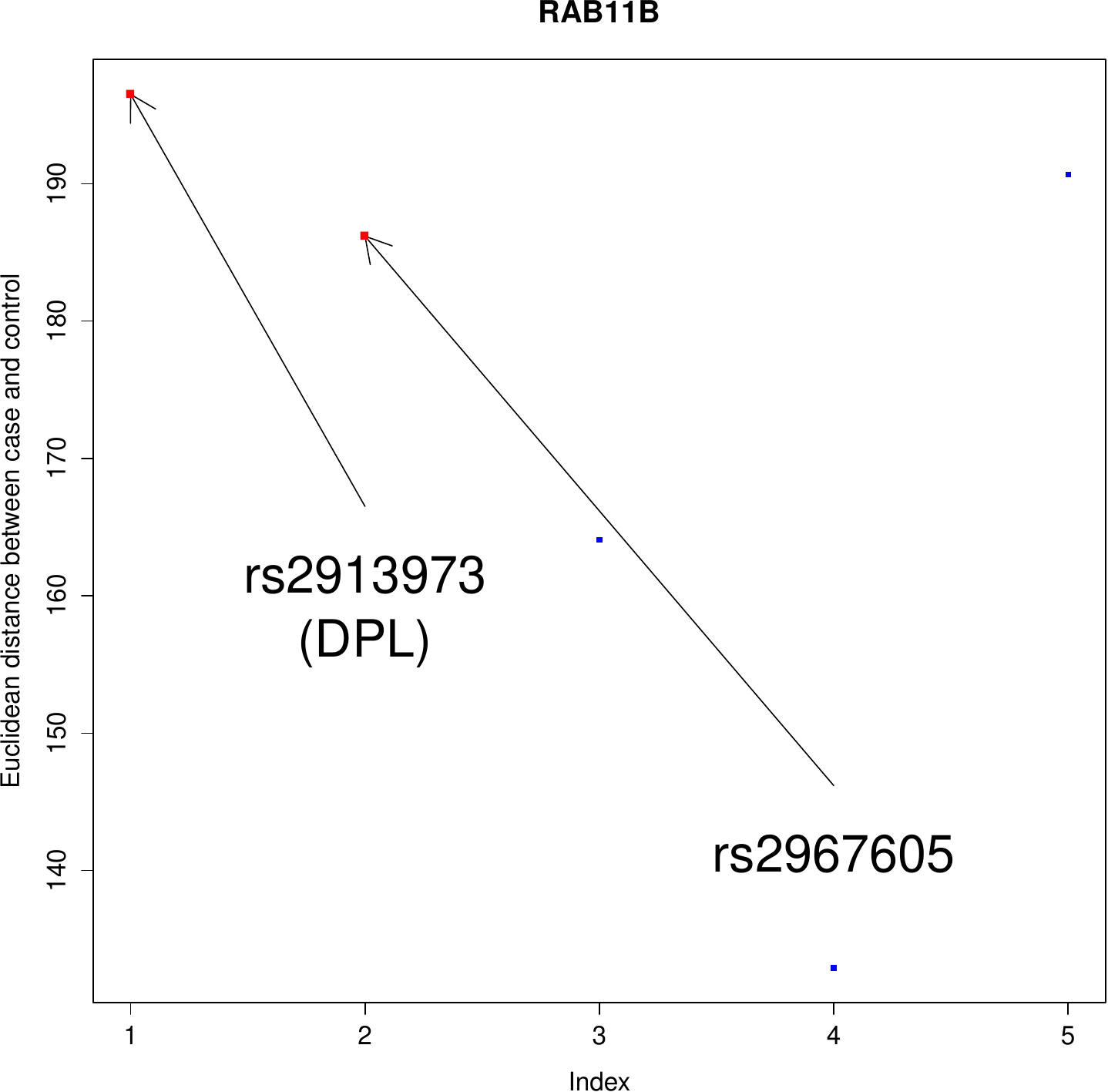}}\\
\vspace{2mm}
\subfigure[DPL of $RP11-136O12.2$.]{ \label{fig:DPL_gene_15}
\includegraphics[width=5cm,height=5cm]{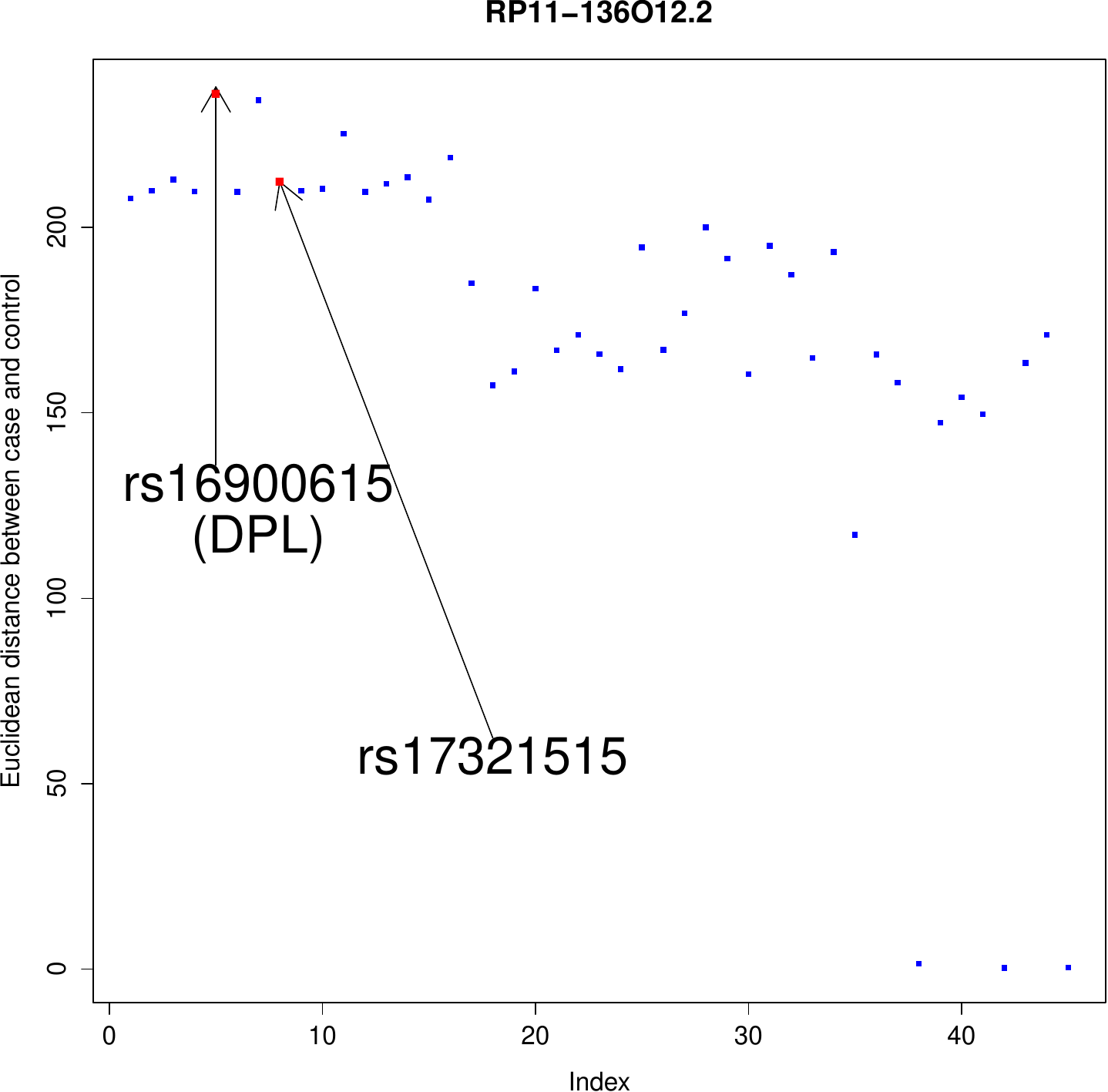}}
\hspace{2mm}
\subfigure[DPL of $ANKS1A$.]{ \label{fig:DPL_gene_16}
\includegraphics[width=5cm,height=5cm]{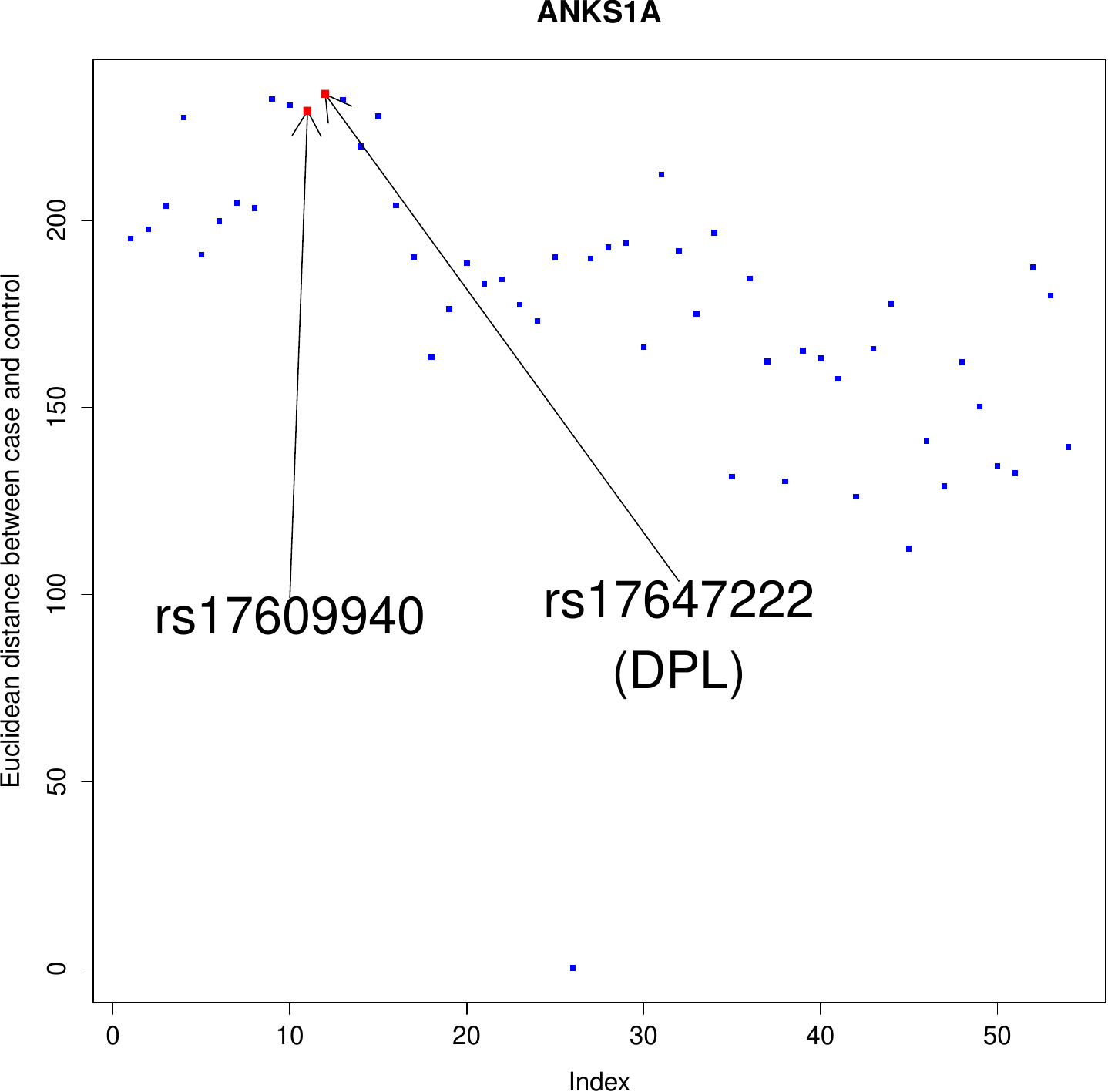}}\\
\vspace{2mm}
\subfigure[DPL of $CELSR2$.]{ \label{fig:DPL_gene_18}
\includegraphics[width=5cm,height=5cm]{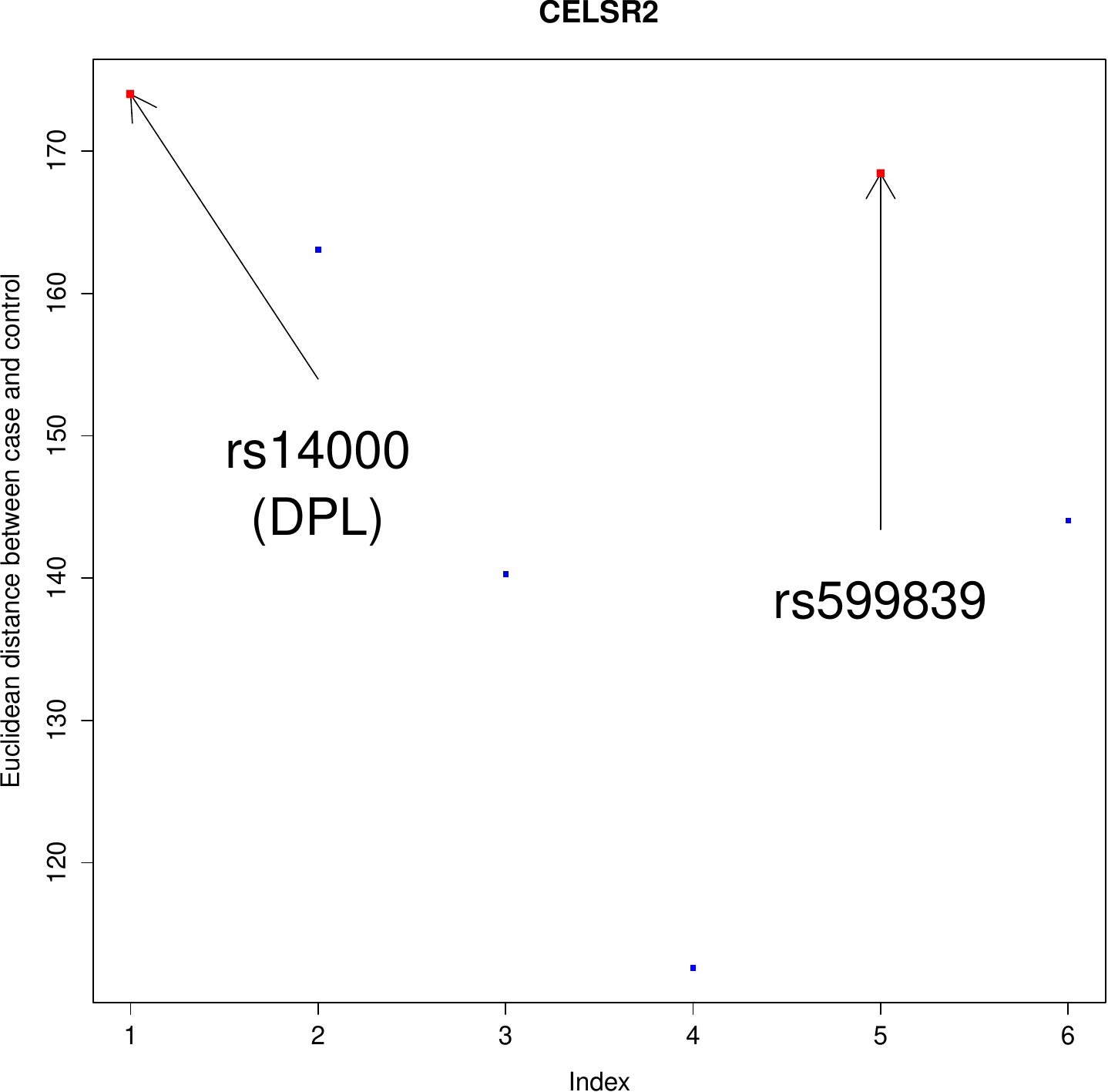}}
\hspace{2mm}
\subfigure[DPL of $GPAM$.]{ \label{fig:DPL_gene_21}
\includegraphics[width=5cm,height=5cm]{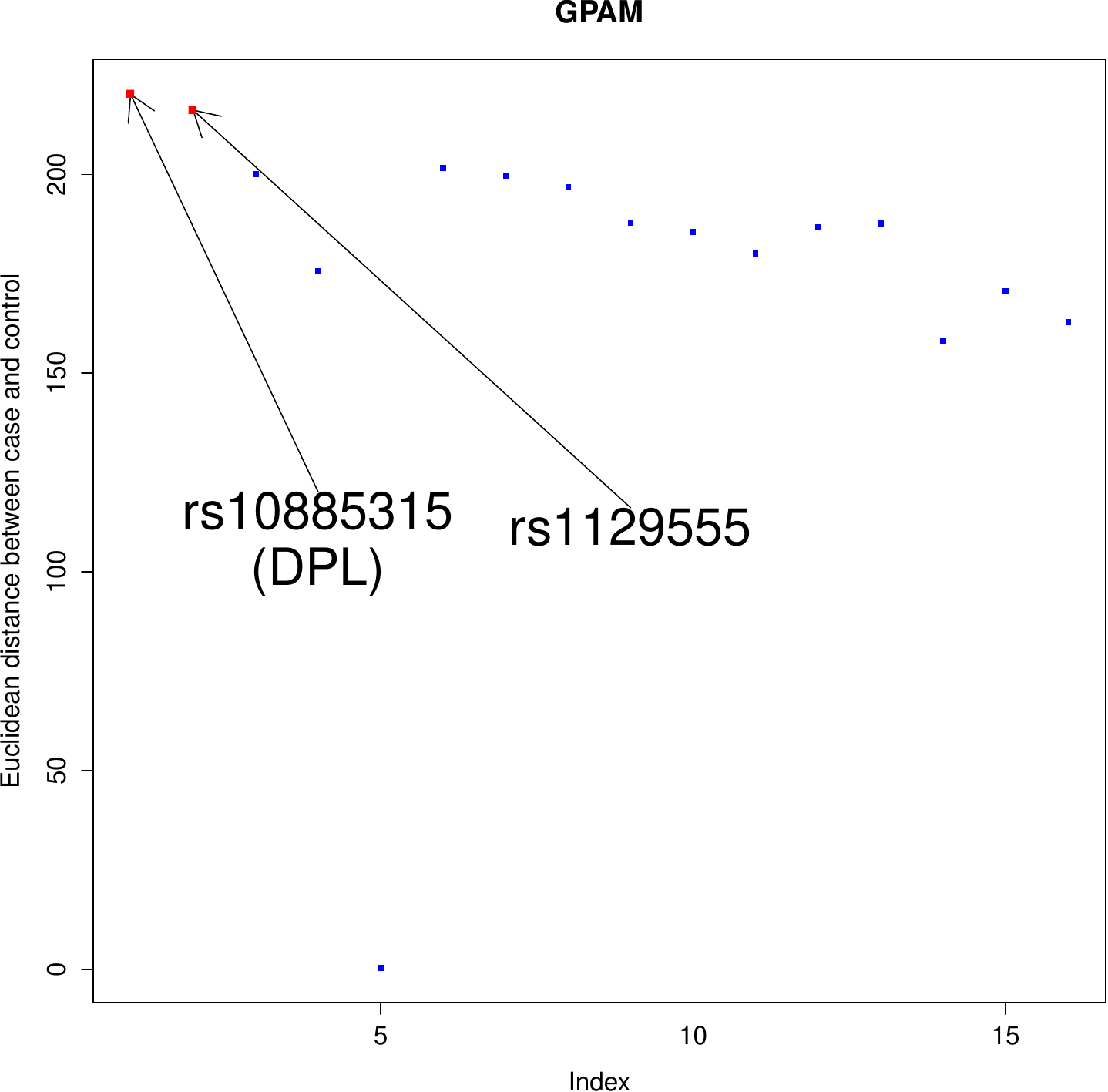}}\\
\caption{{\bf Disease predisposing loci of other influential genes:} 
Plots of the Euclidean distances 
against the indices of the loci of genes $ZNF652$, $RAB11B$, $RP11-136O12.2$, $ANKS1A$, $CELSR2$, $GPAM$.
The figures exhibit adequate agreement of our obtained DPLs and the loci believed to be influential.
}
\label{fig:DPL_others}
\end{figure}

\begin{figure}%[htp]
\centering
\subfigure[DPL of $SLC22A1$.]{ \label{fig:DPL_gene_22}
\includegraphics[width=5cm,height=5cm]{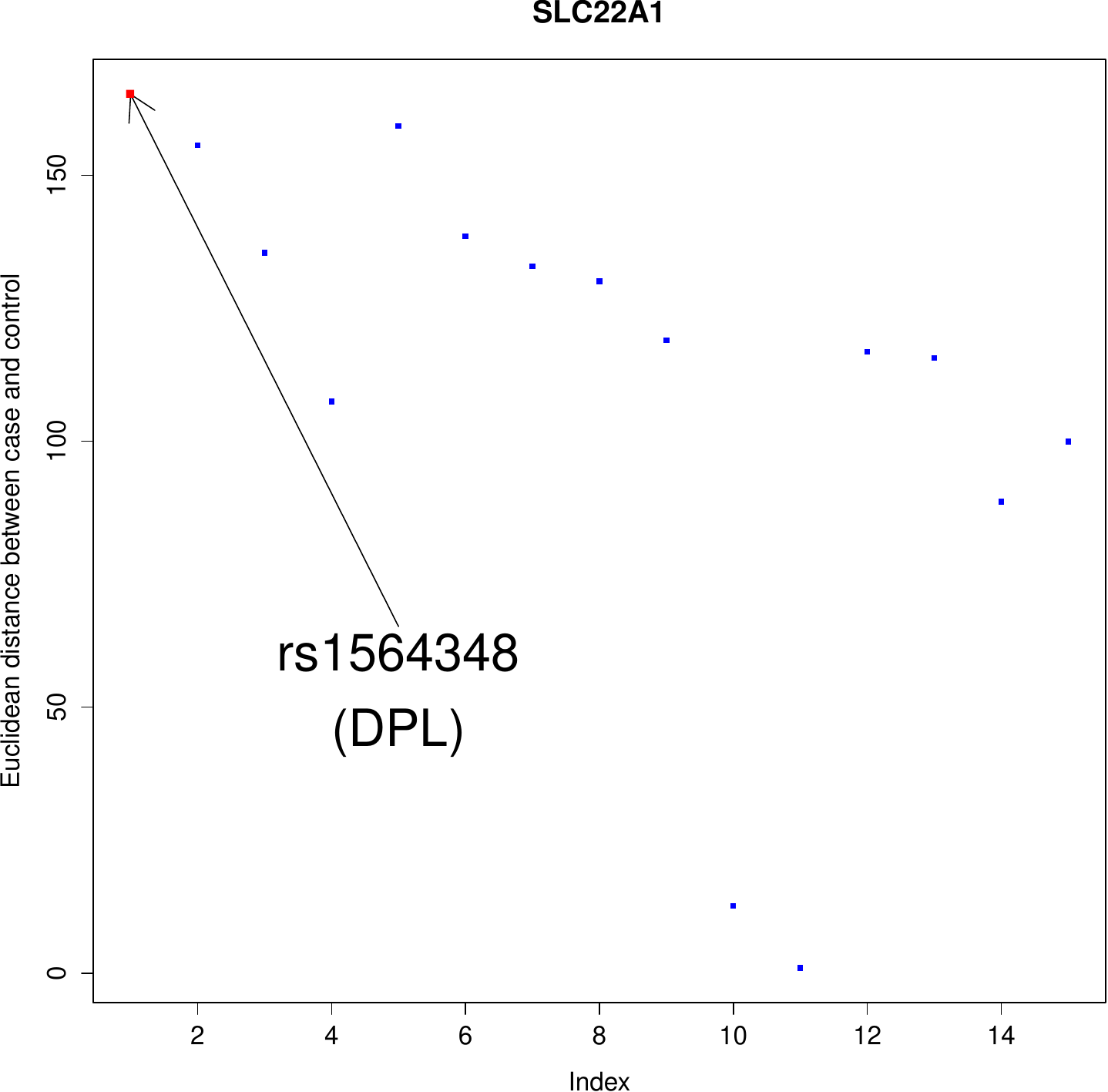}}
\hspace{2mm}
\subfigure[DPL of $BDNF-AS$.]{ \label{fig:DPL_gene_23}
\includegraphics[width=5cm,height=5cm]{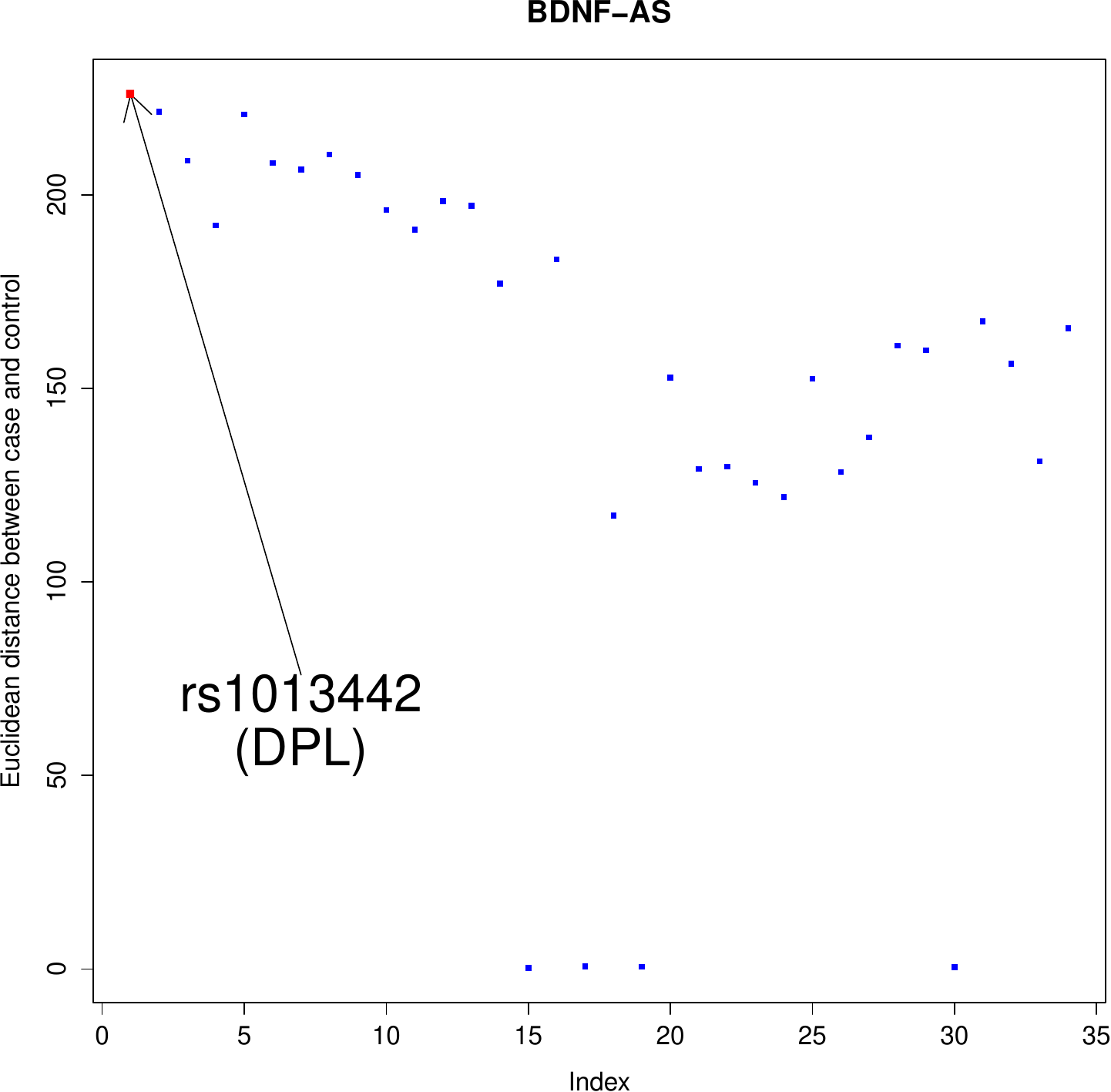}}\\
\vspace{2mm}
\subfigure[DPL of $CDKAL1$.]{ \label{fig:DPL_gene_24}
\includegraphics[width=5cm,height=5cm]{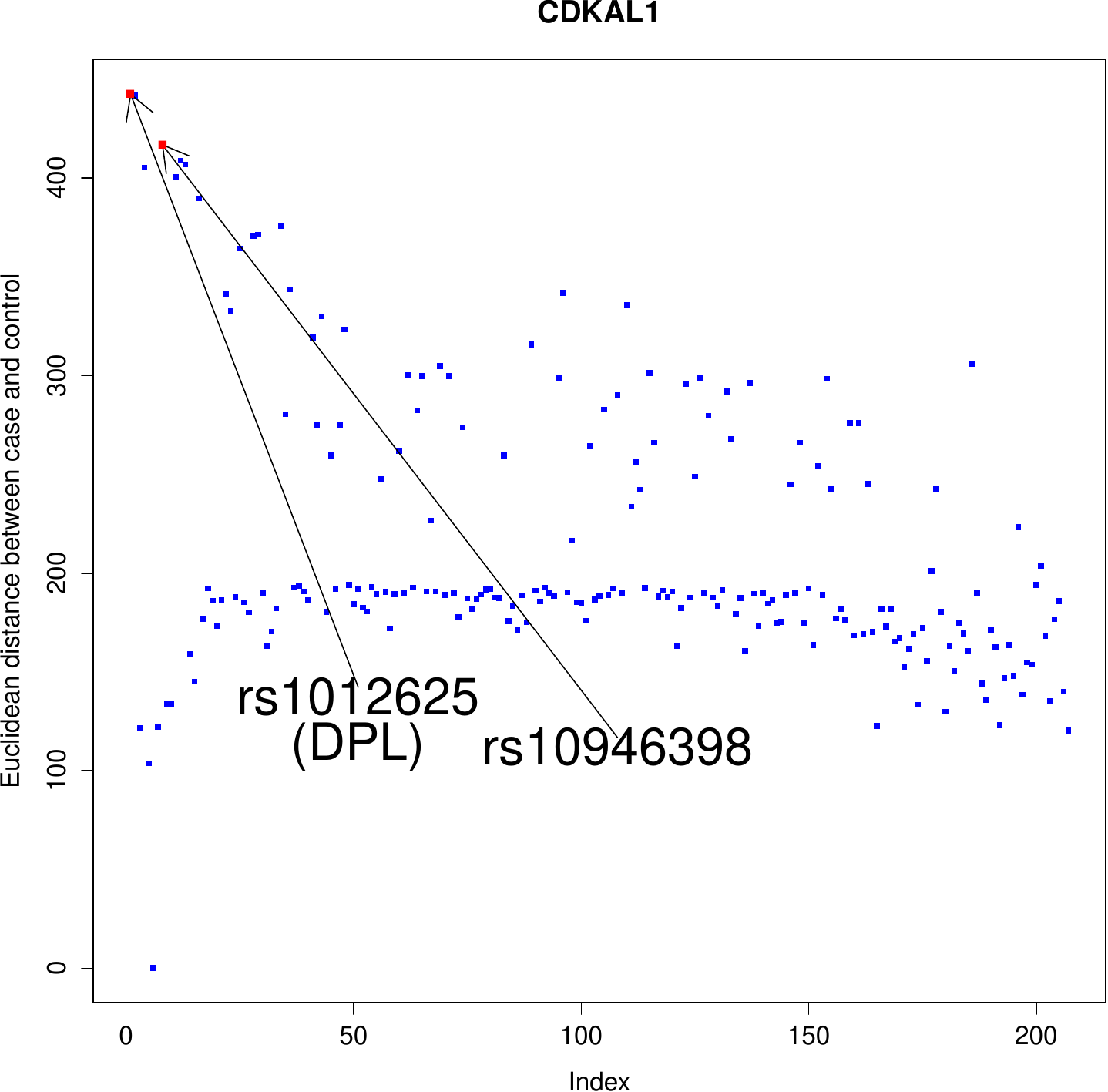}}\\
\caption{{\bf Disease predisposing loci for genes $SLC22A1$, $BDNF-AS$ and $CDKAL1$:} 
Our DPLs remarkably agree with the existing influential SNPs. In fact, for $SLC22A1$ and $BDNF-AS$, 
our DPLs coincide with the existing influential SNPs.}
\label{fig:DPL_others2}
\end{figure}

\begin{figure}%[htp]
\centering
\subfigure[Presence/absence of gene-gene interaction.]{ \label{fig:ggi_indicator_plot}
\includegraphics[width=16cm,height=16cm]{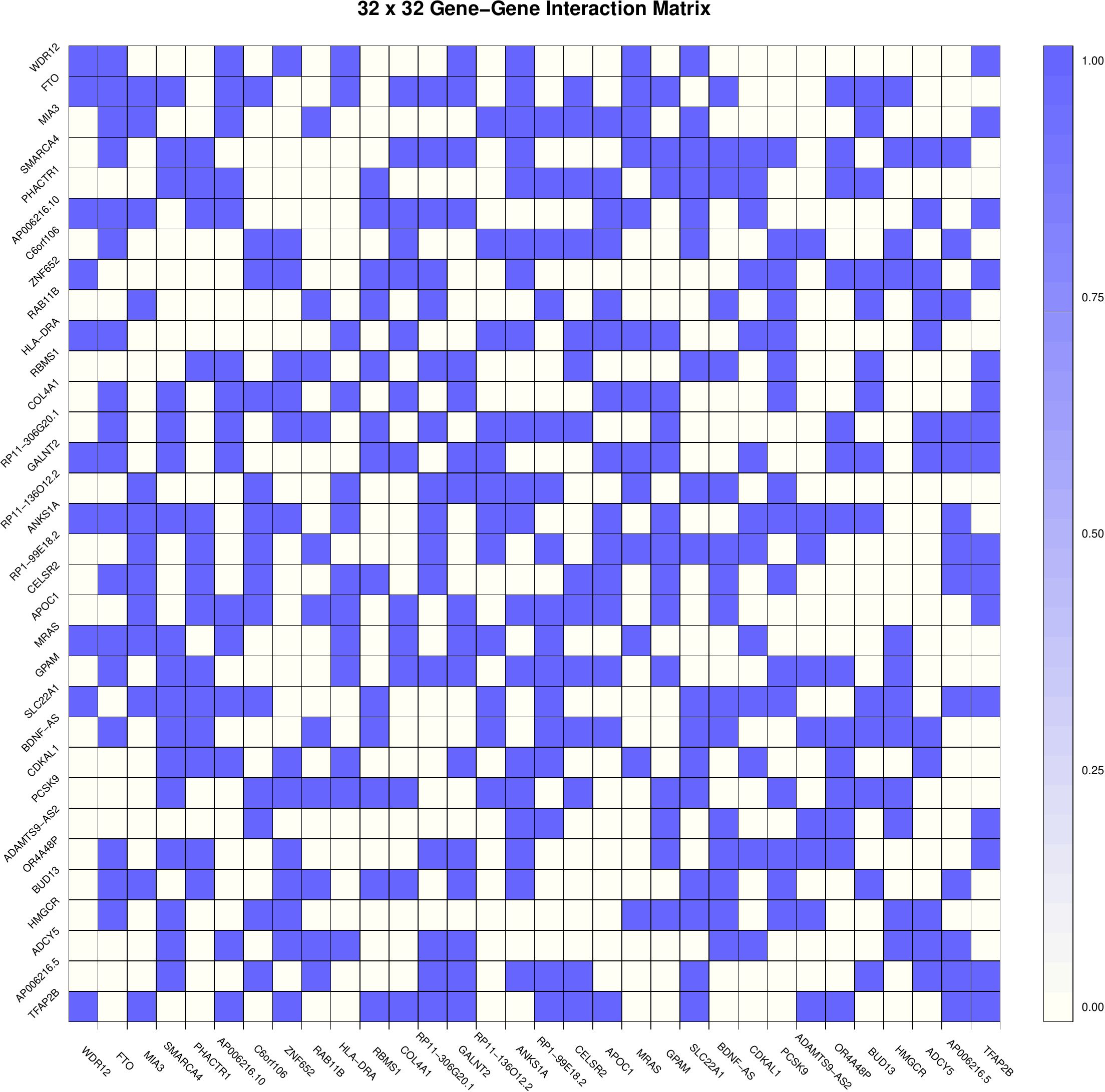}}
\caption{{\bf Presence/absence of gene-gene interactions}: Blue denotes presence and white
represents absence of gene-gene interaction.}
\label{fig:ggi_plots1}
\end{figure}

\begin{figure}%[htp]
\centering
%\subfigure[Posterior of $\tau_{1,0}$.]{ \label{fig:gene_control_1}
%\includegraphics[width=5cm,height=5cm]{plots_realdata/plot_comp/gene_control_eu_1-crop.pdf}}
%\hspace{2mm}
%\subfigure[Posterior of $\tau_{1,1}$.]{ \label{fig:gene_case_1}
%\includegraphics[width=5cm,height=5cm]{plots_realdata/plot_comp/gene_case_eu_1-crop.pdf}}\\
%\vspace{2mm}
\subfigure[Posterior of $\tau_{2,0}$.]{ \label{fig:gene_control_2}
\includegraphics[width=6cm,height=5cm]{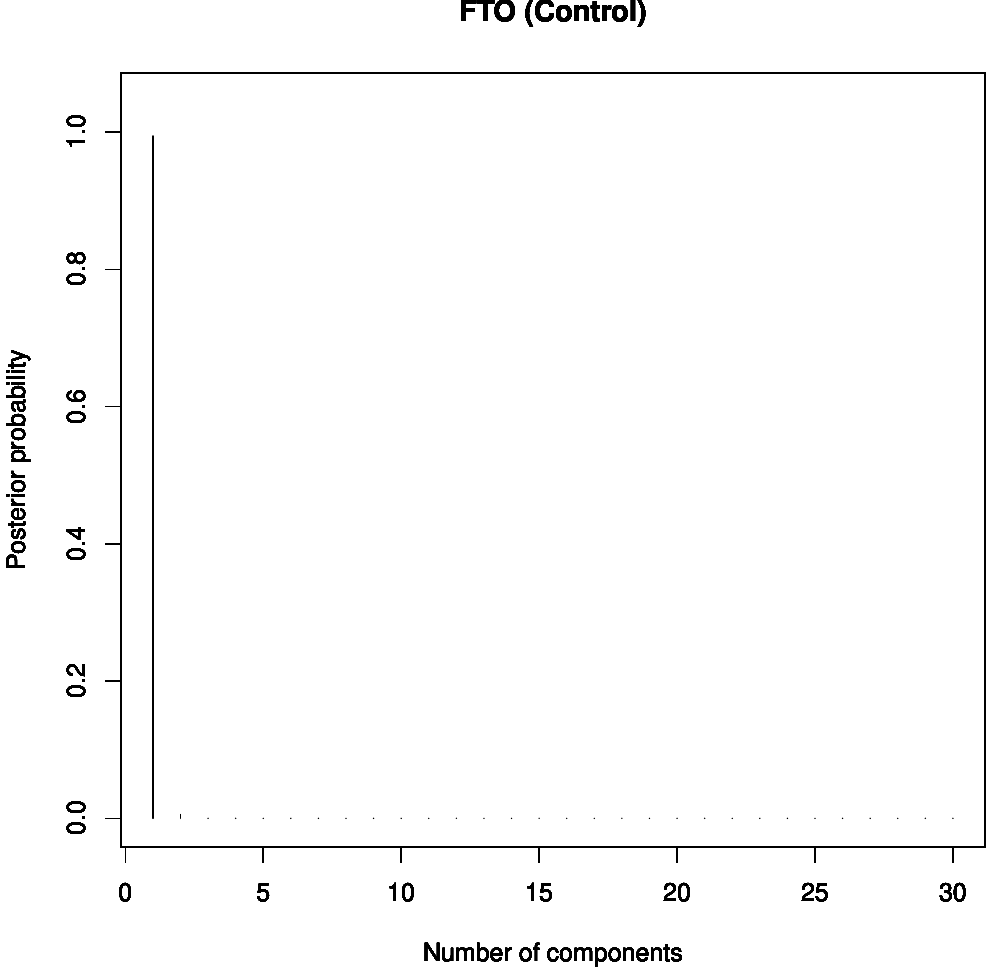}}
\hspace{2mm}
\subfigure[Posterior of $\tau_{2,1}$.]{ \label{fig:gene_case_2}
\includegraphics[width=6cm,height=5cm]{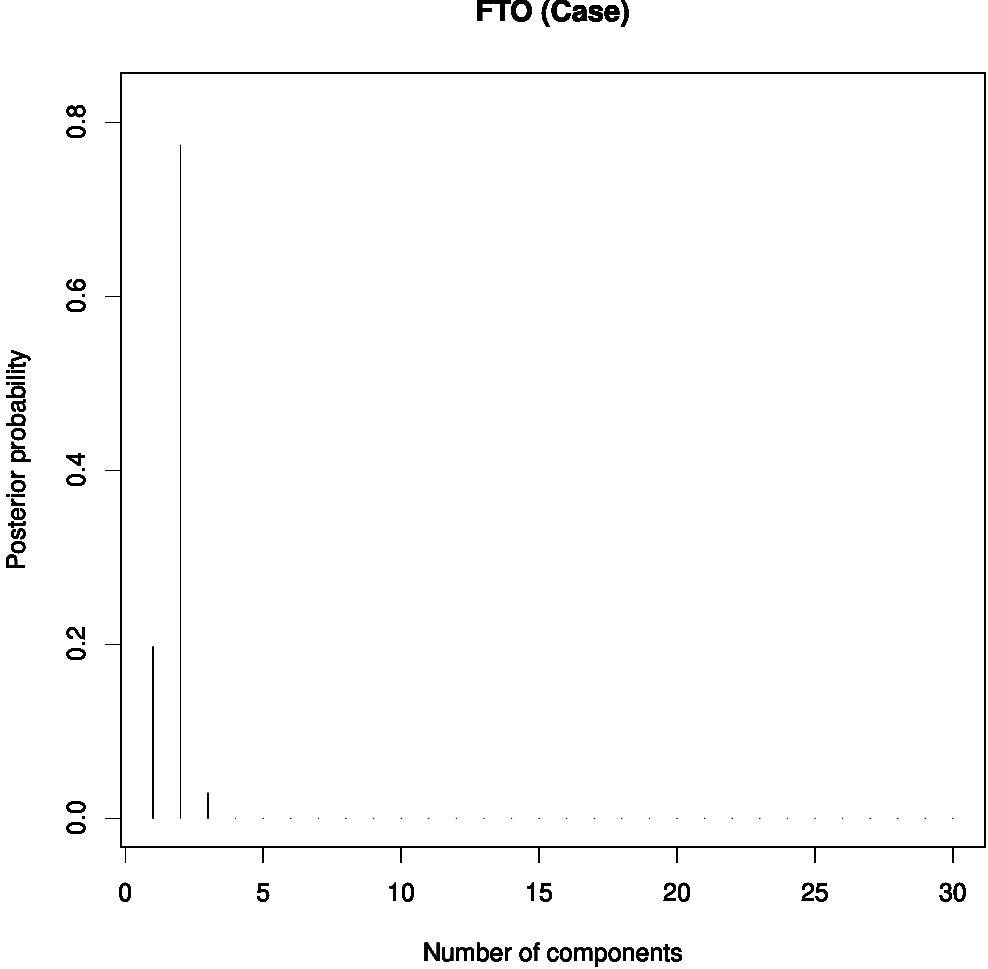}}\\
\vspace{2mm}
\subfigure[Posterior of $\tau_{4,0}$.]{ \label{fig:gene_control_4}
\includegraphics[width=6cm,height=5cm]{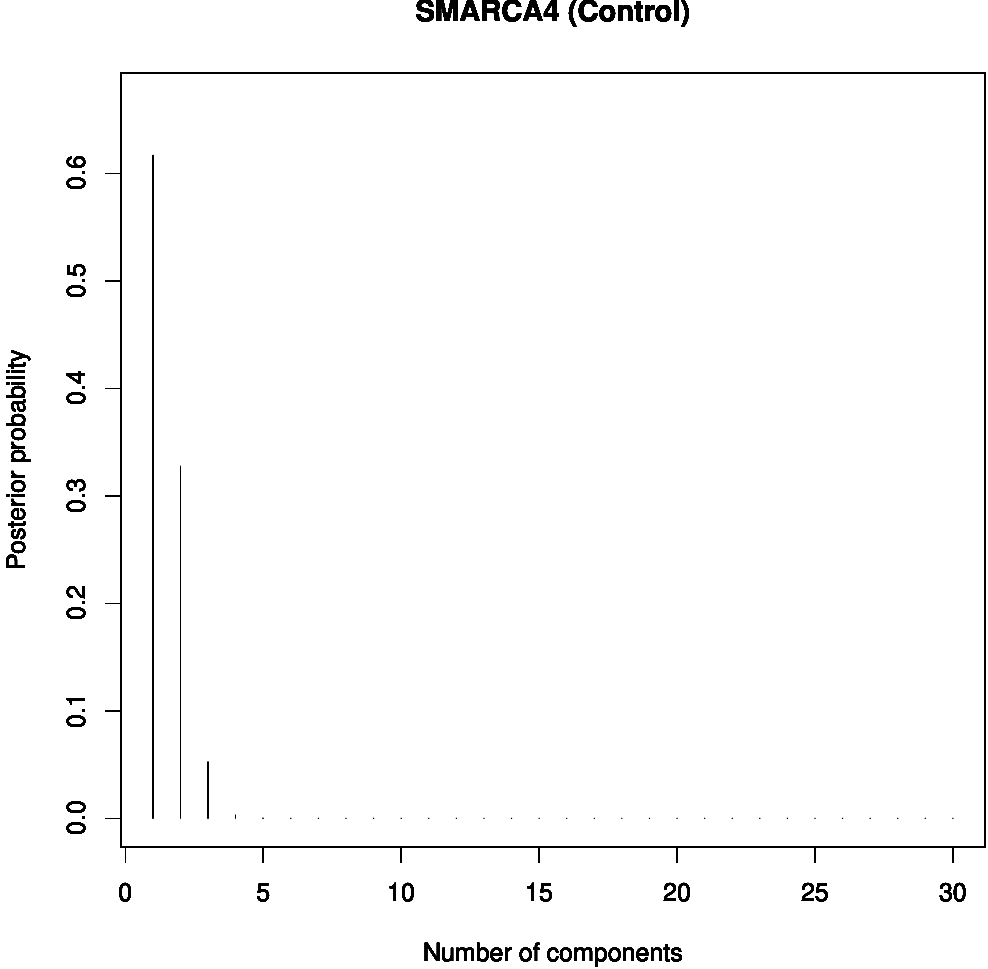}}
\hspace{2mm}
\subfigure[Posterior of $\tau_{4,1}$.]{ \label{fig:gene_case_4}
\includegraphics[width=6cm,height=5cm]{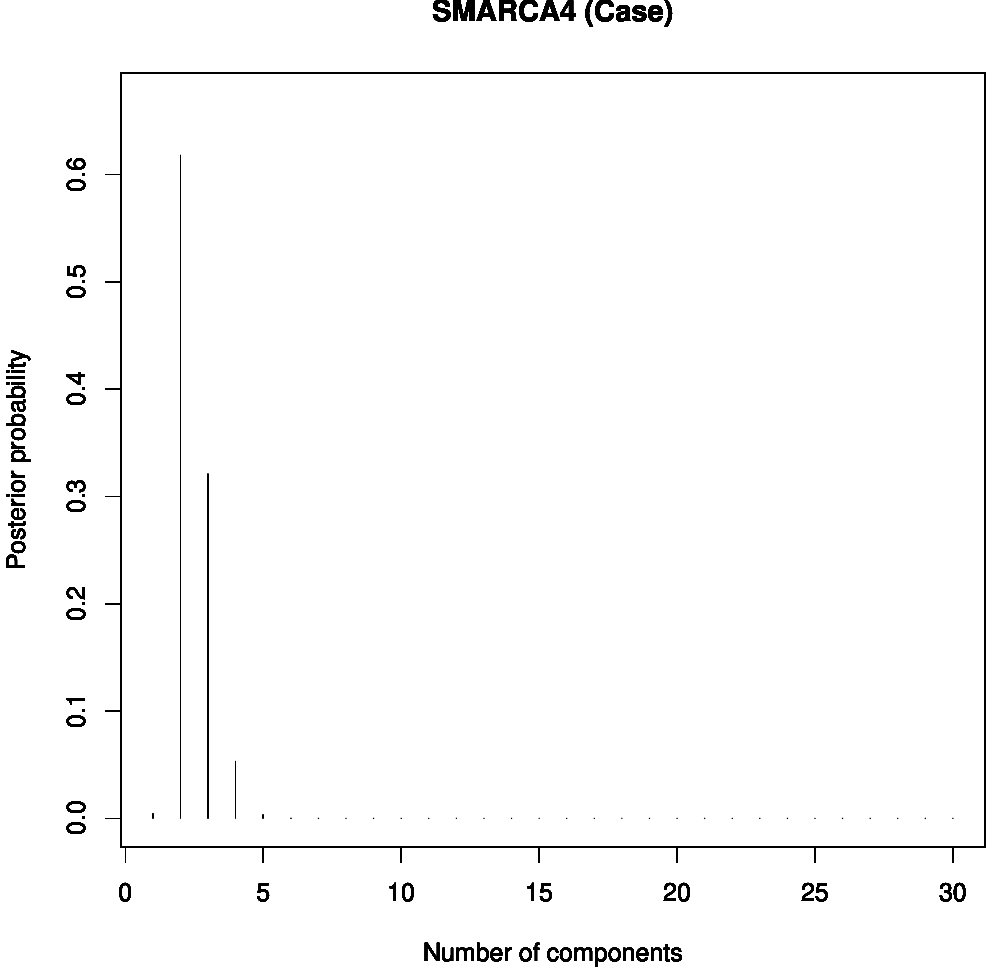}}
\vspace{2mm}
\subfigure[Posterior of $\tau_{8,0}$.]{ \label{fig:gene_control_8}
\includegraphics[width=6cm,height=5cm]{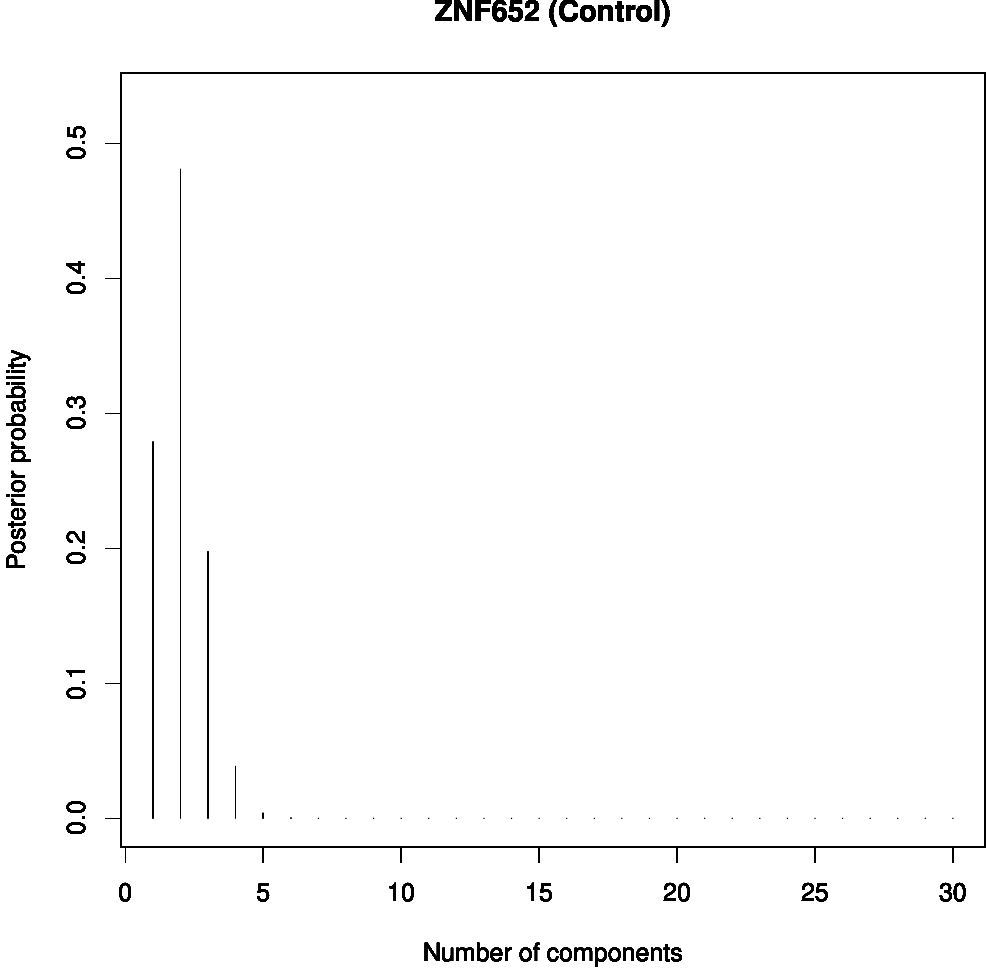}}
\hspace{2mm}
\subfigure[Posterior of $\tau_{8,1}$.]{ \label{fig:gene_case_8}
\includegraphics[width=6cm,height=5cm]{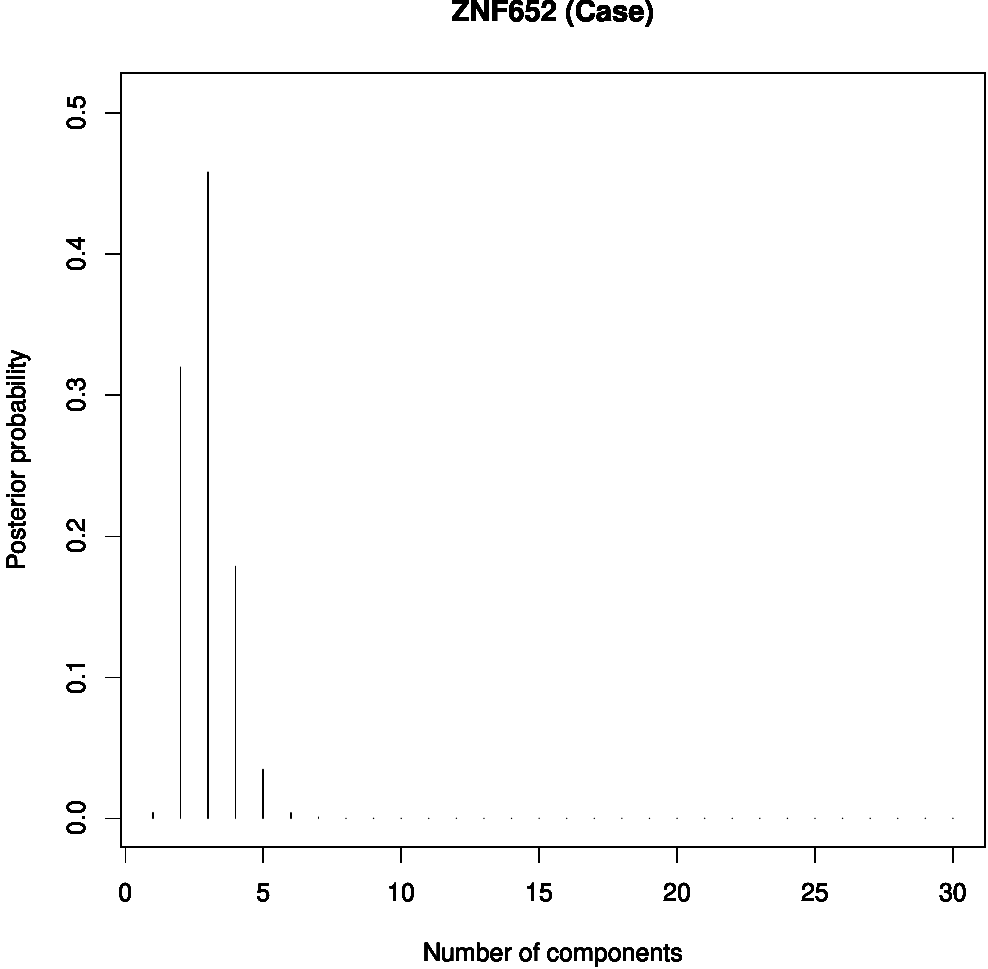}}
\caption{{\bf Posterior of number of components:} Posterior distributions of the number of distinct components $\tau_{j,k}$
for each pair ($j,k$); j=2,4,8; $k=0,1$. %$j=1,2,4,8$; $k=0,1$. 
The left and right panels show the posteriors associated with cases
and controls, respectively.}
\label{fig:ggi_comp_realdata1}
\end{figure}

\begin{figure}%[htp]
\centering
%\subfigure[Posterior of $\tau_{10,0}$.]{ \label{fig:gene_control_10}
%\includegraphics[width=6cm,height=5cm]{plots_realdata/plot_comp/gene_control_eu_10-crop.pdf}}
%\hspace{2mm}
%\subfigure[Posterior of $\tau_{10,1}$.]{ \label{fig:gene_case_10}
%\includegraphics[width=6cm,height=5cm]{plots_realdata/plot_comp/gene_case_eu_10-crop.pdf}}\\
%\vspace{2mm}
\subfigure[Posterior of $\tau_{13,0}$.]{ \label{fig:gene_control_13}
\includegraphics[width=6cm,height=5cm]{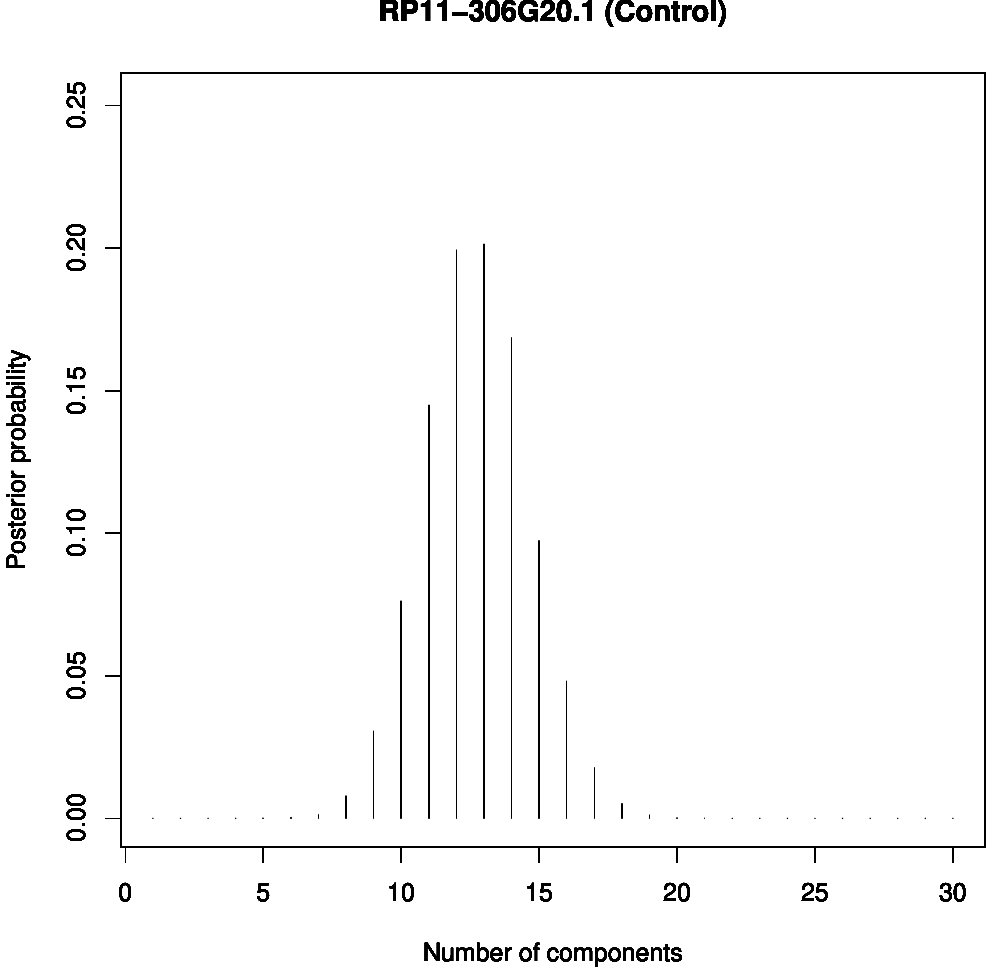}}
\hspace{2mm}
\subfigure[Posterior of $\tau_{13,1}$.]{ \label{fig:gene_case_13}
\includegraphics[width=6cm,height=5cm]{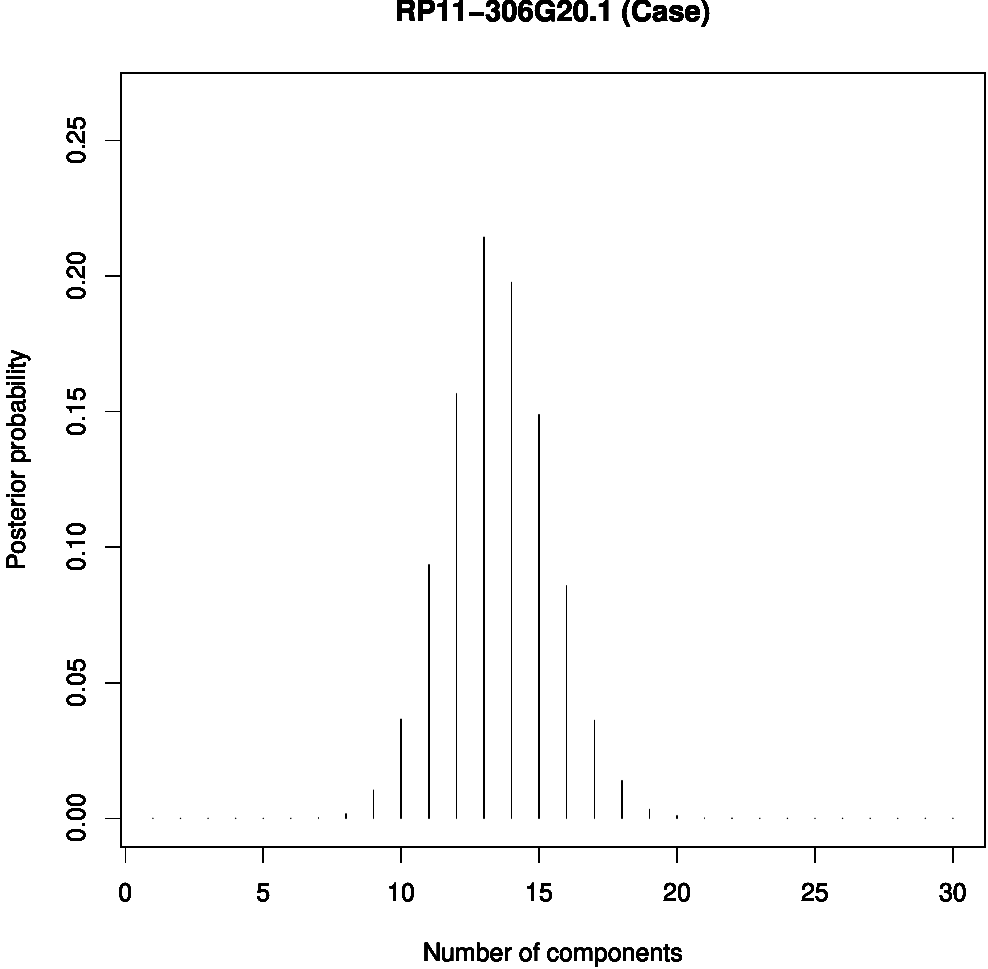}}\\
\vspace{2mm}
\subfigure[Posterior of $\tau_{25,0}$.]{ \label{fig:gene_control_25}
\includegraphics[width=6cm,height=5cm]{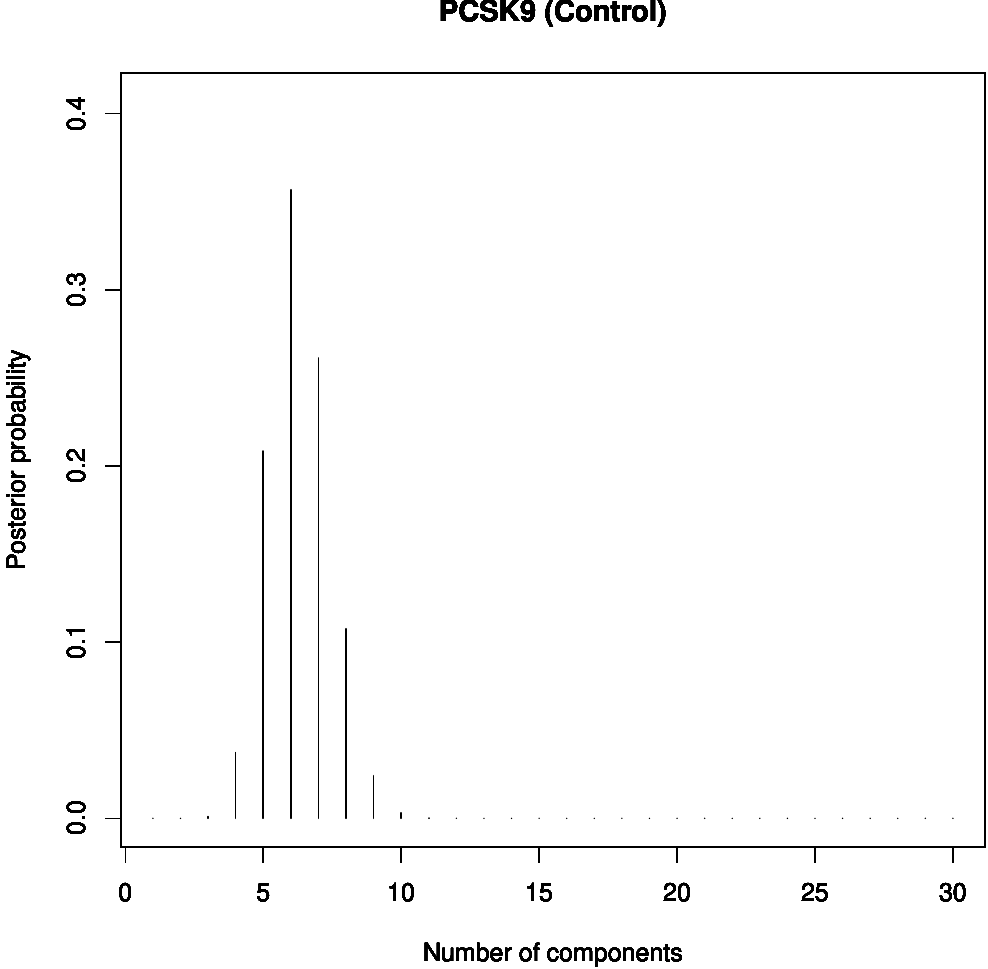}}
\hspace{2mm}
\subfigure[Posterior of $\tau_{25,1}$.]{ \label{fig:gene_case_25}
\includegraphics[width=6cm,height=5cm]{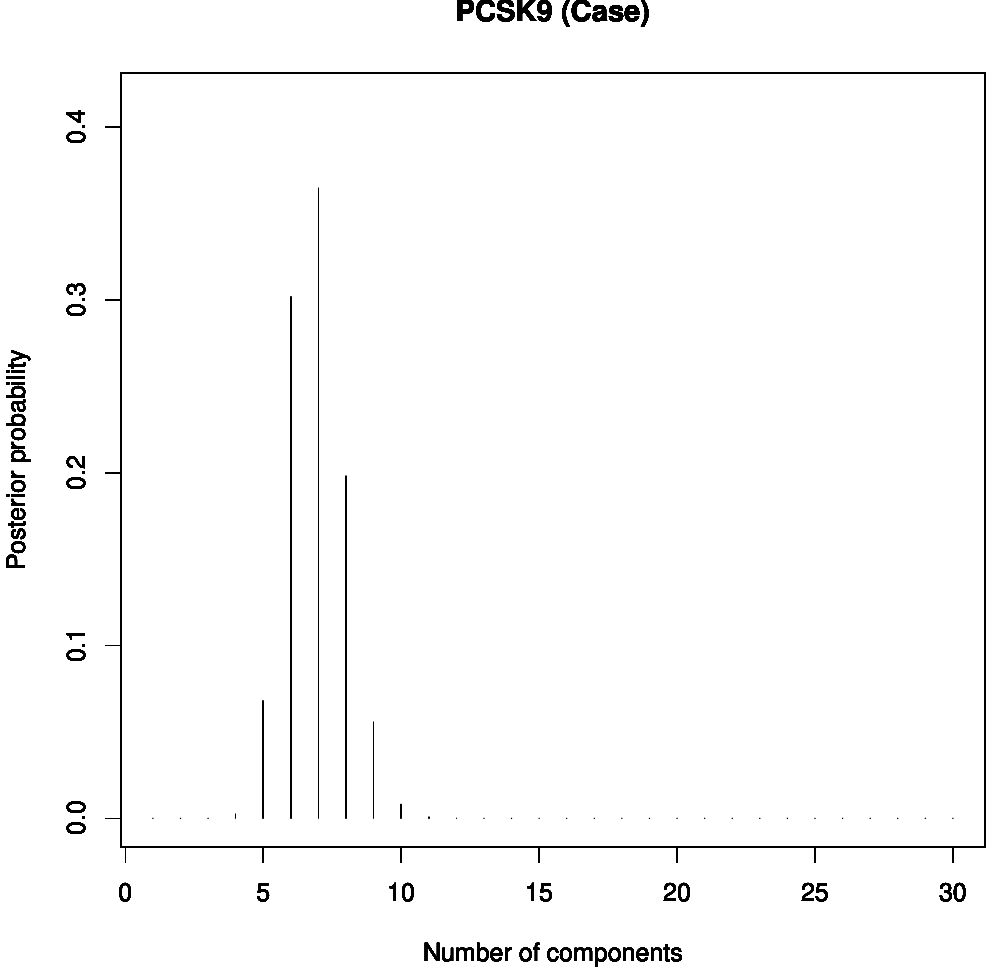}}
\vspace{2mm}
\subfigure[Posterior of $\tau_{30,0}$.]{ \label{fig:gene_control_30}
\includegraphics[width=6cm,height=5cm]{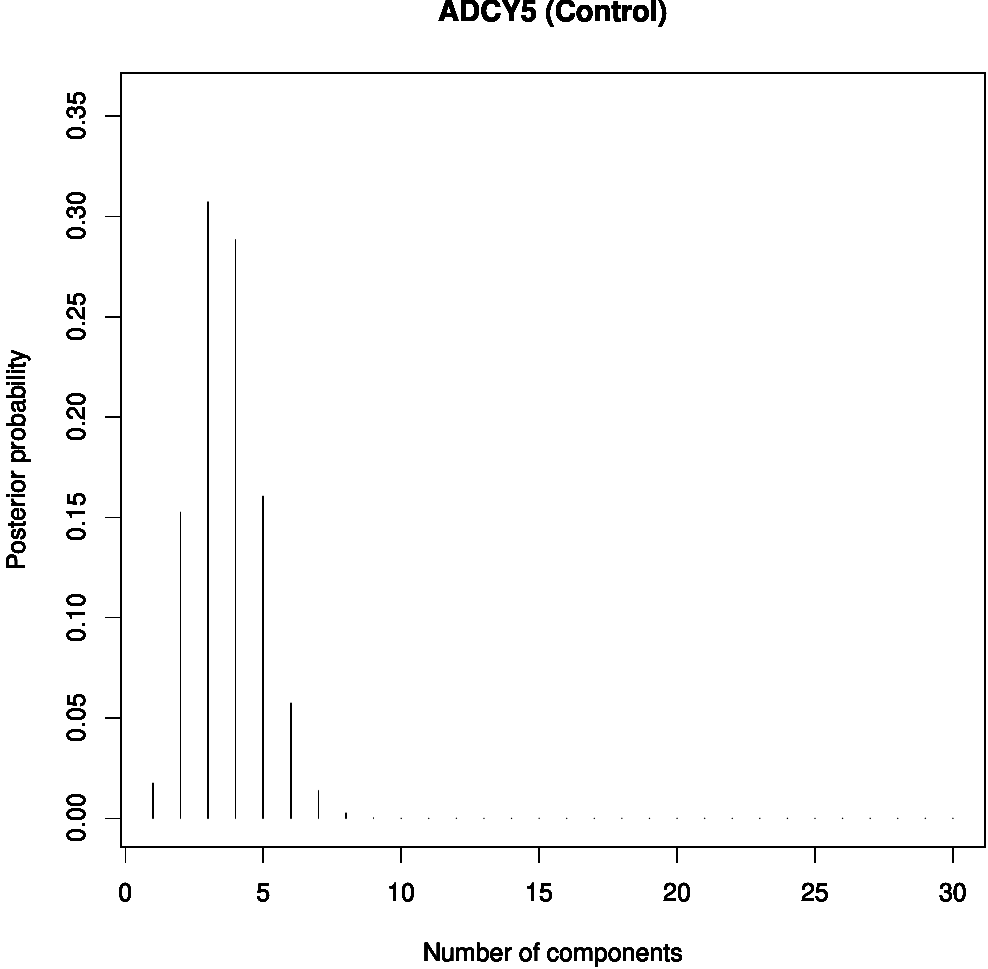}}
\hspace{2mm}
\subfigure[Posterior of $\tau_{30,1}$.]{ \label{fig:gene_case_30}
\includegraphics[width=6cm,height=5cm]{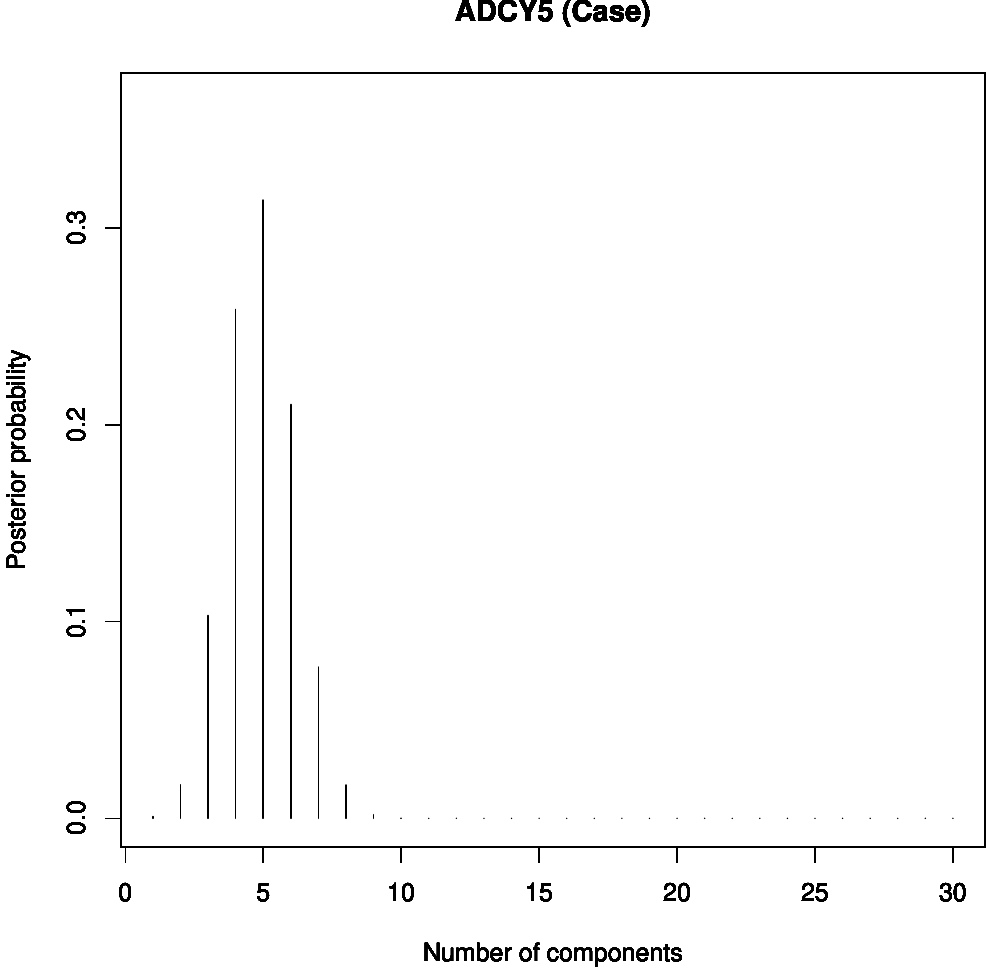}}
\caption{{\bf Posterior of number of components:} Posterior distributions of the number of distinct components $\tau_{j,k}$
for each pair ($j,k$); $j=13,25,30$; $k=0,1$. %$j=10,13,25,30$; $k=0,1$. 
The left and right panels show the posteriors associated with cases
and controls, respectively.}
\label{fig:ggi_comp_realdata2}
\end{figure}

\begin{figure}%[htp]
\centering
\subfigure[Posterior of $\tau_{6,0}$.]{ \label{fig:gene_control_eu_6}
\includegraphics[width=6cm,height=5cm]{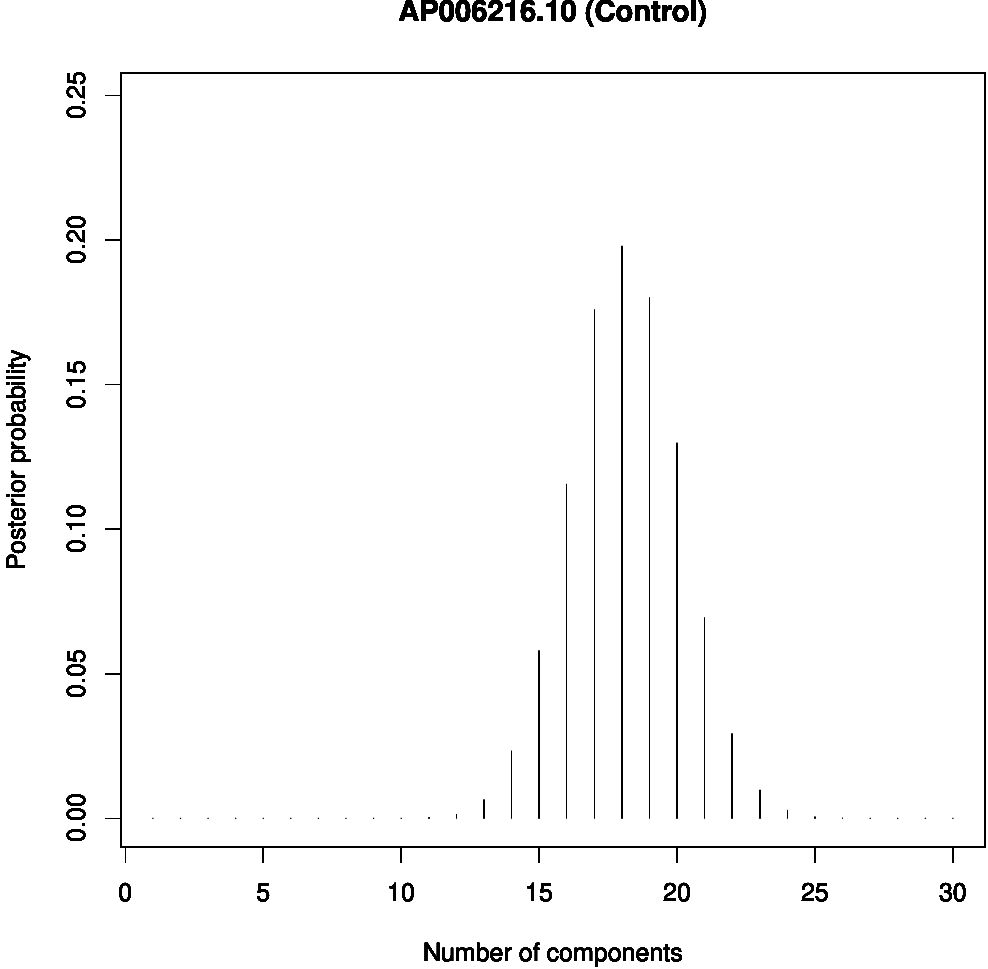}}
\hspace{2mm}
\subfigure[Posterior of $\tau_{6,1}$.]{ \label{fig:gene_case_eu_6}
\includegraphics[width=6cm,height=5cm]{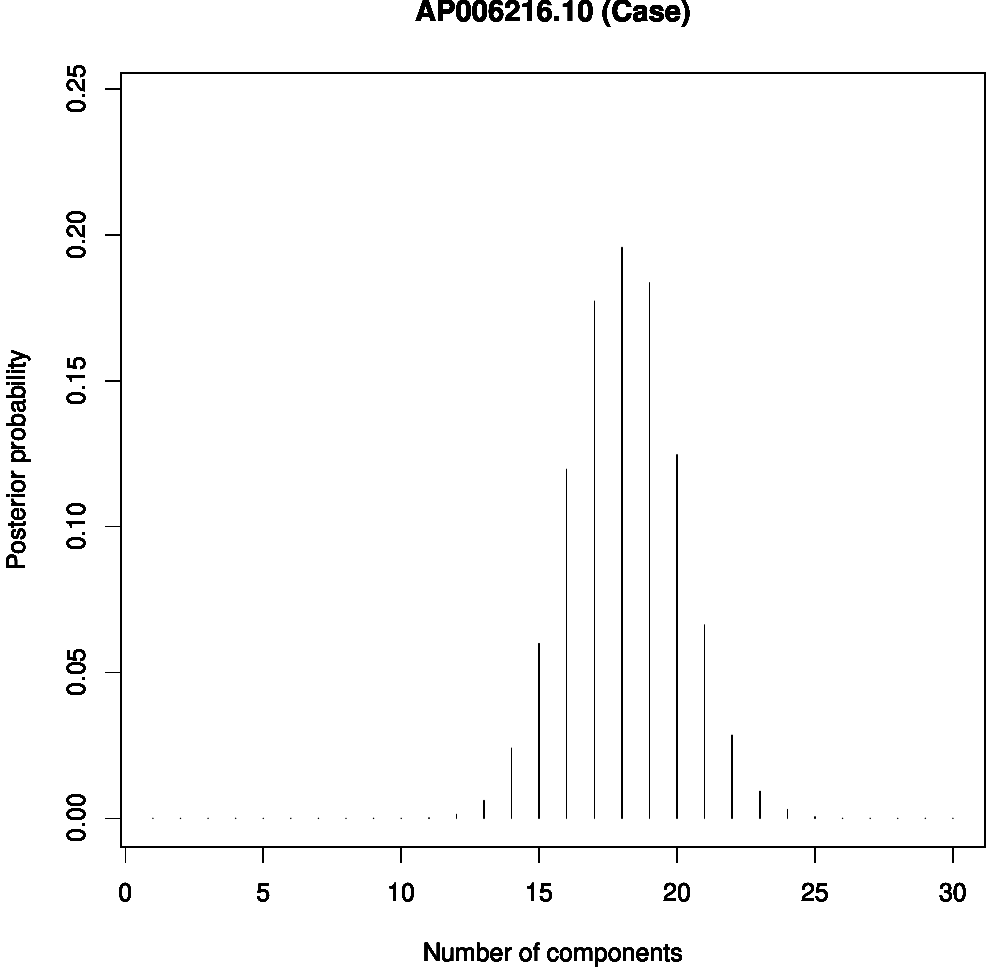}}\\
\vspace{2mm}
\subfigure[Posterior of $\tau_{19,0}$.]{ \label{fig:gene_control_eu_19}
\includegraphics[width=6cm,height=5cm]{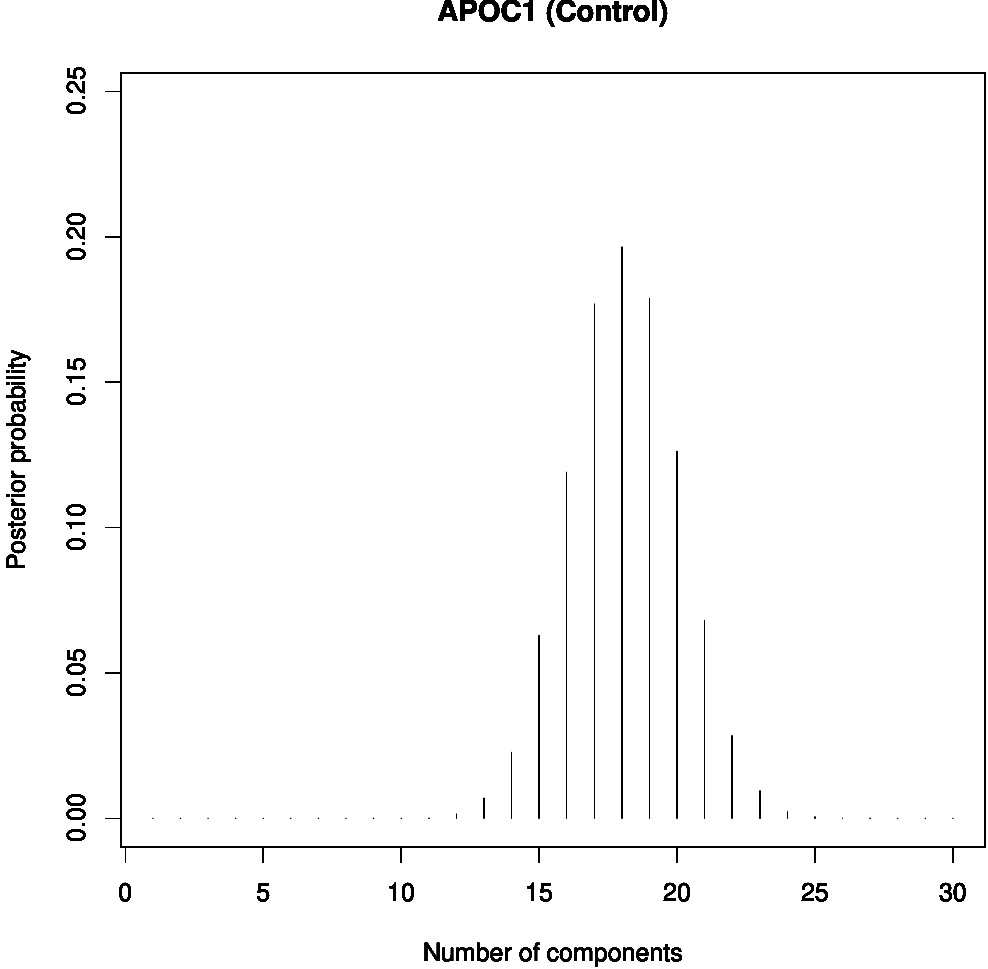}}
\hspace{2mm}
\subfigure[Posterior of $\tau_{19,1}$.]{ \label{fig:gene_case_eu_19}
\includegraphics[width=6cm,height=5cm]{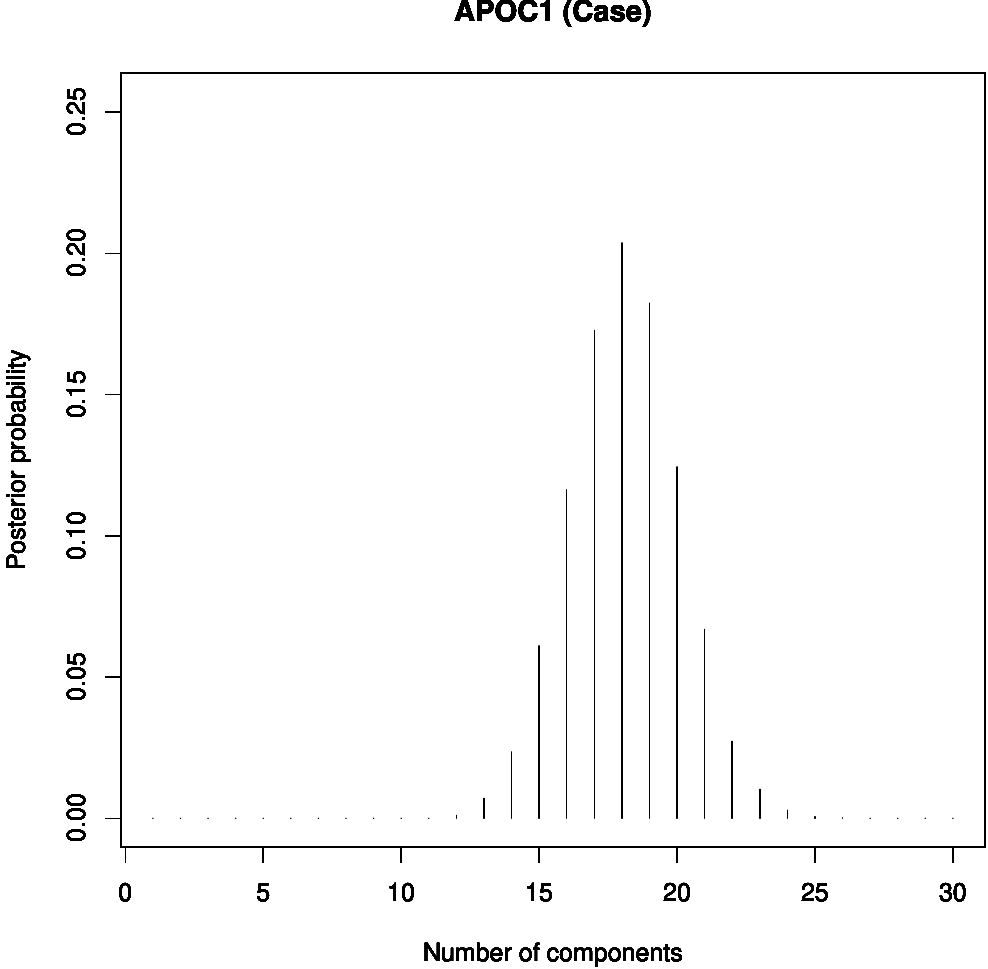}}\\
\vspace{2mm}
\subfigure[Posterior of $\tau_{27,0}$.]{ \label{fig:gene_control_eu_27}
\includegraphics[width=6cm,height=5cm]{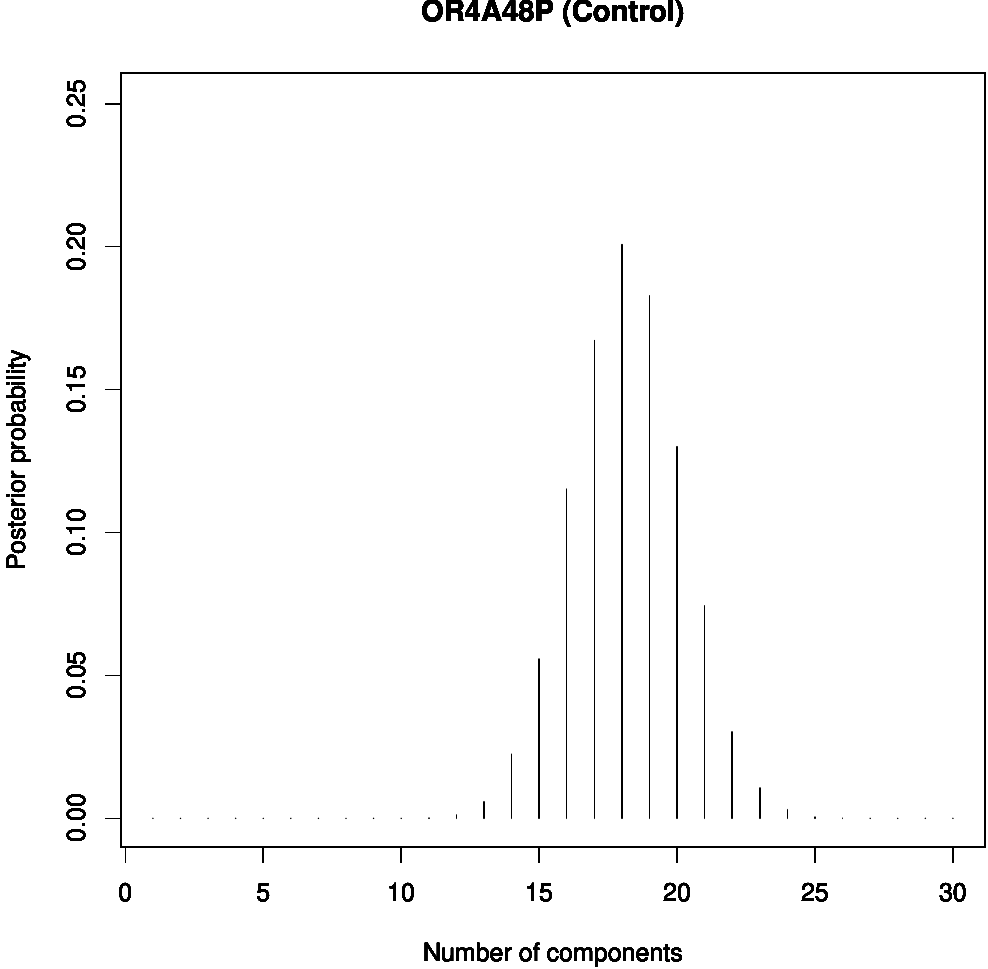}}
\hspace{2mm}
\subfigure[Posterior of $\tau_{27,1}$.]{ \label{fig:gene_case_eu_27}
\includegraphics[width=6cm,height=5cm]{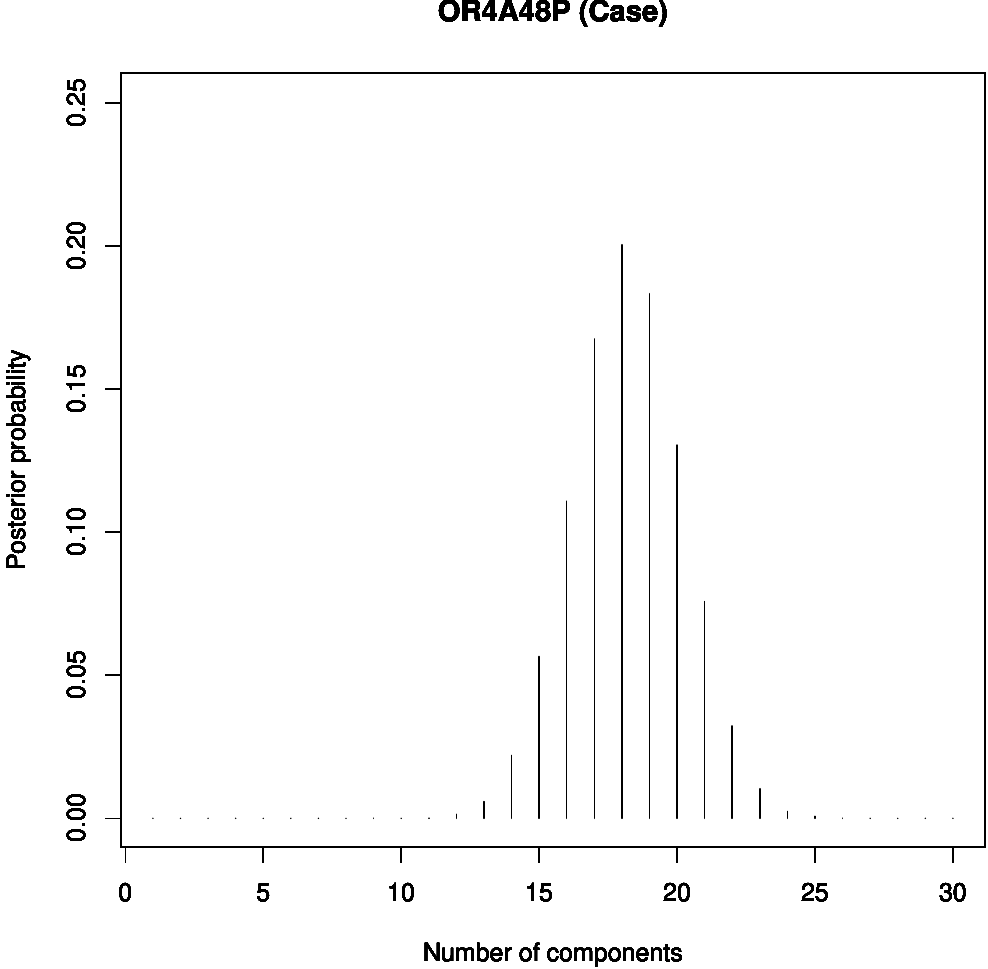}}\\
\caption{{\bf Posterior of number of components:} Posterior distributions of the number of distinct components $\tau_{j,k}$
for each pair ($j,k$); $j=6,19,27$; $k=0,1$. The left and right panels show the posteriors associated with cases
and controls, respectively.}
\label{fig:ggi_comp_realdata3}
\end{figure}

\newpage

\bibliographystyle{ECA_jasa}
\bibliography{irmcmc}

\end{document}